\documentclass[twocolumn]{aastex63}

\usepackage{longtable}
\usepackage{threeparttablex}
\usepackage{footmisc}
\usepackage{color}

\submitjournal{AAS Journals}

\shorttitle{SOAR TESS survey II}
\shortauthors{Ziegler et al.}
\begin{document}

\title{SOAR TESS Survey. II: The impact of stellar companions on planetary populations}

\correspondingauthor{Carl Ziegler}
\email{Carl.Ziegler@sfasu.edu}

\author[0000-0002-0619-7639]{Carl Ziegler}
\affil{Department of Physics, Engineering and Astronomy, Stephen F.
Austin State University, TX 75962, USA}

\author[0000-0002-2084-0782]{Andrei Tokovinin}
\affiliation{Cerro Tololo Inter-American Observatory, Casilla 603, La Serena, Chile} 

\author{Madelyn Latiolais}
\affil{Department of Physics, Engineering and Astronomy, Stephen F.
Austin State University, TX 75962, USA}

\author{C\'{e}sar Brice\~{n}o}
\affiliation{Cerro Tololo Inter-American Observatory, Casilla 603, La Serena, Chile} 

\author[0000-0001-9380-6457]{Nicholas Law}
\affiliation{Department of Physics and Astronomy, The University of North Carolina at Chapel Hill, Chapel Hill, NC 27599-3255, USA}

\author[0000-0003-3654-1602]{Andrew W. Mann}%
\affiliation{Department of Physics and Astronomy, The University of North Carolina at Chapel Hill, Chapel Hill, NC 27599-3255, USA}

\begin{abstract}

We present the results of the second year of exoplanet candidate host speckle observations from the SOAR TESS survey. We find 89 of the 589 newly observed TESS planet candidate hosts have companions within 3\arcsec, resulting in light curve dilution, that, if not accounted for, leads to underestimated planetary radii. We combined these observations with those from Paper I to search for evidence of the impact binary stars have on planetary systems. Removing the quarter of the targets observed identified as false-positive planet detections, we find that transiting planets are suppressed by nearly a factor of seven in close solar-type binaries, nearly twice the suppression previously reported. The result on planet occurrence rates that are based on magnitude limited surveys is an overestimation by a factor of two if binary suppression is not taken into account. We also find tentative evidence for similar close binary suppression of planets in M-dwarf systems. Lastly, we find that the high rates of widely separated companions to hot Jupiter hosts previously reported was likely a result of false-positive contamination in our sample.

\end{abstract}

\keywords{binaries: close; binaries: general; binaries: visual; planets and satellites: detection; planets and satellites: dynamical evolution and stability; planets and satellites: formation}

\section{Introduction} \label{sec:intro}

The Transiting Exoplanet Survey Satellite \citep[TESS,][]{tess} is searching nearly the entire sky for transiting planets around bright stars. Between 2018-19, TESS tiled the Southern ecliptic sky with 13 observed sectors. In mid-2019, TESS began a similar year-long Northern observing campaign. To date,\footnote{As of \today, from the NASA Exoplanet archive, available at \url{https://exoplanetarchive.ipac.caltech.edu/}} 120 planets have been confirmed from TESS detections from over 2500 candidate planetary systems.

The TESS lightcurves are contaminated with light from nearby stars due to the relatively large 21\arcsec\ pixels. This additional light reduces the depth of any real planetary transit, which is used to estimate the radius of the transiting body. A nearby source may also be an eclipsing binary (EB), with a deep transit that, when blended with a brighter star, may appear planetary in nature. Known nearby stars are used to determine a contamination ratio in the TESS input catalog \citep[TIC,][]{tic} which is then applied to correct the radii of planet candidates. This takes into account only stars from seeing-limited surveys, which are generally limited to separations greater than a few arcseconds. As approximately half of solar-type stars have stellar companions \citep{raghavan10}, and the on-sky separation binary distribution generally peaks at less than an arcsecond (e.g., at the median binary separation of $\sim$50 AU and a typical TESS planetary host distance of 200 pc, a binary will have a 0\farcs25 on-sky separation), most binaries are not accounted for in the TIC. Even \textit{Gaia DR2} is sensitive to binaries only at separations greater than $\sim$0\farcs7-1\farcs0 (it is not yet clear how \textit{Gaia EDR3} improves on this separation sensitivity, but final mission limits are simulated to be 0\farcs5 \citep{arenou17}).

Ground-based high-angular resolution imaging is therefore required to search planet candidate hosts for binarity (see \citet{matson19}). With limited resources for TESS follow-up observations (seeing-limited photometry and radial velocities), expedient imaging of planet candidate hosts is a valuable method for identifying potential false positives. Beginning in late 2018, the Southern Astrophysical Research telescope (SOAR) has been performing speckle observations of TESS planet candidates. Speckle imaging on SOAR typically reaches the diffraction limit on bright targets (V$<\sim$12), including most TESS targets, and is efficient, capable of up to 300 observations a night \citep{tokovinin18}. SOAR speckle imaging observes in the visible, in a similar passband to TESS, and can more accurately account for photometric contamination than typical NIR adaptive optics observations. The first results from this survey, covering 542 TESS targets with 117 detected companions, was recently published by \citet[hereafter Paper I]{ziegler20}. An additional 589 TESS targets observed by SOAR are presented in this work.

In addition to being a vital step towards planet confirmation and characterization, careful analysis of the properties and frequency of resolved planet-hosting binaries can provide insight into the role companion stars have in sculpting planetary systems. In particular, the observational evidence for a dearth of close stellar companions around planet-hosting stars is mounting, consistent with expectations from theory via multiple mechanisms \citep[e.g.,][]{quintana07, naoz12, kraus12}. RV surveys of giant planet hosts \citep{knutson13, wang14, ngo16} consistently find low close companion rates. In an AO survey of \textit{Kepler} targets, \cite{kraus16} detected far fewer close binaries than would be expected from field star statistics, a result well modeled by a reduction in binaries by approximately two-thirds at orbital separations below 47 AU. Paper I found similar suppression in close binaries around TESS planet candidate hosts. Recently, \citet{howell21} observed 186 TESS planet candidate hosts in high-resolution using speckle imaging. They also found fewer close binaries than would be expected from field star multiplicity rates, which they interpret as the binary distribution peaking at wider separations. A comprehensive model of binary suppression is being developed based on these observations \citep{moe19}.

\begin{figure*}
    \centering
    \includegraphics[scale=0.62]{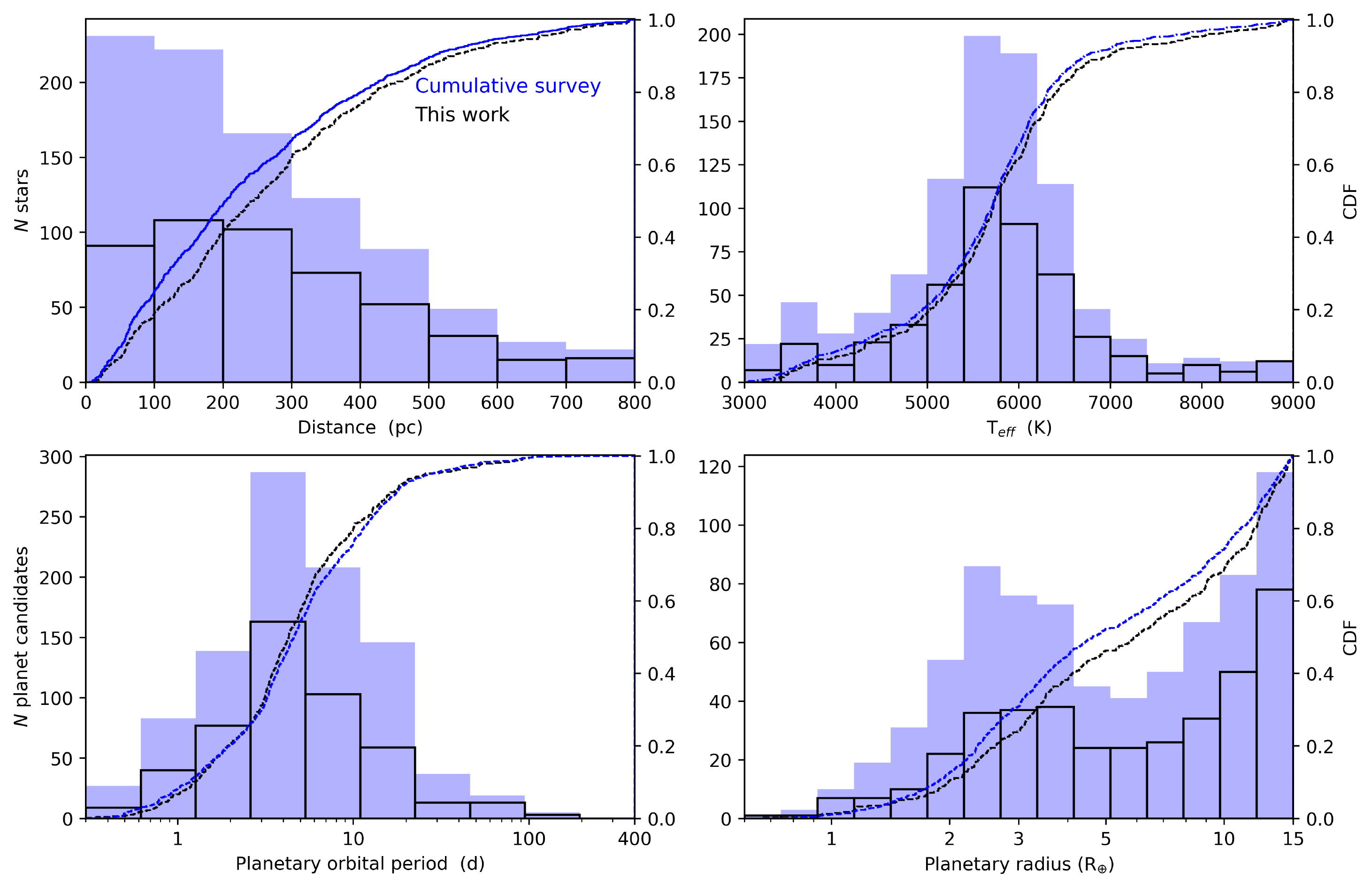}
    \caption{The properties of the 920 TESS planet candidate hosts observed by SOAR in this survey (from Paper I and this work), presented as a blue binned histogram with an overplotted cumulative density function. Targets identified as known or likely false positives, 211 in total, have been removed (see Section \ref{sec:fps}). The properties of only the 493 new, non-false positive TESS targets presented in this paper are also plotted in black. The properties of the host stars and planet candidates observed in Paper I and this work are generally similar.}
    \label{fig:histogram}
\end{figure*}

The close binary suppression reported in Paper I in TESS planet candidate hosts was significant (implying that nearly a quarter of solar-type stars in the galaxy are not able to host planets) but still may have been underestimating the true effect. The sample used for the analysis was not cleaned for known false positives. Approximately one-fourth of the targets in that analysis are now known to be false positives (145 of the 542)\footnote{As of \today, using the TFOPWG disposition for each star on the ExoFOP-TESS website, available at \url{https://exofop.ipac.caltech.edu/tess/}}, many having been identified as such in the interim from concerted community follow-up efforts.

The large set of false positive targets used in the Paper I analysis likely has a similar binary distribution to field stars, and likely contributed many of the observed close binaries. Also, \cite{tokovinin06} found that close eclipsing binaries have a high rate of tertiary companions, suggesting the false positives may have an even higher binary fraction than field stars. In this paper, we perform an analysis into the binary statistics of a cleaned sample of observations from the survey, including targets from Paper I and new observations presented here. 

It has also been suggested that giant close-in planets are preferentially found in systems with wide companions \cite{law14, wang15, ngo16, ziegler18b}, suggesting a dynamical interaction between the planet and nearby star led to planetary orbital migration. \cite{moe19} recently suggested these results were biased by contamination from nearby EBs, which have a high number of wide companions \citep{tokovinin10}. In our analysis in Paper I, we detected a high wide binary fraction for planet candidate hosts compared to field stars. The surplus of wide binaries was found to be solely in systems with the largest planet candidate hosts. In this paper, we investigate whether contamination from false positives was responsible for the observed wide binary fraction.

We begin in Section \ref{sec:observations} by detailing our observations and data analysis. We present the results of the survey in Section \ref{sec:results}, and update our analysis into the impact binaries have on the TESS planets in Section \ref{sec:analysis}. We discuss the results further in Section \ref{sec:discussion}. Finally, we conclude in Section \ref{sec:conclusions}.

\section{Observations and Analysis}\label{sec:observations}

\subsection{Target selection}

The targeted TESS planet candidates hosts (TESS objects of interest, or TOIs) were selected from the data releases available online at the TESS data release portal.\footnote{\url{https://tev.mit.edu/toi/}, account required for access.} Faint stars (typically, T$_{mag}>$13 mag) that are not well suited for speckle observations were not targeted; this magnitude limit reduces the number of late-type stars that are observed in this survey. The magnitude limit on the observations also biases our results towards more binaries (systems with two stars will be brighter than single systems); this bias will be accounted for in the forthcoming analysis. Previously confirmed planet hosts, primarily from the WASP \citep{wasp} and HATS \citep{hats} surveys, were targeted with lower priority, as these systems have been well studied in the past \citep[e.g.,][]{ngo15, evans16, evans18}. Lastly, stars that at the time had follow-up dispositions of false positive on TESS ExoFOP were not targeted.

To increase observing efficiency, target acquisition was improved using precise target coordinates, determined for each night with proper motions from Gaia DR2 \citep{gaia}, when available, and from the TIC \citep{tic} otherwise. As previously noted in \citet{arenou18}, we find that many targets with only two-parameter astrometric solutions in Gaia DR2 are actually close binaries.

The properties of the host stars and planet candidates observed are plotted in Figure \ref{fig:histogram}. The systems presented in this paper have similar properties to those observed in Paper I: a Kolmogorov-Smirnov test gives greater than 95\% probability that the distributions for stellar distance, effective temperature, planetary orbital period, and planetary radius are drawn from the same population.

\begin{figure*}
    \centering
    \includegraphics[scale=0.52]{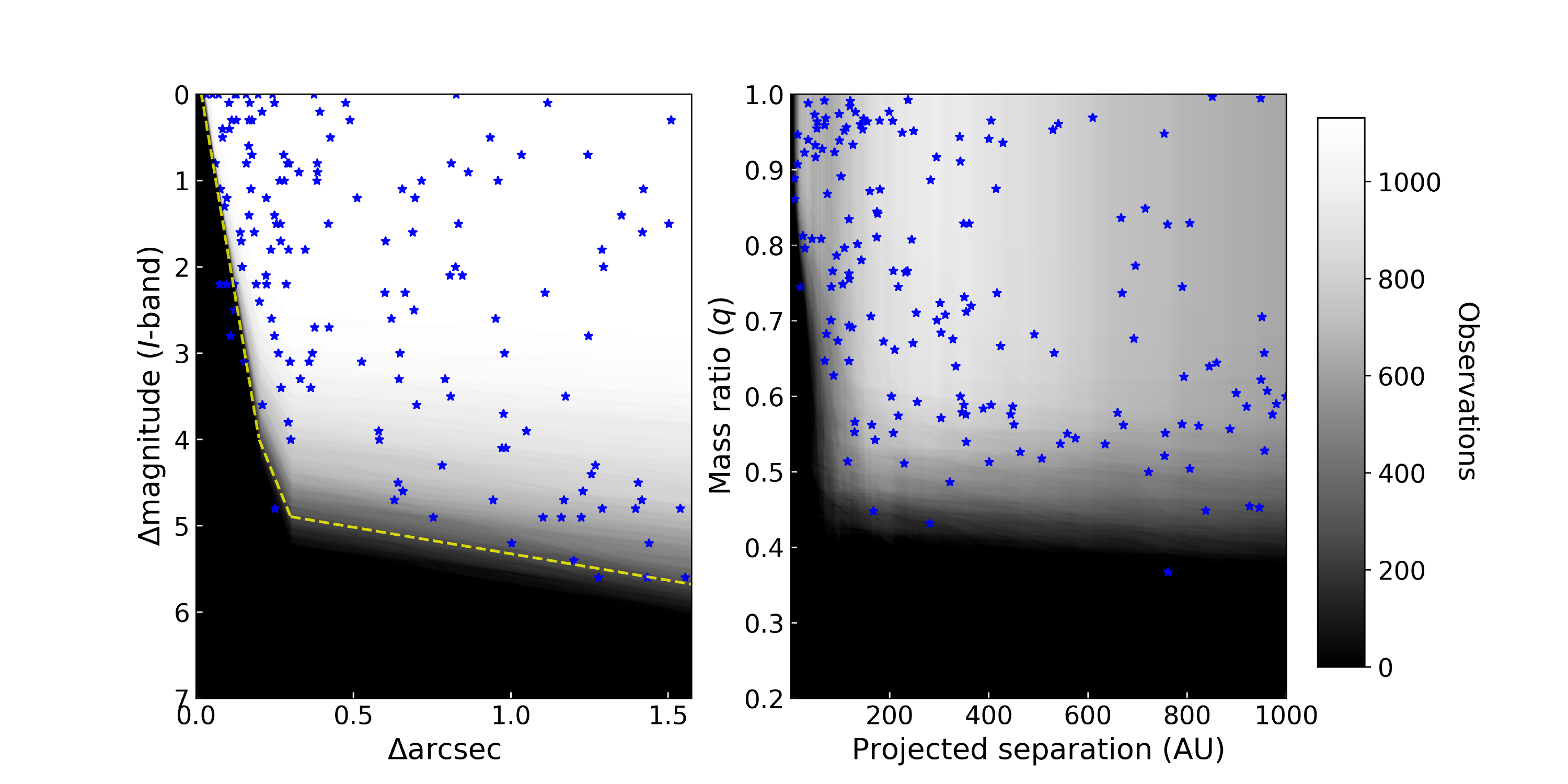}
    \caption{In the left panel, close companions ($\rho <$1\farcs55) to TESS planet candidate hosts detected by SOAR speckle imaging in Paper I and this work, plotted by the $I$-band magnitude difference and separation from the primary star. The average 5-$\sigma$ detection limits of the observations are plotted, trending from black (no observations are sensitive to binaries with that combination of separation and contrast) to white (all 1131 observations are sensitive to that combination). The yellow dashed line shows the median sensitivity of the survey. In the right, are companions at all separations detected with SOAR plotted using the projected separation from the host star and the inferred mass ratio assuming the companion is bound using the observed contrast.}
    \label{fig:binariesplot}
\end{figure*}

\subsection{SOAR observations and data reduction}

We observed 589 TESS planet candidate hosts with the high-resolution camera (HRCam) speckle imager on the 4.1-m SOAR telescope over eight nights in 2019-2020. The observation procedure and data reduction are described in \citet{tokovinin18} and in Paper I. Briefly, each observation consists of 800 frames split between two data cubes. Each frame is a 200$\times$200 binned pixels region of interest centered on the target star (6\farcs3 on a side at the pixel scale of 0\farcs01575 and 2$\times$2 binning), taken in approximately 11 s with an Andor iXon-888 camera. The resulting data cube is processed by a custom IDL script, which computes the power spectrum; a resolved multiple stellar system is revealed by characteristic fringes. Binary parameters (separation, position angle, and magnitude difference) are determined by modeling the power spectrum. Secondary stars also appear as mirrored peaks in the speckle auto-correlation function (ACF), the Fourier transform of the power spectrum, at the separation and position angle of the companion. To remove the 180-degree ambiguity inherent to the classical speckle interferometry, our pipeline also computes the shift-and-add (SAA) images, centered on the brightest pixel in each frame (this is sometimes called ``lucky imaging"). Relatively bright binaries with $\Delta m > 0.5$ mag often have their companions visible in the SAA images, allowing us to select the correct quadrant (these measurements are marked by the flag `q' in Table \ref{tab:whitelist} in the Appendix). In all other cases, the position angles are determined with a 180\degr   ambiguity. Figure 4 in \citet{tokovinin18} gives an example of typical speckle data, including the SAA image.  Observations were taken in the $I$-band ($\lambda_{cen}$=824 nm, $\Delta\lambda$=170 nm), which is approximately centered on the TESS bandpass.

Between Paper I and this work, 95$\%$ (1079) of the 1136 bright (T$<$13) planet candidates (those not identified as a false positives, see Section \ref{sec:fps}) that are accessible from the South (declination $<$ +20$^{\circ}$) have been observed at high angular resolution in the SOAR TESS survey.

We detail the observations in Table \ref{tab:whitelist} in the Appendix. The cumulative 5-$\sigma$ detection sensitivities are plotted in Figure \ref{fig:binariesplot}, along with the derived physical properties, projected separation from the host star and mass ratio (assuming the secondary star is bound to the primary), of the companions. The SOAR observations are generally sensitive to candidates with mass ratios down to 0.5 at projected separations of 10 to 500 AU. The sensitivity decreases at wider projected separations as the speckle function at the corresponding angular separations becomes decorrelated.

In addition to the TESS planet candidates, 167 suspected eclipsing binaries, flagged by the TESS TOI working group, were observed on 2019.03.03 UT. These observations were processed similarly to the TESS planet candidate hosts. Details of the observed eclipsing binaries and resolved companions are available in Table \ref{tab:ebs} in the Appendix. 

\subsection{Planet radius corrections}\label{sec:radiuscorrections}

The observed depth of a transit in a TESS light curve will be slightly smaller if additional light from a nearby source is present. This dilution of the transit results in an underestimated radius for the planet candidate. As described in detail in Paper I, we compute correction factors to the radius estimates derived from the TESS light curves for two scenarios: 1) the planet orbits the target star; and 2) the planet orbits the secondary star which is bound to the primary star. A third scenario not investigated here, where the two stars are unbound, requires an estimate of the radius of the foreground or background star. For the first scenario, the radius correction of the planet depends simply on the fraction of the light in the photometric aperture coming from the host star. For the second scenario, the radius of the companion star must be estimated. Assuming the star is bound, we use the radius of an appropriate fainter star within the Dartmouth stellar models \citep{dotter08}, using the $I$-band contrast SOAR observed in as a proxy for the TESS band observation.

The TIC includes a contamination ratio that takes into account stars within 10 TESS pixels of the target. This includes stars typically down to the limiting magnitude of the 2MASS \citep{skrutskie06} and APASS \citep{Henden09} catalogs ($T\sim17-19$). Using the list of detected close binaries to \textit{TESS} planet candidate hosts and their binary parameters, a custom Python script crossmatched each of their coordinates to stars in the TIC catalog. We find 30 companions  in the TIC had similar positions relative to the primary as was found in SOAR imaging ($\Delta\rho<0\farcs5$ and $\Delta \theta <$20\degr, or $[\Delta \theta\pm180^{o}]<$20\degr). The properties of these systems are available in Table \ref{tab:ticbinaries} in the Appendix.

We provide a correction factor for hosts, as in some cases the crossmatch between the TIC and the SOAR binary is ambiguous; however, we caution that the correction should be used judiciously. For all other systems, the contamination ratios reported should be used in addition to the TIC contamination ratio. In practice, the reported radius estimates of TESS planet candidates on the TESS data release portal and ExoFOP typically take into account flux contamination. The additional correction due to binaries detected by SOAR is the product of the original radius estimate and the radius correction factor reported in this work.

\subsection{System parameters}

The properties of the planet candidates were drawn from the ExoFOP TOI table, which is sourced from the NASA SPOC pipeline \citep{jenkins16}, which analyzes the 2-minute cadence data, or the MIT quick-look pipeline \citep{qlp}, which searches the full-frame images with 30-minute cadence. The host star properties are drawn from the TICv8 \citep{tic}. For 35 of the 1131 observed stars without distances in the TIC (which themselves are generally sourced from Gaia DR2 data), a crossmatch with the Gaia DR2 \citep{gaia2016} catalog was performed and the distances reported by \citet{bailerjones18} were used.

For 46 of the 98 systems without planetary radius estimates, we used the radius estimates generated by EXOFAST \citep{exofast} available on the ExoFOP website. The majority of the remaining systems without planetary radius estimates (42 of 52) have T$_{eff}$ estimates that are consistent with early-type stars (T$_{eff}>$7440), and would not have been used in the subsequent analysis which is focused on solar-type stars.

\subsubsection{Gaia DR2 Crossmatch}

A crossmatch between the detected companions from SOAR was performed with the Gaia DR2 catalogue \citep{gaia}, yielding 32 matches to SOAR detected companions. The parameters used to confirm the match between the nearby stars were similar to those used in the TIC crossmatch described in Section \ref{sec:radiuscorrections}.

These companions are all widely separated ($\rho>1\arcsec$). The separations measured by SOAR have a mean and median difference of 16 and 17 mas, respectively, compared to those reported in Gaia DR2. The disparity between the SOAR and Gaia separations increases slightly with separation, suggesting this is caused by measurement error, not orbital motion of the component stars. The average disparity in magnitude differences of binary components between SOAR and Gaia DR2 (Gp band) is 0.36 mag, and is likely due to the different passbands.  The majority of the fainter companion stars do not have Bp and Rp magnitudes from Gaia, and also exhibit larger magnitude errors than similarly faint single-stars due to blending with the primary star. The separation and contrast disparity is similar to those found in the target sample from Paper I. The properties of these systems are available in Table \ref{tab:gaiamatches} in the Appendix.

If both the primary and secondary stars have similar proper motions, it is highly likely that they would indeed be bound. Unfortunately, for all but a few of the companions which are present in the Gaia catalogue, no proper motion is reported for the secondary star. We expect this to improve with future releases. In addition, follow-up SOAR speckle imaging with several year temporal baselines for the earliest TESS planet candidates are on-going to detect proper motion shifts which can validate association.

\subsection{Identifying false positives} \label{sec:fps}

To improve our sample for analysis, we searched for TESS planet candidate hosts which were observed with SOAR but subsequently were flagged as planetary false positives (FPs). The TESS follow-up efforts are on-going, and typically planet candidates are identified as false positives due to the detection of a nearby EB through ground-based photometry or from a large radial velocity variation of the purported host star, inconsistent with an orbiting planetary-mass body.

Of the 1131 TOIs observed in Paper I and in this work, 211 (19$\%$) have a TESS follow-up working group disposition of FP.\footnote{As of \today.} In Paper I, the majority of observed targets had not yet been observed by ground-based facilities and the number of flagged false positives was significantly smaller (only 24 out of 542 targeted systems, or 4.4$\%$). In the interim, 91 additional contaminating FPs have been identified in the Paper I sample. Since flagged FPs are not observed, the false-positive contamination in the SOAR survey is slightly lower than that for all Southern accessible (declination $<$$+$10\degr) year-one TOIs (230 FPs out of 1092 TOIs, or 21.0$\%$). Southern TOIs from TESS year-three observations (beginning mid-2020) have not had sufficient time for ground vetting, and have a much lower FP rate (21 FPs out of 367 TOIs, or 5.7$\%$). Observations of known FPs  are removed from the subsequent analysis.  

Extremely large planet candidates are significantly more likely to be an eclipsing binary \citep{moe19}. Of the 168 systems hosting planet candidates with R$_{p}>15R_{\oplus}$, 82 (49$\%$) have been identified as false-positives. The validation tool \texttt{triceratops} \citep{triceratops} found that for a set of unclassified planet candidates with R$_{p}>8R_{\oplus}$, an order-of-magnitude more systems have false positive probabilities $>$0.95 (likely EB) than $<$0.05 (likely planet).\footnote{Conversely, the ratio of likely real small planets to false positives was approximately 7 to 1.} We, therefore, identify and remove the 86 remaining (non-FP) large planet candidate systems (R$_{p}>15R_{\oplus}$) from subsequent analysis due to the high likelihood of these systems being false positive EBs. 

\section{Results}\label{sec:results}

We detected 71 and 100 companions within 1\farcs5 and 3$\arcsec$ around 67 and 89 TESS planet candidate hosts, respectively, out of a total of 589 newly observed with speckle imaging on SOAR. The implied companion rates within 1\farcs5 and 3$\arcsec$ are 11.1$\pm$1.4$\%$ and 14.8$\pm$1.6$\%$, significantly lower than the rates found in Paper I (16.2$\pm$1.7$\%$ and 23.2$\pm$2.0$\%$ within 1\farcs5 and 3$\arcsec$, respectively).

Removing 211 flagged false positives from the combined survey sample (Paper I and this work), as described in Section \ref{sec:fps}, we find 130 companions within 1\farcs5 of 121 targets (a 13.1$\pm$1.1$\%$ companion rate) and 190 companions within 3$\arcsec$ of 169 targets (18.3$\pm$1.4$\%$). The companion rates for the 211 false-positive targets are significantly higher than the planet candidate hosts: 17.0$\pm$3.1$\%$ (36 companions) and 26.5$\pm$3.4$\%$ (58 companions around 56 stars) within 1\farcs5 and 3$\arcsec$, respectively. The companion rates for the 167 suspected EBs observed are similar to the false positives: 17.3$\pm$3.3$\%$ (29 companions around 28 stars) and 22.2$\pm$3.7$\%$ (39 companions around 36 stars) within 1\farcs5 and 3$\arcsec$, respectively.

\tabcolsep=0.12cm
\begin{tiny}

\begin{ThreePartTable}
\begin{TableNotes}
\footnotesize

\end{TableNotes}

\begin{longtable*}{ccccccccccc}
\caption{Nearby stars detected by SOAR to TESS planet candidate hosts \label{tab:binaries}}\\
\hline
\hline
\noalign{\vskip 3pt}  
\text{TOI} & \text{Separation} & \text{P.A.} & \text{Contrast} & \text{T$_{eff}$} & \text{Distance} & \text{Proj. Sep.}\tnote{a} & \multicolumn{2}{c}{Radius correction factor} & Prev. & WDS DD\\ 
 & \text{(\arcsec)} & (deg) & \text{($I$-band)} & \text{(K)} &  \text{(pc)} & \text{(AU)} & \text{(primary host)} & \text{(secondary host)} & det.? &  \\ [0.1ex]
\hline
\noalign{\vskip 3pt} 
\endfirsthead

\multicolumn{11}{c}
{\tablename\ \thetable\ -- \textit{Continued}} \\
\hline \hline
\noalign{\vskip 3pt} 
\text{TOI} & \text{Separation} & \text{P.A.} & \text{Contrast} & \text{T$_{eff}$} & \text{Distance} & \text{Proj. Sep.}\tnote{a} & \multicolumn{2}{c}{Radius correction factor} & Prev. & WDS DD\\ 
 & \text{(\arcsec)} & (deg) & \text{($I$-band)} & \text{(K)} &  \text{(pc)} & \text{(AU)} & \text{(primary host)} & \text{(secondary host)} & det.? &  \\ [0.1ex]
\hline
\noalign{\vskip 3pt}  
\endhead
\endfoot
\hline
\endlastfoot

212 & 0.9585 & 292.1 & 1.0 & 3332 &  &  & 1.182 & 1.478 & 1 \\ 
212 & 0.255 & 358.0 & 0.5 & 3332 &  &  & 1.277 & 1.268 & 1 \\ 
319 & 0.1196 & 76.3 & 2.2 & 4349 & 182 & 21 & 1.064 & 2.821 & 1 \\ 
330 & 2.7559 & 269.9 & 2.8 & 4314 & 200 & 551 & 1.037 & 3.705 & 3\\ 
479 & 2.2772 & 177.2 & 5.0 & 5600 & 194 & 441 & 1.005 & 9.611 & 3 & WSP 49AB\\ 
489 & 0.9428 & 139.6 & 4.7 & 6096 & 506 & 477 & 1.007 & 8.627 & 1 \\ 
641 & 2.8132 & 282.6 & 5.3 & 5646 & 134 & 376 & 1.004 & 11.022 & 1 \\ 
641 & 0.37 & 188.8 & 0.1 & 5646 & 134 & 49 & 1.383 & 1.385 & 1 \\ 
641 & 3.5923 & 86.7 & 4.4 & 5646 & 134 & 481 & 1.009 & 7.317 & 1 \\ 
728 & 2.1892 & 309.9 & 1.5 & 5648 & 172 & 376 & 1.119 & 2.134 & 1 \\ 
739 & 0.12 & 40.6 & 2.5 &  & 340 & 40 & 1.049 & 3.264 & 1 \\ 
781 & 1.2562 & 45.2 & 4.4 & 5227 & 341 & 428 & 1.009 & 7.523 & 1 \\ 
801 & 1.9672 & 129.2 & 7.1 & 6283 & 71 & 139 & 1.001 & 25.164 & 1 \\ 
936 & 0.7154 & 131.7 & 1.0 & 4320 & 238 & 170 & 1.182 & 1.843 & 1 \\ 
940 & 2.5473 & 131.7 & 1.8 & 5100 & 157 & 399 & 1.091 & 2.358 & 3 \\ 
940 & 0.1097 & 55.3 & 0.1 & 5100 & 157 & 17 & 1.383 & 1.366 & 1 \\ 
963 & 3.6555 & 222.7 & 4.1 & 5814 & 203 & 742 & 1.011 & 6.026 & 1 \\ 
981 & 1.5537 & 82.4 & 5.6 & 6711 & 297 & 461 & 1.003 & 12.644 & 1 \\ 
1001 & 1.5971 & 1.1 & 5.3 & 7070 & 295 & 471 & 1.004 & 10.56 & 1 \\ 
1006 & 0.2547 & 341.3 & 1.5 & 6616 & 182 & 46 & 1.119 & 2.13 & 1 \\ 
1006 & 3.5697 & 300.5 & 0.4 & 6616 & 182 & 649 & 1.301 & 1.492 & 1 \\ 
1032 & 1.3517 & 276.5 & 1.4 & 10395 & 587 & 793 & 1.129 & 1.99 & 3 & B 1166 \\ 
1043 & 1.6637 & 332.8 & 2.5 & 6902 & 208 & 346 & 1.049 & 3.256 & 3 \\ 
1057 & 1.8025 & 24.5 & 5.6 & 5599 & 98 & 176 & 1.003 & 12.643 & 3 \\ 
1060 & 0.269 & 166.7 & 3.4 & 5687 & 128 & 34 & 1.022 & 4.774 & 1 \\ 
1081 & 0.2932 & 149.6 & 1.8 & 6027 & 354 & 103 & 1.091 & 2.398 & 1 \\ 
1095 & 2.014 & 194.7 & 2.2 & 7066 & 582 & 1172 & 1.064 & 2.685 & 3 \\ 
1099 & 7.6379 & 310.5 & 2.3 & 4867 & 23 & 175 & 1.058 & 2.999 & 3 \\ 
1101 & 3.5573 & 235.7 & 2.5 & 5803 & 262 & 932 & 1.049 & 2.991 & 3 \\ 
1114 & 0.8099 & 94.8 & 0.8 & 8048 & 174 & 140 & 1.216 &  & 1 \\ 
1208 & 5.6059 & 298.2 & 1.6 & 5626 & 134 & 751 & 1.109 & 2.215 & 1 \\ 
1215 & 1.7263 & 84.7 & 0.1 & 3751 & 34 & 58 & 1.383 & 1.342 & 1 \\ 
1217 & 1.0489 & 244.8 & 3.9 & 6354 & 463 & 485 & 1.014 & 5.69 & 1 \\ 
1263 & 2.6501 & 116.2 & 3.8 & 5098 & 46 & 121 & 1.015 & 5.509 & 3 \\ 
1894 & 0.8258 & 42.0 & 0.0 & 5611 & 178 & 146 & 1.414 & 1.414 & 3 \\ 
1943 & 3.6514 & 114.3 & 4.1 & 5742 & 129 & 471 & 1.011 & 6.233 & 3 \\ 
1986 & 0.1427 & 54.5 & 1.7 & 8613 & 1228 & 175 & 1.1 &  & 1 \\ 
1986 & 2.359 & 94.1 & 3.6 & 8613 & 1228 & 2896 & 1.018 &  & 2 & RST 415AB\\ 
1990 & 3.7585 & 320.9 & 4.8 & 6075 & 171 & 642 & 1.006 & 9.273 & 3 \\ 
1992 & 3.184 & 202.2 & 0.5 & 6736 & 375 & 1194 & 1.277 &  & 3 \\ 
2193 & 1.885 & 124.0 & 3.8 & 6079 & 337 & 635 & 1.015 & 5.748 & 3 \\ 
2195 & 3.3534 & 210.2 & 4.8 & 5296 & 174 & 583 & 1.006 & 8.654 & 3 \\ 
2201 & 1.0028 & 159.0 & 5.2 & 6252 & 252 & 252 & 1.004 & 10.526 & 1 \\ 
2221 & 1.7586 & 127.9 & 1.5 & 3588 & 9 & 15 & 1.119 & 2.106 & 1 \\ 
2226 & 1.2289 & 23.5 & 4.6 & 5828 & 164 & 201 & 1.007 & 7.556 & 1 \\ 
2231 & 1.5088 & 49.4 & 0.3 &  & 175 & 264 & 1.326 &  & 2 & I 1423\\ 
2232 & 2.4207 & 239.8 & 0.1 &  & 571 & 1382 & 1.383 &  & 2 & I 1043\\ 
2307 & 2.9343 & 205.8 & 4.4 & 5465 & 470 & 1379 & 1.009 & 7.374 & 3 \\ 
2310 & 0.6874 & 334.5 & 1.6 & 5266 & 552 & 379 & 1.109 & 2.348 & 1 \\ 
2311 & 0.2232 & 130.9 & 2.2 & 5104 & 264 & 58 & 1.064 & 2.733 & 1 \\ 
2312 & 0.128 & 135.2 & 0.0 & 5590 & 490 & 62 & 1.414 & 1.414 & 1 \\ 
2314 & 1.2239 & 16.8 & 4.9 & 6140 & 540 & 660 & 1.005 & 9.45 & 1 \\ 
2326 & 0.3829 & 131.3 & 1.0 &  & 154 & 58 & 1.182 &  & 2 & RST 2216\\ 
2335 & 3.8176 & 54.8 & 3.5 & 6004 & 439 & 1675 & 1.02 & 4.903 & 1 \\ 
2362 & 2.2124 & 259.0 & 2.0 & 6104 & 634 & 1402 & 1.076 & 2.66 & 3 \\ 
2387 & 0.6202 & 340.2 & 2.6 & 5813 & 396 & 245 & 1.045 & 3.119 & 1 \\ 
2409 & 0.0759 & 156.2 & 1.1 & 5017 & 187 & 14 & 1.168 & 1.884 & 1 \\ 
2417 & 1.8554 & 285.3 & 2.9 & 5485 & 257 & 476 & 1.034 & 3.789 & 3 \\ 
2419 & 1.72 & 331.4 & 3.4 & 5924 & 245 & 421 & 1.022 & 4.641 & 3 \\ 
2422 & 0.8641 & 140.9 & 0.9 & 6159 & 100 & 86 & 1.199 & 1.785 & 2 & RST 4182\\

\end{longtable*}
\textbf{Table 1 Notes. -- }Columns (1-4) gives the properties of companions to TOIs detected by SOAR. Uncertainties for these measurements and the observation epoch are provided in Table \ref{tab:whitelist}. Columns (5-6) gives the effective temperature and distance to the TOI given in the TIC \citep{tic}. Column (7) gives the projected separation of the companion (assuming it is physically associated with the primary), derived from the on-sky separation measured by SOAR and the distance to the star. Columns (8-9) give the radius correction factor for hosted planets in each system due to the contamination from the detected star in the scenario where the primary is the planetary host and the scenario in which the physically associated secondary is the planetary host. Column (10) is a flag denoting a previous detection of each companion. The flags are \textbf{(1)}: new pair, contamination not included in the TIC; \textbf{(2)}: known pair, contamination not included in the TIC; \textbf{(3)}: known pair, contamination included in the TIC. Column (11) provides the discoverer designation code if the companion is in the Washington Double Star Catalog maintained by the USNO. Explanations for codes are available at \url{https://www.usno.navy.mil/USNO/astrometry/optical-IR-prod/wds/WDS}. \\
\end{ThreePartTable}
\end{tiny}

The properties of the detected companions presented in this work are plotted in Figure \ref{fig:binariesplot}, along with the average detection sensitivities from all new observations which are detailed in Table \ref{tab:whitelist} in the Appendix. We include the radius correction factors for planets in these systems, whether they orbit the primary or secondary star. The auto-correlation functions of resolved systems showing the position of the companions are shown in Figures \ref{fig:grid1}, \ref{fig:grid2},  \ref{fig:grid3}, and \ref{fig:grid4} in the Appendix. Resolved binaries to known false positives are detailed in Table \ref{tab:fp_binaries} in the Appendix.

The results of each night's observations were processed within a week and posted on the TESS Exoplanet Follow-up Observing Program (ExoFOP) webpage\footnote{\url{https://exofop.ipac.caltech.edu/tess/}} to aid in confirmation of the planet candidates.

\subsection{Effect of binary dilution on TESS planet candidates}

The presence of a nearby star means that the planet candidate, even if real, is larger than the original estimate based on the diluted TESS light curve. In general, the identity of the host star is also ambiguous. A method for deriving correction factors for either host scenario was presented in Section \ref{sec:radiuscorrections}, and the individual radius corrections for each resolved system are given in Table \ref{tab:binaries}. Twenty of the companions presented in this work were present in the TIC, and therefore are already included in the contamination ratio used to correct the planetary radius estimate. These binaries are listed in the Appendix in Table \ref{tab:ticbinaries}. We provide correction factors for these systems in case the crossmatch is incorrect, however, these values are not used in the subsequent analysis. 

The mean radius corrections for the 89 systems with companions within 3\arcsec (excluding those with companions in the TIC) are 1.11 and 2.75 for the primary host and secondary host scenarios. Both values are similar to those found in Paper I (1.11 and 2.55, respectively), which are in line with the correction factors found from \textit{Kepler} planet candidates \citep{ciardi15, hirsch17, ziegler18a}.

In the combined sample with the 297 known and likely (R$_{p}>15R_{\oplus}$) false positives removed, we find an average radius correction of 1.08 for the primary host scenario, and 3.54 for the secondary host scenario. The latter value is slightly higher than for the contaminated sample, due to the companion stars in the FP systems being on average slightly brighter than for planet candidate systems. It is unclear why this is the case, but it may be due to the high number of tertiary companions found to close EBs, as discussed further in Section \ref{sec:analysis}. 

For \textit{Kepler} planetary systems, a relatively homogeneous stellar sample located in a small section of the sky, a separation of 1\arcsec was determined to be a useful discriminator between likely bound and likely unassociated nearby stars \citep{horch14}. Finding a similar separation cutoff for likely bound stars to TESS systems may prove elusive, as the host stars lie in fields across the entire sky with a wide range of stellar densities (leading to more or fewer nearby unassociated stars). However, the proximity of the TESS systems would, on the whole, increase the average separation of bound companions. Using the \textit{Kepler} discriminator value of 1\arcsec of our targets, we find radius corrections for the planet candidates of 1.09 and 2.45 for the primary and secondary hosts scenarios. Since these values are slightly lower than those found for our entire sample, this suggests that a larger fraction of the high-contrast companions at wider separations are not bound compared to close companions.

\section{Stellar companions impact on transiting planets}\label{sec:analysis}

A close binary star system presents many theoretical challenges to planetary formation and survival. In Paper I, a significant reduction in close companion rates for the planet candidate systems was found compared to field stars. Conversely, many more wide binaries were found orbiting planet candidate hosts. We revisit the analysis here with a significantly larger sample that has a higher fraction of real planet-hosting stars (from the identification and removal of known and suspected false positives) to determine the strength or even validity of these previously observed trends.

\begin{figure*}
    \centering
    \includegraphics[scale=0.52]{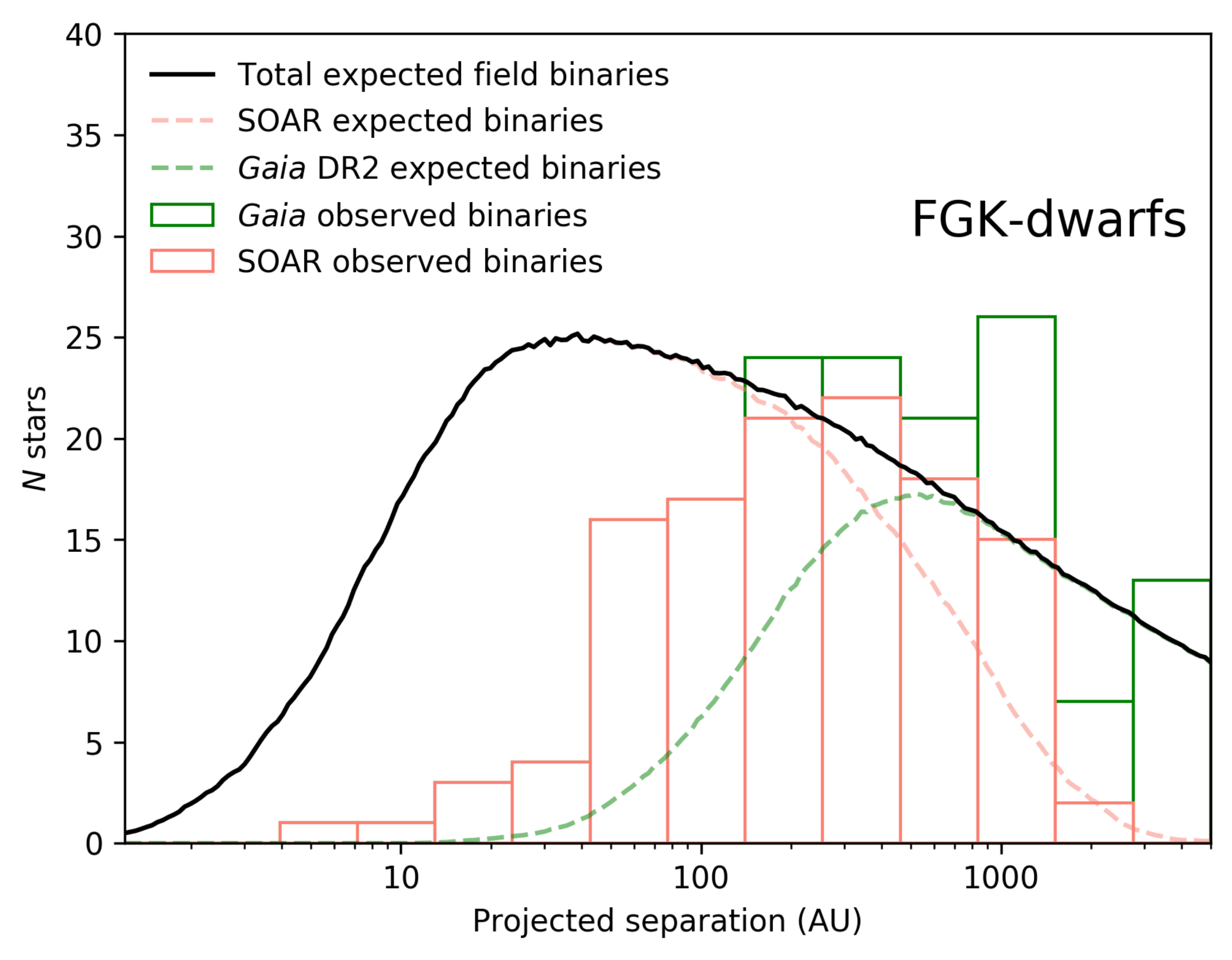}
    \caption{In red and green, the number of observed companions from SOAR and in Gaia DR2 for solar-type TESS planet candidate hosts in logarithmic bins of projected separation of 0.25 dex width. Companions found by both SOAR and Gaia are included in the SOAR sample. In black is the expected distribution from a multiplicity study of field stars \citep{raghavan10}, combining both field binaries that would be detected by SOAR and Gaia. The expected binaries from SOAR and Gaia, individually, are also plotted. These distributions take into account the detection sensitivity of both SOAR and Gaia. The observed distribution shows a clear paucity of TESS planet candidate host binaries at projected separations less than $\sim$40 AU compared to the field stars, and are consistent with field expectations at wider separations.}
    \label{fig:tessvsfield}
\end{figure*}

\subsection{Preparation of sample}\label{sec:prepseample}

For this analysis, we must cull the sample of 1131 observed TOIs using several parameters. First, we remove the 297 stars that are either flagged false positives or giant planet candidates, as discussed in Section \ref{sec:fps}. Second, we remove 117 stars with T$_{eff}$ in the TIC inconsistent with an FGK-type star (i.e., T$_{eff}>7400$ K and $<3900$ K), using the relations of \citet{pecaut12}. We also remove the 10 FGK stars that host planet candidates without radius estimates. Finally, we remove binaries with contrasts indicating mass ratios $q<0.4$ (approximately equivalent to $\Delta I>$5.1). Removing the lowest mass companions has a minimal effect on the analysis as generally these systems are too faint for detection (see Figure \ref{fig:binariesplot}), however, we do caution that our analysis is only strictly valid for stellar companions with $q>0.4$. These systems, with high magnitude differences, are significantly more likely to be chance alignments based on the analysis in Paper I. The final sample consists of 655 stars, compared to a sample of 455 targets used in the Paper I analysis (which included 91 targets later identified as false positives).

As the SOAR speckle imaging is incomplete at large angular separations (typically $\rho>$1\farcs5), we supplement the SOAR observations with common proper motion pairs found in Gaia DR2 \citep{gaia}, which is generally complete within this separation range \citep{roboaogaia}. We search the catalog around the target star to an on-sky angular separation star consistent with a projected binary separation of 5000 AU. Again, binaries with contrasts consistent with mass ratio $q<0.4$ were removed. The properties of these companions are provided in the Appendix in Table \ref{tab:gaiabinaries}.

%
%

\subsection{Multiplicity of solar-type TESS planet candidate hosts}\label{sec:multiplicity}

As in Paper I, our analysis strategy is to compare the frequency of planet candidate hosting binaries at different separations to that seen in field stars. To do that, we populate binaries found in a simulated survey of field stars using the properties of the observed TOIs. We use the binary properties for FGK stars found by \citet{raghavan10}: a flat eccentricity distribution, a log-normal period distribution (with a mean of log $P_{day}$ = 5.03, corresponding to an orbital semi-major axis of approximately 50 AU, and $\sigma_{log P}$ = 2.28), and a nearly uniform mass ratio distribution (with a sharp increase of near-equal mass ratio pairs). We follow the procedures of \citet{kraus16} to account for projection effects, Malmquist bias, and the detection limits of our survey (i.e., we use the measured contrast curves, seen in Figure \ref{fig:binariesplot}, to determine whether SOAR could detect each simulated binary).

For each solar-type star observed in our survey, a Monte Carlo model was constructed to determine the expected number of binary companions at a range of projected separations between 1-5000 AU. In each of 10$^{5}$ iterations, there was a 33\%$\pm$2\%, 8\%$\pm$1\%, and 3\%$\pm$1\% probability that one, two, or three companion stars would be populated, respectively (the observed multiplicity of solar-type stars). Since binaries are over-represented in flux-limited surveys \citep{schmidt68}, we correct for Malmquist bias by adjusting this probability by an additional factor equal to the fractional volume excess in binaries due to their relative brightness, V$_{bin}$/V$_{single}$. The period, eccentricity, and mass ratio of these binaries were drawn from the distributions reported in \citet{raghavan10}. The period was converted to a semi-major axis using the TIC estimated stellar masses. We select uniformly distributed values for the cosine of inclination, the position angle of the ascending node, the longitude of periastron, and the time of periastron passage. Finally, the instantaneous separation was projected to the distance to the primary star as reported in Gaia DR2. The mass ratio was converted to an approximate magnitude contrast using the relations in \citet{kraus07}, and possible detection by SOAR speckle imaging and Gaia DR2 was determined using the measured sensitivity limits and the companion's contrast and separation. The simulated companion was considered detected or not detected if it fell above or below the measured 5$\sigma$ contrast curve for the given target observation, as shown in Figure \ref{fig:binariesplot}. We use the ratio of non-detected binaries to the total number of binaries at each separation to determine a completeness correction due to limitations in the ability to resolve close or wide companions. The model was also run for the 297 systems removed from the planet candidate sample (known and likely false positives) and for the set of 167 targeted EBs. Both datasets and simulations are limited to mass ratios $q>0.4$, similar to the planet candidate hosting sample.

The resulting distributions of observed binaries among TESS planet candidate hosts from SOAR and Gaia compared to the expected number derived for field stars are shown in Figure \ref{fig:tessvsfield}. The uncertainty in the expected number of observed binaries at each separation range is derived from the spread of binaries in the simulated surveys, which propagates the field binary rate uncertainties reported by \citet{raghavan10}. The observed companion rate to the TESS planet candidates as a function of projected separation was determined by dividing the number of observed binaries by the total number of stars observed. Few close binaries are found and the number of wide binaries is consistent with field rates, a departure from the enhanced wide binary fraction seen in Paper I. We address each of these separation ranges in turn.

\subsubsection{Suppression of planet-hosting close binaries}\label{sec:closebinarysuppression}

\begin{figure}
    \centering
    \includegraphics[scale=0.32]{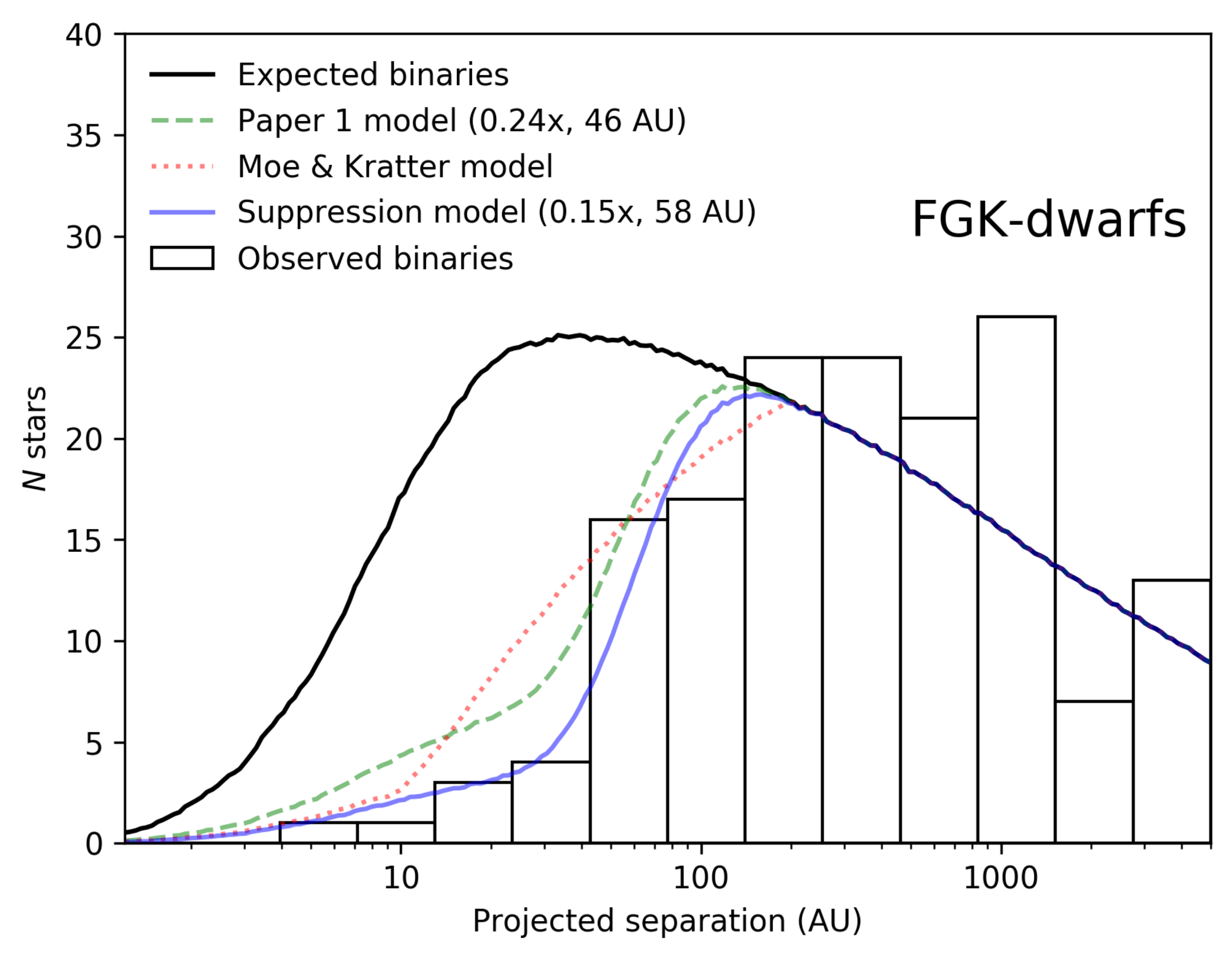}
    \caption{The observed, with SOAR and Gaia, and expected number of binaries hosting planet candidates with three suppression models. First, is the suppression model from Paper I, fit to a sample with significant false-positive contamination. Next is the model adopted by \citet{moe19}, fit to the results from several studies (including Paper I). Finally, the best fit two-parameter model to the data, signifying a reduction in binaries by a factor of 0.15  at physical separation less than 58 AU.}
    \label{fig:suppression}
\end{figure}

Close binary suppression has been observed in \textit{Kepler} targets by \citet{kraus16} and was observed for TESS planet candidates in Paper I. In the larger sample of planet candidates with a significant number of false positives removed, the lack of binaries at close separations is even more stark. Without removing known and likely FPs, 32 binaries are detected at a projected separation of less than 50 AU. After removal, just 11 such close binaries are detected. From our model using field star statistics, we would expect 83$\pm$8 close binaries to be detected, a nearly 9$\sigma$ discrepancy with observations of planet candidate hosting stars.

In Paper I, we modeled the close binary suppression as a reduction in the companion rate for planet candidate hosts by a factor of 0.24$\times$ cutting on at physical separations below 46 AU. This was in agreement with the model for \textit{Kepler} targets by \citet{kraus16}. Recently, \citet{moe19} combined results from multiple studies and adopted a functional form for binary suppression, S$_{bin}(a)$.  In this model, binaries hosting S-type planets (i.e., not circumbinary) with a$<$1 AU are fully suppressed. S$_{bin}(a)$ increases to 0.15 at 10 AU and to 1.0 (no suppression) at 200 AU, interpolated with respect to log $a$.

\begin{figure}
    \centering
    \includegraphics[scale=0.4]{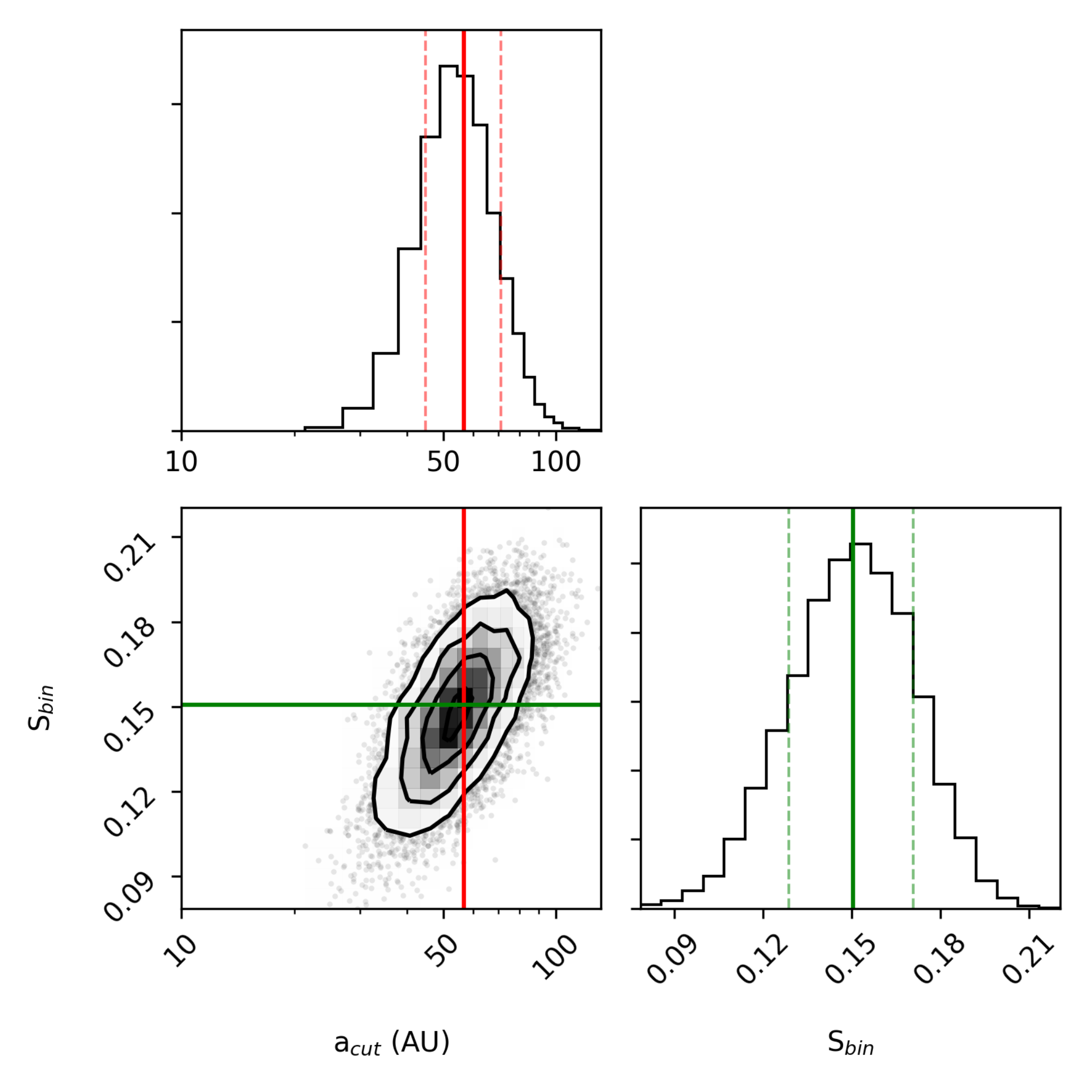}
    \caption{The distributions of suppression factors and semi-major axis cuts from 10$^{6}$ chains of an MCMC analysis to model the observed close binary suppression seen in systems with TESS planet candidates. Solid lines indicate the median value of each distribution (a$_{cut}$=58 AU and S$_{bin}$=0.15) and dashed lines mark the 68$\%$ confidence interval. The two parameters are correlated, such that less suppression is required if the semi-major axis cutoff is smaller, and vice versa.}
    \label{fig:suppressioncorner}
\end{figure}

We find that neither the model from Paper I or the \citeauthor{moe19} model qualitatively matches the observed suppression in our data (see Figure \ref{fig:suppression}). The Paper I model predicts 31$\pm$5 and 54$\pm$6 binaries to be detected at projected separations less than 50 AU and 100 AU, respectively. The \citet{moe19} model predicts 28$\pm$4 and 51$\pm$6 binaries at those separation ranges. Both estimates are significantly more than what was observed at close projected separations (11 binaries within 50 AU) and slightly more than was observed within 100 AU (33 binaries).

To determine a model that better fits the observed data, we begin, as in Paper I, with modifying our two-parameter step model. The choice of this model is due to its simplicity. As noted by \citet{kraus16} and \citet{moe19}, the step model is only a useful construction and certainly not physically valid. With only two free parameters, the closeness of its fit to the observed data provides a useful statistical model as well as insight into the magnitude of binary suppression of planet formation within this close separation range.

We performed a Markov chain Monte Carlo (MCMC) analysis to explore 10$^{6}$ possible values for $S_{bin}$ and $a_{cut}$, seeking to reduce the $\chi^{2}$ goodness of fit to the observed distribution. The resulting distributions are shown in Figure \ref{fig:suppressioncorner}. We find the optimal values for the suppression model with 68$\%$ credibility ranges to be $S_{bin}$=15$^{+3}_{-2}\%$ and $a_{cut}$=58$^{+11}_{-10}$ AU, shown in Figure \ref{fig:suppression}. The values of $S_{bin}$ and $a_{cut}$ are strongly correlated (e.g., a smaller value of $S_{bin}$ corresponds to a larger value of $a_{cut}$).

Applying the best-fit step suppression model to the expected binary distribution results in 11$\pm$1 and 32$\pm$2 binaries expected within projected separations of 50 and 100 AU, in agreement with observations. This model also predicts a peak binary separation for planet hosts at approximately 100 AU, in agreement with the findings of \citet{howell21} for TESS targets.

\subsubsection{The closest binaries hosting planet candidates}\label{sec:closestbinaries}

This suppression model results in a substantial quenching of binaries with projected separations of less than 40 AU. Nevertheless, 8 companions were observed within this range, although projection effects mean their true separations may be much greater. The properties of these systems may provide insight into the conditions needed for planets to survive in close binary systems. In this section, we, therefore, discuss each of these systems in order of increasing projected separation. We note that the real separation for these systems will be higher than in projection, on average by a factor of 1.26 for systems with no orbital information \citep{fischer92}.

TIC235037761 (TOI-131) is the system with the closest binary companion, a $\Delta$I=1.0 at an angular separation of 0\farcs07. At its distance of 58 pc, this results in a projected separation of 4.1 AU (a separation less than Jupiter's orbital distance). The planet candidate has a distinct V-shaped transit, increasing the likelihood that this in fact a triple star system, with the primary being a close eclipsing binary star. Its current TFOPWG disposition is anomalous planet candidate (APC).

TIC321068176 (TOI-2409) has a $\Delta$I=1.1 companion at 0\farcs75 on-sky separation, which at its distance of 187 pc corresponds to a projected separation of 14 AU. Little follow-up has been done on this target, but the TESS transit lightcurve is flat-bottomed and the expected radius is consistent with a planet even with dilution corrections due to the nearby star: 7 R$_{\oplus}$ if orbiting the primary or 11.5 R$_{\oplus}$ if orbiting the secondary.

TIC307610438 (TOI-831) has a 0\farcs167 companion (14 AU projected separation at its distance of 87 pc) with $\Delta$I=1.4. The transits show significant chromaticity in transit depth, varying from less than 0.1$\%$ in V band to 1$\%$ in the near-infrared Y-band. While its disposition is APC, TOI-831 is most probably a blended EB.

TIC277683130 (TOI-138) has a $\Delta$I=1.1 companion at 0\farcs10 separation (a projected physical separation of 19 AU at 191 pc distance). It has a large expected planetary radius (13$R_{\oplus}$) and a V-shaped transit, making it probable this is, in fact, an eclipsing binary system. It's TFOPWG disposition is APC.

TIC389753172 (TOI-319) has a $\Delta$I=2.2 companion at 0\farcs12, corresponding to 22 AU projected separation at 182 pc. It has a depth of 0.5$\%$ in Sloan-g and 2.7$\%$ in Sloan-zs, a significant enough difference for it to be considered a blended EB by the seeing-limited photometry follow-up working group.

TIC300293197 (TOI-211) has a near-equally bright companion at 24 AU projected separation (0\farcs20 on-sky separation at 121 pc distance). This target is disposed as an APC, but is almost certainly a blended EB due to significant chromaticity in its transit depths.

TIC420049884 (TOI-462) has a companion at a projected separation of 34 AU (0\farcs167 at 205 pc). This $\Delta$I=0.3 mag system has been observed with the ShARCS AO system on the 3-m Shane telescope, and the AstraLux lucky imager 2.2-m CAHA telescope. It is not clear if the companion was detected in these observations.  The MuSCAT2 team observed this target on 2019.09.24 UT in three bands and observed significant chromaticity in the transit depth, perhaps indicative of an EB. An LCO 1-m telescope at McDonald Observatory observed TOI-462 on 2019.12.16 UT and detected a potential nearby EB on a 15\arcsec neighbor.

The last binary with projected separation less than 40 AU is TIC101230735 (TOI-1060), which has a $\Delta$I=3.4 mag companion at a projected separation of 34 AU (0\farcs269 at 128 pc). The transit of the 3 R$_{\oplus}$ planet candidate has been confirmed on target,  and the field has been cleared for nearby EBs.

Resuming, of the eight close systems, it is probable that only two, TOIs 2409 and 1060, host a real transiting planet. Perhaps tellingly, the candidate system  most likely to host a \textit{bona fide} transiting planet, TOI-1060, has a low-$q$ companion (consistent with an M-dwarf), while the six likely FPs have solar-type companion stars. Nevertheless, it seems the best-fit suppression model, which already reduces binaries by nearly an order of magnitude at  separations below 58 AU, may still underestimate the effect of binaries on planet survival due to residual FP contamination.

\subsubsection{Bias against detecting planets in binary stars}

A possible alternative explanation for the large suppression in planet candidate hosting close binaries is that many planets are not detected by TESS due to the transit dilution from a companion star. In Paper I, we estimated the number of single systems in our sample that would have planets not detected if a binary were present. We found that approximately 2$\%$ of planets would not be detected due to binary dilution. This suggests that the population of missed planets due to binary dilution is small, with negligible impact on the results of the Paper I analysis.

In this work, the culled sample of TESS planet candidates shows a significantly higher rate of close binary suppression compared to \textit{Kepler} targets. The photometric precision of \textit{Kepler} was far superior to TESS, perhaps making it less susceptible to the effect of binary dilution on planet detections. To investigate this, we seek to estimate the number of planets in binary systems that TESS did not detect that would have been detected if the stellar host were single.

We begin by simulating a planetary population around the stars observed by TESS. We use the candidate target list from the TESS Input Catalog v8 \citep{tic} for the population of planet host stars. The TIC provides stellar parameters, including TESS magnitude, stellar mass, radius, effective temperature, and coordinates, for each of 9.4 million stars. An estimate of the spectral type for each star was made using the temperature thresholds of \citet{pecaut}. The TESS observing baseline is equal to 27.4 days multiplied by the number of observed sectors for each star, which was determined using the TESS pointings.

The host stars were populated with planets using the occurrence rates of \cite{fressin13} for AFGK-type stars and \cite{dressing13} for M-type stars. Both studies provide the number of planets around each star as a function of orbital period and planetary radius. Each star had planets populated at random using these occurrence rates.

The orbital parameters of each simulated planet were estimated using the methods of \cite{cooke18}. In brief, the orbital separation was calculated using Kepler's third law, periastron angle was drawn from a uniform distribution between -$\pi$ and $\pi$, inclination $i$ derived from cos$i$ with a uniform distribution between 0 and 1, and eccentricity was drawn from a beta distribution with $\alpha$=1.03 and $\beta$=13.6 found by \cite{vaneylen15}. Transit durations ($T_{dur}$) were calculated using the relation in \citep{barclay18},

\begin{equation}
\label{eq:t_dur}
T_{dur} = \frac{P}{\pi} \times \arcsin\left({\frac{R_{\star}}{a} \times \frac{\sqrt{1 + \frac{r_p}{R_{\star}} - b^2}}{\sqrt{1 - \cos^2{i}}}}\right).
\end{equation}

\noindent
where $P$ is the orbital period, $R_{\star}$ is the stellar radius, $a$ is the semimajor axis, and $r_p$ is the planetary radius. The probability of transit was determined from the impact parameter using the relation from \cite{winn10},

\begin{equation}
\label{eq:impact_param}
b = \frac{a\cos{i}}{R_{\star}} \times \frac{1-e^2}{1+e\sin{\omega}},
\end{equation}

\noindent
where a transit is defined as occurring for $|b|<1$.

For transiting planets, an initial transit time, $T_0$, was drawn randomly from a uniform distribution between 0 and $P$. The planet transits during the TESS observation if $T_0 + n\times P$, where $n$ is an integer, occurs within the TESS observation baseline of the host star. The number of subsequent transit observations by TESS can also be determined using $P$. Some TESS systems have additional observations from the extended TESS mission. In some cases, additional observed transits resulted in sufficient S/N for a planet candidate detection, regardless of potential binary contamination. The probability that a star was observed by TESS in the extended mission observations at the time of the SOAR observation is estimated using the reported release dates of the sector observations. To determine whether TESS can detect the transit, we estimate the S/N using the equation of \cite{barclay18}, 

\begin{equation}
\label{eq:SNR}
S/N = \frac{\delta_{eff}}{\sigma_{1hr}}\sqrt{\frac{T_{dur}}{\Delta T}}\sqrt{n}
\end{equation}

\noindent
where $\delta_{eff}$ is effective transit depth, $\sigma_{1hr}$ is the photometric precision in 1 hour of TESS data, $n$ is number of observed transits, $T_{dur}$ is transit duration in hours and $\Delta T$ is observing cadence. We estimate the effective transit depth using the equation,

\begin{equation}
\label{eq:delta}
\delta_{eff} = \left(\frac{R_p}{R_{\star}}\right)^2 \times \frac{1}{1+C},
\end{equation}

\noindent
where C is the contamination ratio from background stars provided in the TIC. The photometric precision for TESS was estimated using the relations provided in \cite{tic}. A transit was determined to be detected around the single star if the S/N$>$7.3, a threshold generally used by the SPOC pipeline \citep{jenkins16}. The S/N threshold of the MIT Quick Look Pipeline \citep{huang20}, which searches for planet candidates in the TESS full-frame images, is not defined, but is likely of a similar magnitude.

If the planet was detected, a companion star was then populated using the binary statistics of \citet{raghavan10}, as described in Section \ref{sec:multiplicity}. The photometric contamination from this star was added to the TIC contamination ratio and the transit depth recalculated with the additional dilution. If the S/N of the previously detected planet was now less than 7.3 when in the binary system, and the projected separation of the binary was less than 21\arcsec (i.e., the size of a TESS pixel, so the companion would likely fall within the photometric aperture of the planet host), the planet was considered missed due to binary dilution. The model was run 10$^2$ times to estimate the number of planet detections and non-detections due to binaries.

We find from 1922$\pm$140 simulated planet detections (comparable to the TESS planetary yield simulations of \citet{sullivan15} and \citet{barclay18} and the actual number of TESS detections), that  only 130$\pm$19 planets (6.8$\%$) would not be detected due to binary dilution. A similar number of non-detections is likely to come from the TESS full-frame images (FFIs), based on the ratio of FFI planets to CTL planets found by \citet{barclay18}. As  expected, the characteristics of the missed planets fell at the limits of the sensitivity of TESS. The average missed planet was small (R$_{avg}$=3.3R$_{\oplus}$), had a relatively long period (P$_{avg}$=11.5 d), with generally only 2 or 3 detected transits, and is hosted by stars fainter than average TESS targets (T$_{mag,avg}$=12.4). 

\begin{figure}
    \centering
    \includegraphics[scale=0.30]{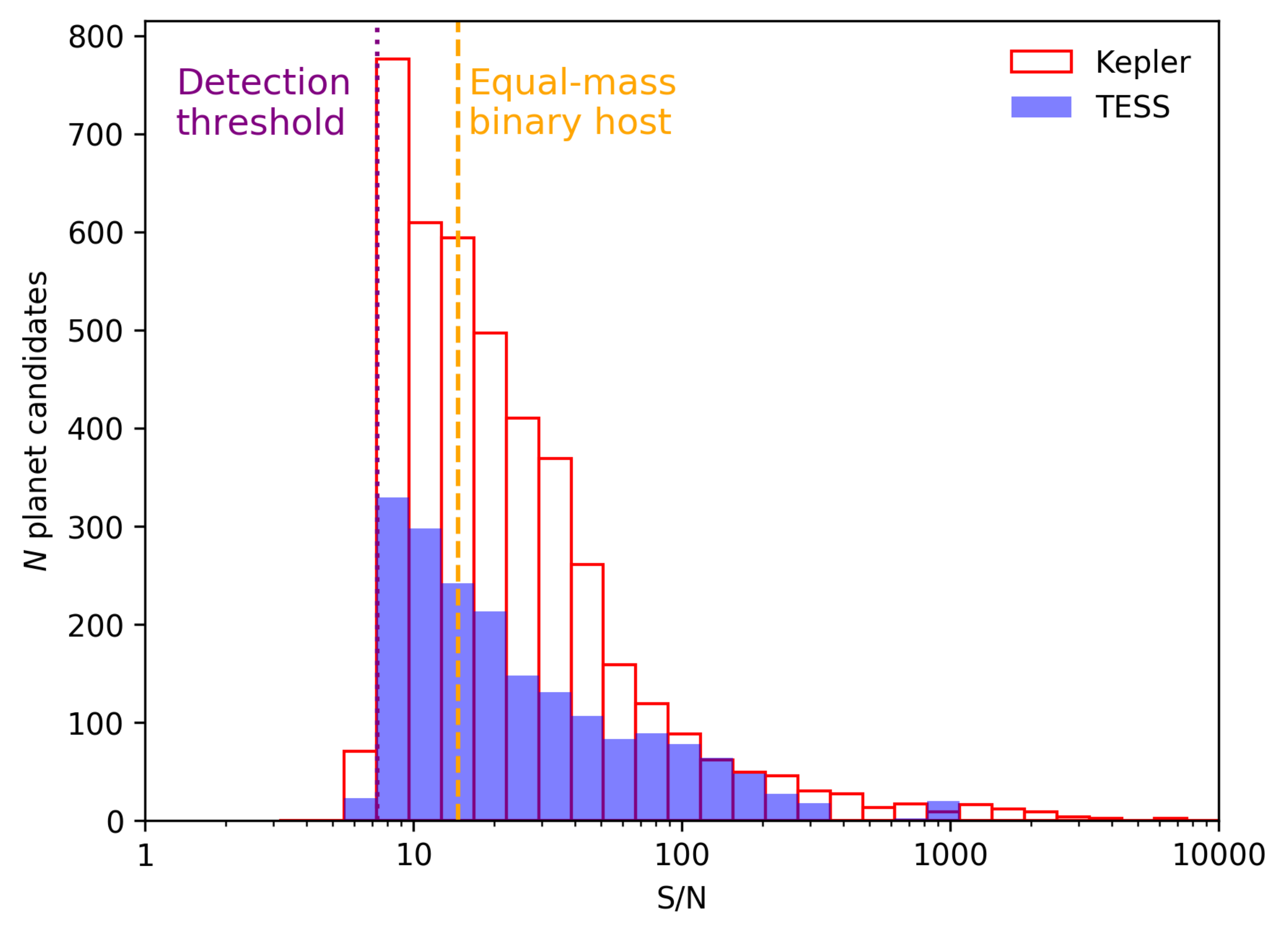}
    \caption{The distribution of reported S/N ratios of TESS and \textit{Kepler} planet candidate detections. The majority of detections were near the S/N threshold of 7.3 (a small number of known planets fell below this threshold). In the presence of an equal-mass binary, the S/N would need to be twice this threshold (S/N$>$14.6) to be detected. For both missions, slightly less than half of the detected planet candidates would have been missed in the presence of an equal mass companion.}
    \label{fig:tessnsrs}
\end{figure}

The number of expected non-detections due to binaries is significant, and the distribution of the S/N ratios of TESS planet candidate detections thus far suggests it is reasonable (see Figure \ref{fig:tessnsrs}). The majority of TESS planet candidates were low S/N detections, only slightly above the formal S/N $>$ 7.3 threshold.\footnote{One TESS system, HD 219134b and c, was a known system and was likely flagged despite a S/N$<$7.3.} An equal-mass binary (maximum binary dilution) would reduce the transit depth by a factor of 2, which would reduce the S/N ratio of a detection by the same amount (see Equation \ref{eq:SNR}). We find that 777 of the 1992 TOI planet candidates (40.4$\%$) would be susceptible to non-detection if in an equal-mass binary. For comparison, for \textit{Kepler} planet candidates, 1768 of 4251 (41.6$\%$) would not be detected in the presence of an equal-mass stellar companion.\footnote{The \textit{Kepler} pipeline S/N ratio threshold was slightly lower at 7.1, which may slightly increase the number of planet candidates that would be susceptible to binary dilution.}

If we simply assume that half of the low S/N ratio planet candidates from TESS are in binaries, and 30$\%$ of these have mass ratios that potentially could lead to transit dilution ($q>$0.75), we would expect that approximately 5$\%$ of the known TESS planet candidates would not be detected due to binary dilution. This is comparable to the fraction of non-detections in the simulated TESS planet population.

The number of planet hosting binaries that were not detected by TESS due to dilution of the transit is relatively small, reducing the binary fraction by approximately one or two percentage points, and had a negligible effect on our analysis. The properties of the companion stars that lead to planet non-detections due to dilution likely do not bias our results, either. These stars generally have a large $q$ ($q_{avg}$=0.85) and lie at physical separations similar to the field population ($a_{avg}$=64 AU).

\subsubsection{Implications for planet occurrence rates}

The harsh environment for planets in close binaries means a significant fraction of solar-type stars is not able to host planets. \citet{kraus16} estimated that, with a suppression factor of 0.34 cutting separations below 47 AU, 19$\%$ of solar-type stars could not host planets. \citet{moe19} used updated binary statistics that account for the low number of WD and late-M companions in \citet{raghavan10} and a model that includes the nearly complete suppression at close separations found in RV surveys. They estimate that a third (33$\pm$4$\%$) of FGK stars are not able to host planets, a fraction which increases to nearly half in magnitude limited, visible-light surveys due to Malmquist bias resulting in an over-representation of binaries.

The suppression found in this analysis is significantly larger than that found by \citet{kraus16} or the model adopted by \citet{moe19} (see Figure \ref{fig:suppression}), and the suppression turns on at a slightly closer physical separation. We use our suppression model to estimate the fraction of solar-type stars that are disallowed from having planets in the galaxy. This fraction is equal to the number of planet candidate hosting binaries subtracted from the number of field binaries divided by the total number of stars. For this analysis, we increase the number of companions with a$<$50 AU from 21$\%$ in the \citet{raghavan10} survey to 40$\%$, for reasons described above and in \citet{moe19}. As our model already accounts for Malmquist bias (which results in a larger number of binary companions in a magnitude-limited survey due to the increased brightness of a multiple star system compared to a single star system), the resulting fraction is therefore relevant for a volume-limited sample. We can also disable this correction in the model, which reduces the number of binaries, to estimate the fraction for magnitude limited samples.

We also note that our sample only uses systems with a mass ratio of $q>0.4$. It is possible that suppression is less significant with lower mass companions, however, as discussed in Section \ref{sec:high_low_q}, we see no evidence for that in our sample. For our estimate of how binary stars impact the planet occurrence rates, we assume suppression is uniform across all mass companions.

We estimate that around a third (34$\pm$4$\%$) of solar-type stars in the galaxy will not be able to host planets due to binary interference. In a magnitude-limited transit survey performed in the visible, we estimate that nearly half (48$\pm$5$\%$) of observed solar-type stars will not be able to host planets.

\subsection{Wide binary enhancement}

In Paper I, we detected a large number of wide binaries hosting planets: 119 observed binaries with projected separations greater than 100 AU, compared to the expected number of 77$\pm$7. The enhancement in wide companions was exclusively attributable to the systems hosting large planet candidates (radii greater than 9 R$_{\oplus}$), similar to the results of previous surveys \citep{ngo16, ziegler18b, fontanive19}. This was interpreted as possible evidence of planetary orbital migration caused by dynamical interactions with the companion star.

\citet{moe19} suggested that the observed wide binary enhancement is exclusively due to false-positive contamination. That is, many of the targets with wide companions are in fact eclipsing binaries with observed transits consistent with planetary bodies due to binary dilution. Close binaries, with periods less than 7 days (an orbital period typical of the TESS planet candidates), are significantly more likely to have tertiary companions \citep{tokovinin06}. In the revised analysis of Paper I  with false positives and giant planet candidates removed, they found no enhancement in binaries at separations between 100-2000 AU (17$\pm4\%$ companion rate compared to 16$\pm3\%$ expected).

\begin{figure}
    \centering
    \includegraphics[scale=0.31]{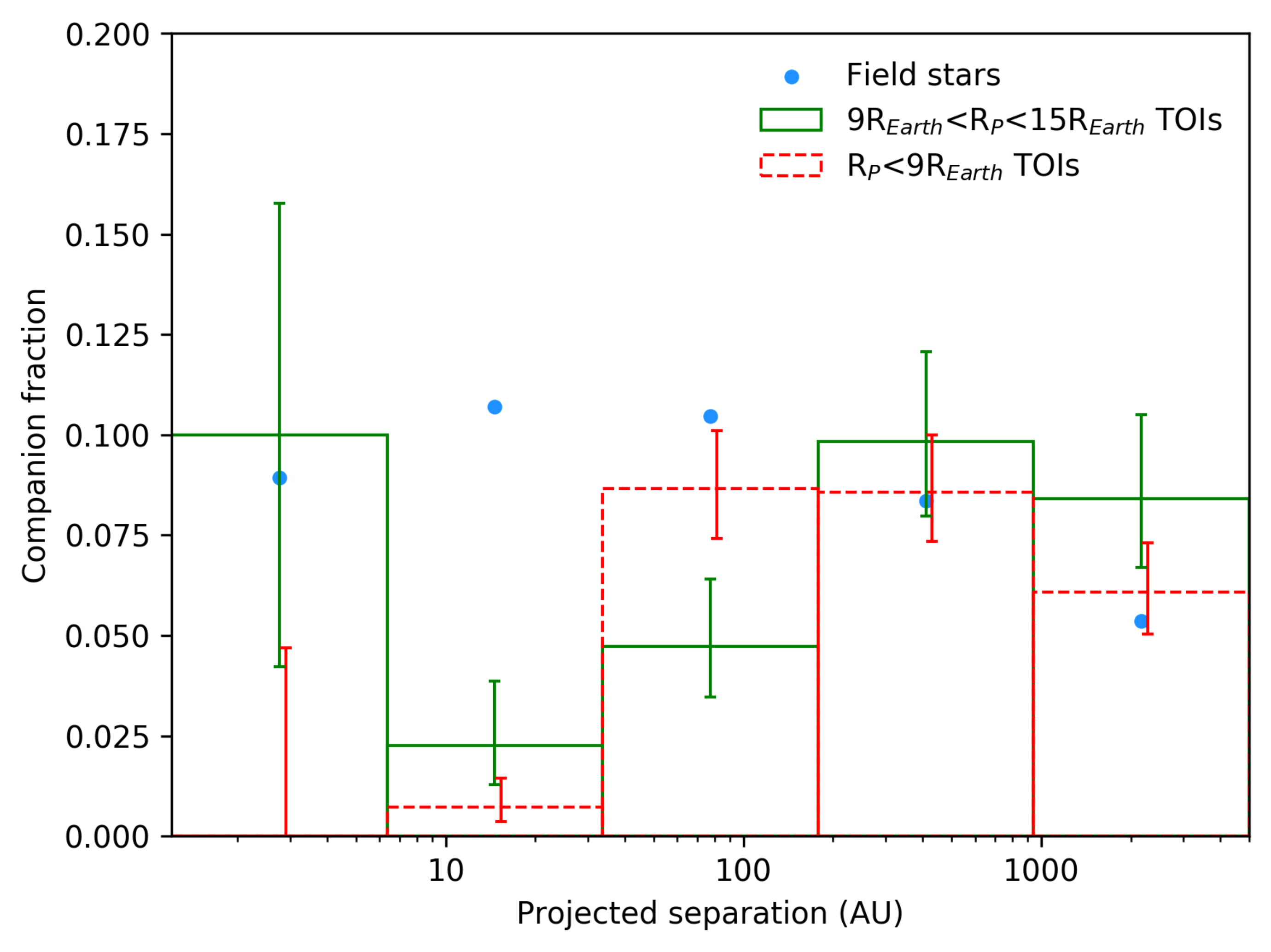}
\caption{The completeness corrected companion fraction per bin in projected separation for small and large TESS planet candidate hosts observed in this survey. For reference, the separation distribution of field binaries from \citet{raghavan10} is included. Both populations of TESS planet hosts have suppressed rates of close binaries (the companion fraction for large planets in the lowest separation bin is consistent with field rates, but due to the small number of detectable close systems, the uncertainty is high). At wide separations, both small and large planets have a companion rate consistent with field stars.}
    \label{fig:planetradius}
\end{figure}

If this enhancement is due entirely to false positives, we would expect to see a significant reduction in the companion rates in this analysis using the culled sample. We can also check the companion fractions of the removed targets and the targeted EBs for wide-binary enhancement.

The completeness-corrected companion rates for the 430 small (R$_{P}<$9 R$_{\oplus}$) and 225 large (R$_{P}=$9-15 R$_{\oplus}$) planet candidate hosting systems are shown in Figure \ref{fig:planetradius}. For systems with multiple planet candidates, the radius of the largest planet was used. Both populations show similar suppression of close binaries. We use the radius corrections for systems with companions under the assumption that the primary is the host. At wider separations, both populations are consistent with field star rates. For comparison to the rates above, the companion fraction in the range 100-2000 AU (with 100 AU being the beginning of binary suppression and the 2000 AU being the outer limit for dynamical interactions) is 38/214, or 17.7$\pm$3$\%$, consistent with the expectation value of 16$\pm3\%$.

\begin{figure}
    \centering
    \includegraphics[scale=0.32]{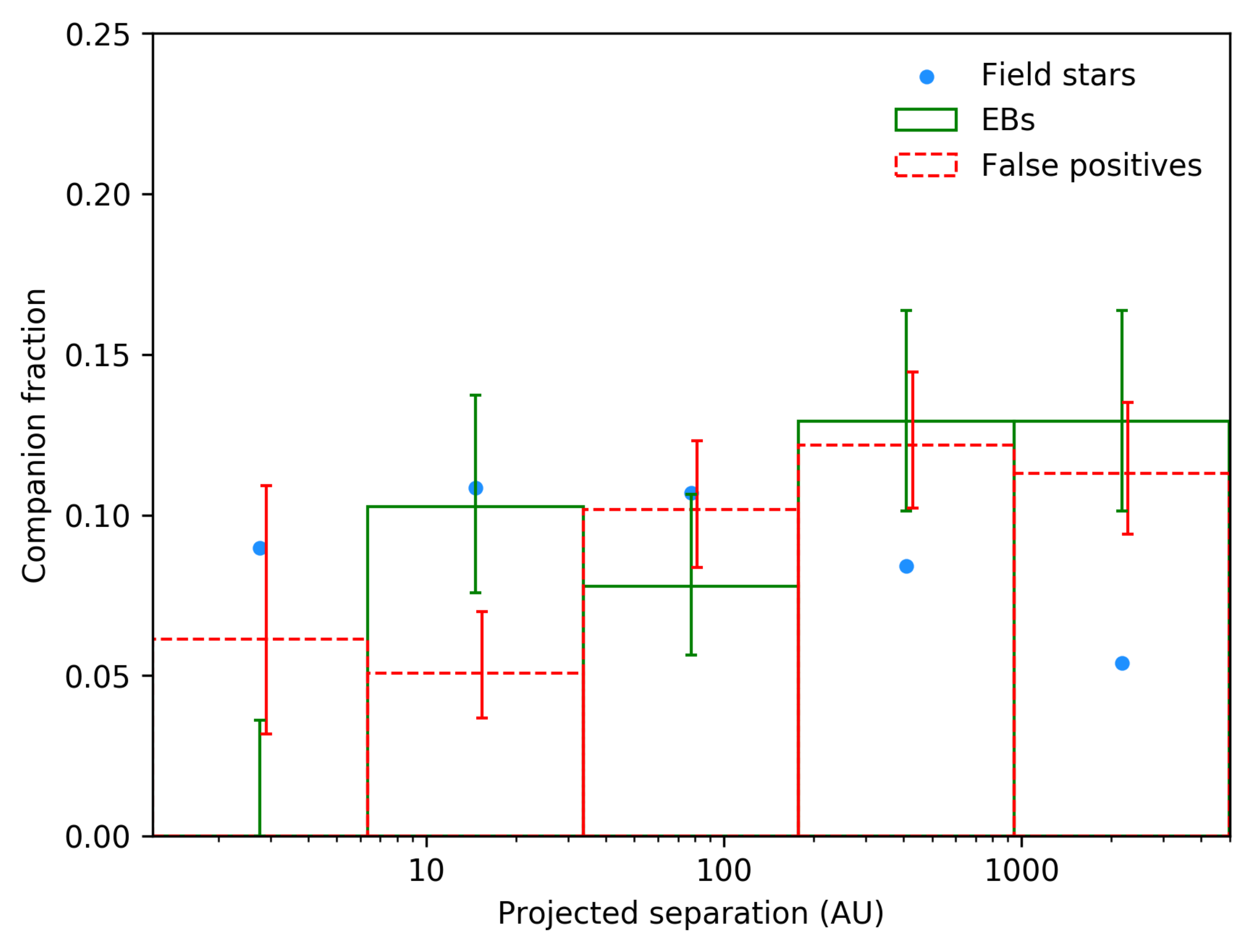}
    \caption{The observed companion rates for the 297 known and likely false positives removed from the planet candidate sample, compared to field star companion rates. Also plotted is the observed companion rate of the 167 targeted EBs. At small separations, the FP companion rates are similar to field stars, however, wide binaries are detected at a high rate, perhaps a population of wide tertiary companions to the diluted EBs that are the source of the false-positive transit signal. The targeted EBs likewise show a similar enhanced binary rate at wide separations. }
    \label{fig:fps}
\end{figure}

The companion rates for known and suspected FPs and targeted EBs is shown in Figure \ref{fig:fps}. We find at close separations, FPs have similar companion rates to field stars. EBs have a 2$\sigma$ lower companion rate at close separations ($\rho=$1-6 AU), perhaps evidence of dynamical instability at close separations for tertiary companions \citep{tokovinin06}. At wide separations, both the FPs and EBs show significant enhancement, with companion rates approximately twice the field rates at $\rho$=1000-5000 AU, in line with the high rate of tertiary companions to close period binaries found by \citet{tokovinin06}.

We conclude that the high wide binary fraction observed in Paper I was due largely to false-positive contamination, as suggested by \citet{moe19}. The binary fractions for the culled sample, which is near zero at close separations and is consistent with field rates at wide separations, suggest that the majority of false positives have been removed.

\begin{figure}
    \centering
    \includegraphics[scale=0.31]{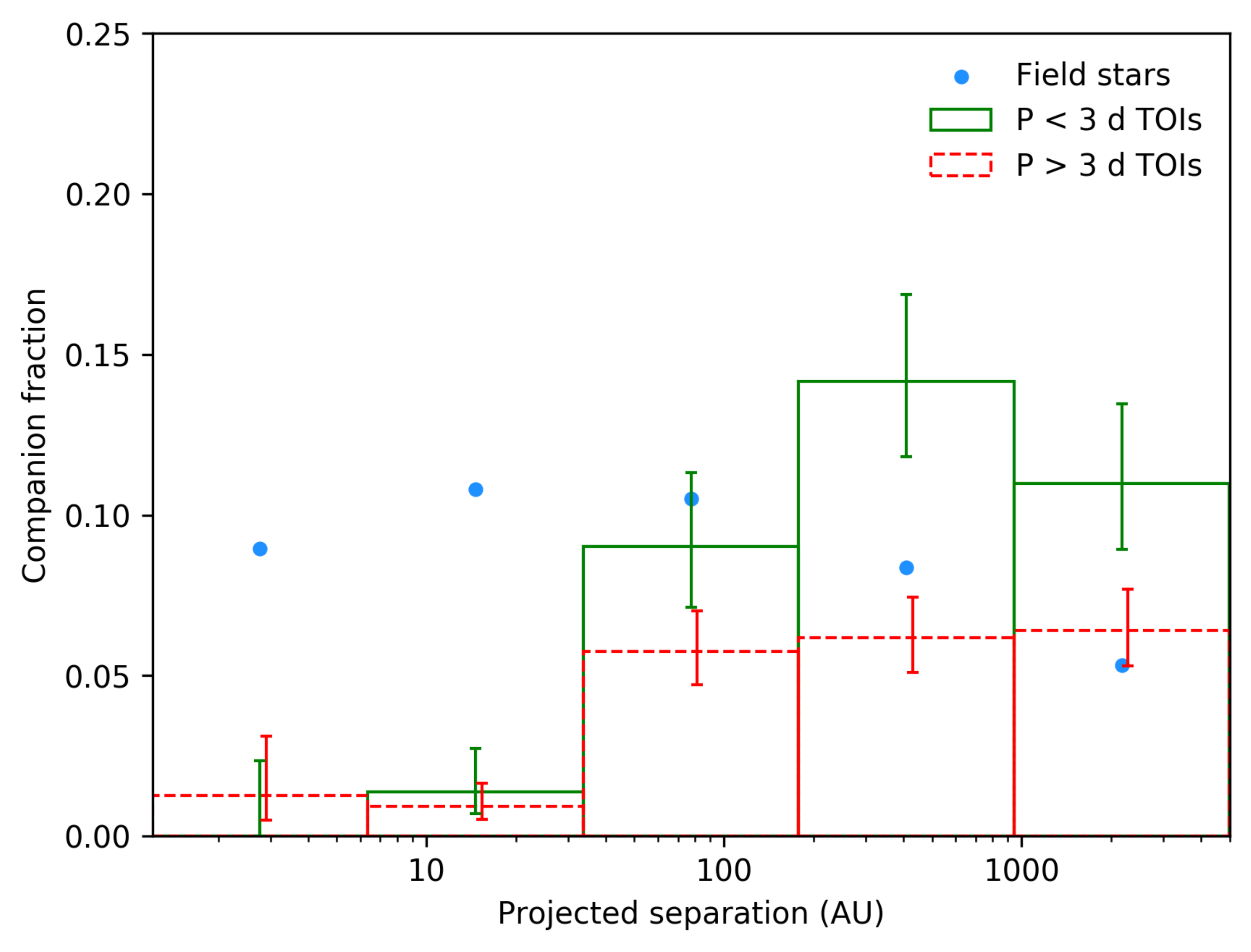}
    \caption{The completeness corrected companion fraction per bin of projected separation for the 191 shorter (P$<$3 d) and 464 longer (P$>$3 d) TESS planet candidate hosts observed in this survey. For reference, the separation distribution of field binaries from \citet{raghavan10} is included. Both populations show suppression of binaries at close separations. At wide separations, while the longer period planets have similar companion rates as field stars, the short period planets have significantly higher companion rates.}
    \label{fig:period}
\end{figure}

\subsubsection{Planetary period and wide binaries}

There is some evidence that the binary fraction for close period planets (P$<$3 d) is significantly higher than that for longer period planets \citep{ngo16, ziegler18b}. A similar trend, however, is observed in the triple rate for close stellar systems: \citet{tokovinin06} found nearly all ($\sim$96$\%$) P$<$3 d binaries have additional companions, a rate which steadily decreases with period and at P=6-12 d is nearly equivalent to the field companion rate.

From our culled sample of 655 planet candidate systems, 191 (29$\%$) have inner planets with periods less than 3 d (the remaining 464 systems have inner planets with periods greater than 3 d). In Figure \ref{fig:period}, we compare the completeness corrected companion fractions for short and long period populations. The completeness corrected companion fraction is significantly higher for close planets (35$\pm$5$\%$) than for longer period planets (20$\pm$4$\%$). While both populations display close binary suppression, only long-period systems have similar wide binary rates to field stars. Short-period planets display an increase in binary rate at wide separations ($\rho=$1000-5000 AU). 

\begin{figure}
    \centering
    \includegraphics[scale=0.32]{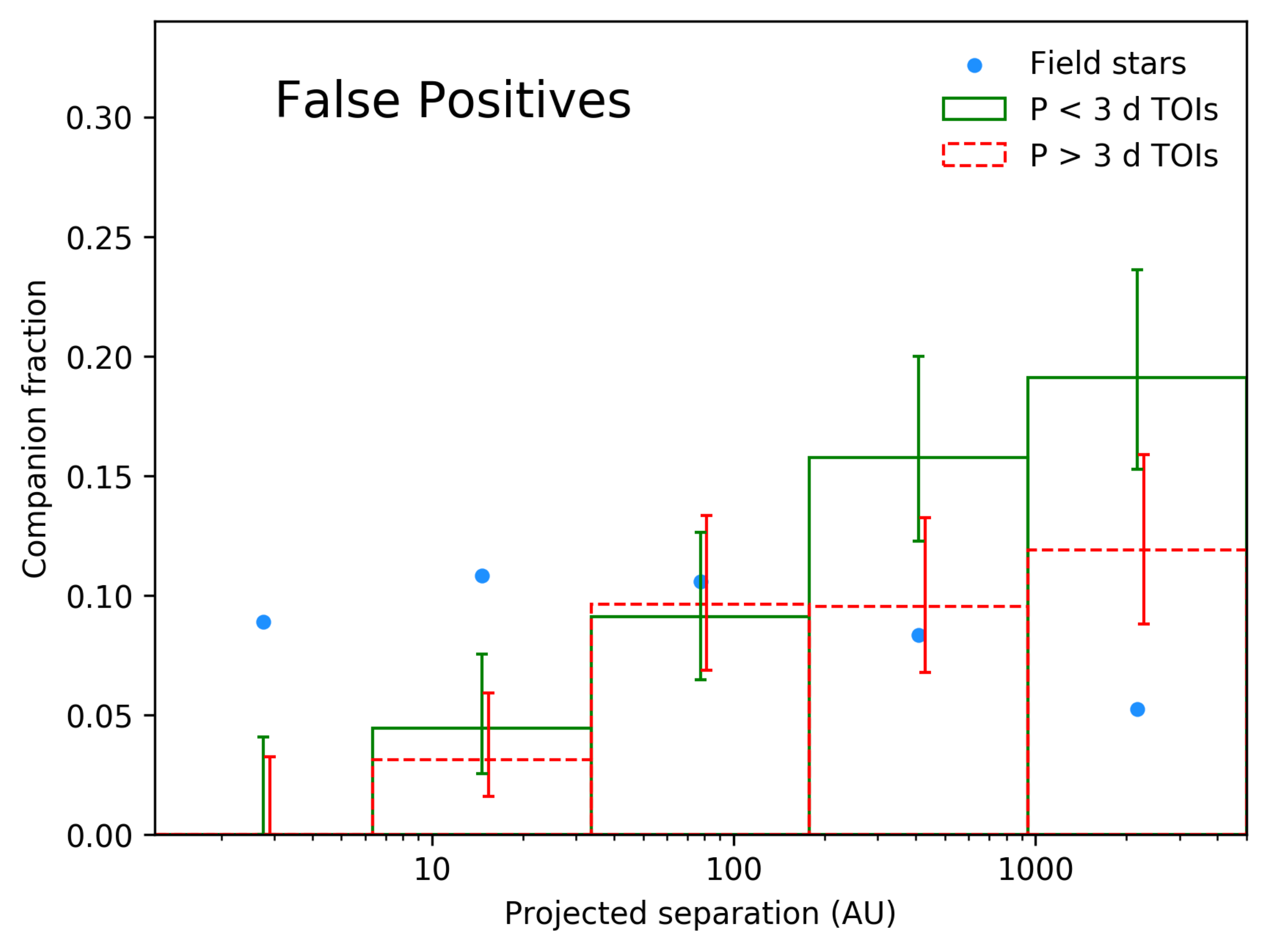}
    \caption{Similar to Figure \ref{fig:period} but with 90 shorter (P$<$3 d) and 90 longer (P$>$3 d) period false positives, which are generally eclipsing binaries. For reference, the separation distribution of field binaries from \citet{raghavan10} is included. Similar to the short period planet candidates, the short period FPs have a sharp increase in companion fraction at wide separations.}
    \label{fig:periodfps}
\end{figure}

For comparison, Figure \ref{fig:periodfps} shows the companion fraction for known false positives. The fraction of P$<$3 d FPs (50$\%$) is significantly higher than planet candidates. The short period FPs display a significant increase in companion fraction at wide separations ($\rho=$200-5000 AU). For P$>$3 d period FPs, the cumulative companion fraction for projected separations $\rho=$1-5000 AU is 36$\pm$4$\%$. In the same separation range, P$<$3 d period FPs have a 49$\pm$6$\%$ companion fraction. These are lower than the rates from \citet{tokovinin06}, likely because many of these targets are actually single stars that are blended with nearby eclipsing binaries.

While \citet{tokovinin06} found companion rates for close binaries to be high, the period distribution for P$<$7 d was similar to field stars, without an anomalously high wide binary fraction. It is likely that in many of these FP systems, the transit signal comes from the wide companion, which is an eclipsing binary. It is not clear, however, why these systems are found preferentially at wide separations. It may be that these systems are the most likely to be identified as FPs, with large centroid shifts in the blended PSF during transits, or are separated enough for seeing-limited photometry to identify the nearby EB. These systems will also be preferentially at wide separations due to largely being chance-alignments.

\subsection{Binary companions to M-dwarf planet candidate hosts}\label{sec:mdwarfs}

\begin{figure}
    \centering
    \includegraphics[scale=0.32]{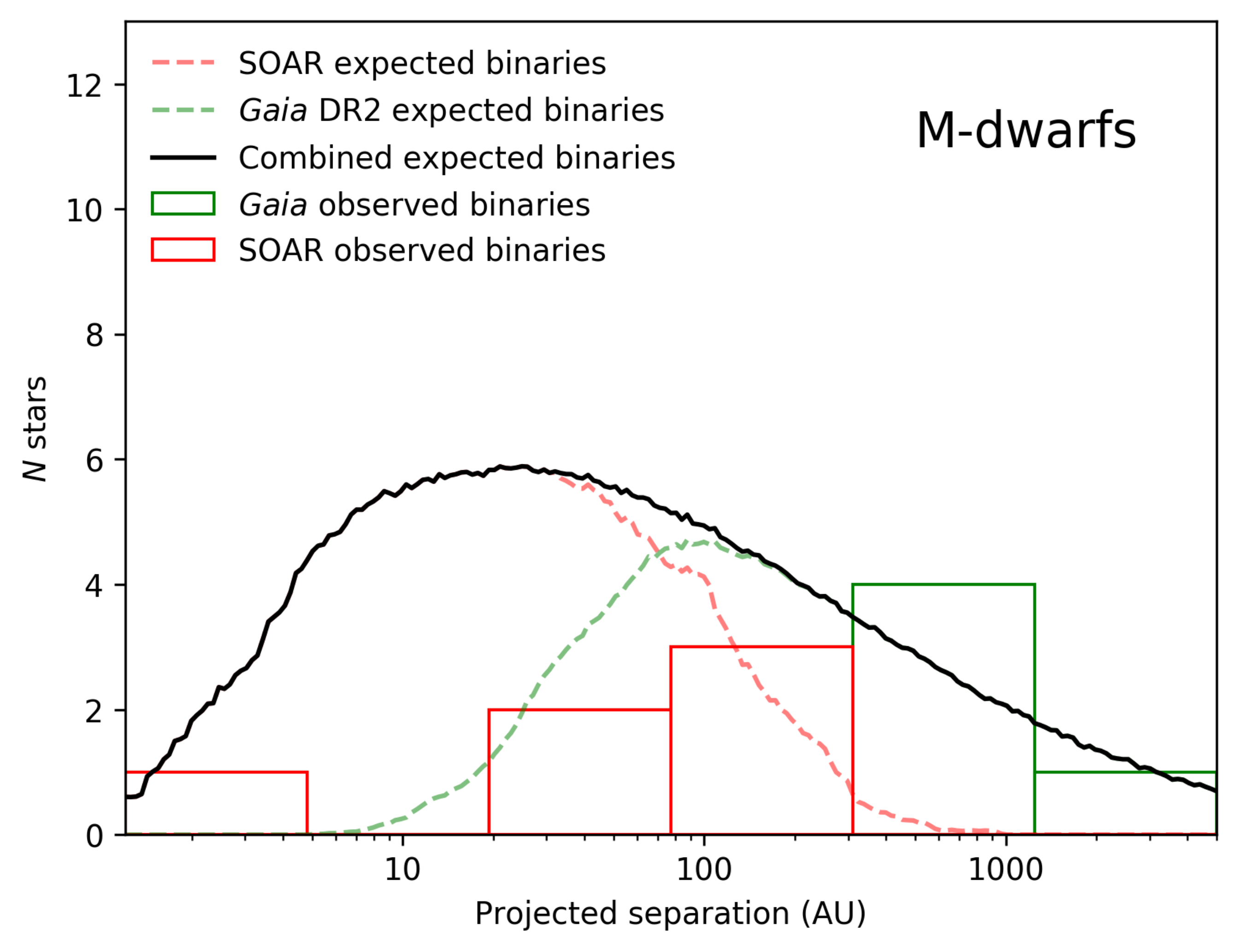}
    \caption{The distribution of observed companions to TESS planet candidates around M-dwarf hosts resolved in SOAR speckle imaging and in Gaia DR2. Known and suspected false positives have been removed. The expected distribution from field M-dwarf statistics \citep{winters19} is also plotted. Close binary suppression at projected separations less than 20 AU is apparent, with a low significance.}
    \label{fig:mdwarfs}
\end{figure}

In Paper I, we found tentative evidence for binary suppression in M-dwarf planet candidate hosts compared to the field star statistics of \citet{winters19}. The M-dwarf systems detected by TESS (with T$_{eff}<$3900 K, \citet{pecaut12}) have a higher false-positive rate than solar-type stars, 83 of the 157 M-dwarf targets observed at SOAR (53$\%$) are known or likely FPs. The observed M-dwarf binary distribution compared to expectations based on field rates are shown in Figure \ref{fig:mdwarfs}.

The field binary distribution peaks at separations of approximately 20 AU. The TESS targeted M-dwarfs are relatively close, at a median distance of 36 pc, and SOAR imaging is sensitive to close binaries within 20 AU for every M-dwarf target. We find one and two systems with projected separations less than 20 and 50 AU, respectively. The numbers observed in both separation ranges are significantly less than the field expectations of 7$\pm$2 and 11$\pm$3 binaries within 20 and 50 AU, respectively. The low companion rate at close separations is similar to that seen for planet candidate around solar-type stars, suggesting that suppression occurs in M-type stars as well.

One of these close M-dwarf systems, TOI-224 (TIC70797900) with a companion at a projected separation of 4 AU, has been identified as a likely EB.

The other close M-dwarf binary system is TOI-455 (TIC98796344), which was originally reported in Paper I as a binary with a projected separation of 7.1 AU, seemingly making it an outlier in the observed binary distribution. The system is actually a hierarchical triple blended in a single TESS pixel, and \citet{winters19b} confirmed the planet LTT 1445Ab around the primary, not the close BC components observed by SOAR. This system has been included in this analysis using the separation from A to the barycenter of the BC component at a 7\farcs10  ($\sim$34 AU) from the most recent observation in 2017 according to the Washington Double Star catalog \citep{mason09}.

\subsection{Mass ratios of planet candidate host binaries}\label{sec:qs}

The mass ratio, or $q$, distribution of solar-type binary systems was found to be nearly uniform by \citet{raghavan10}, with an increase for near-equal mass binaries. \citet{winters19} found a similar distribution at high-$q$ for M-dwarfs. In Paper I, we reported a large number of low-$q$ companions, similar to the findings of \citet{ngo16}, who noted that mass ratio distribution for hosts of hot Jupiters was heavily weighted towards low-$q$ companions. \citet{tokovinin06} found that tertiary companions to EBs also generally have low mass ratios, so this may be a byproduct of false-positive contamination. We would expect then that with the culled sample in this analysis, the number of low-$q$ companions will decrease. 

\begin{figure}
    \centering
    \includegraphics[scale=0.32]{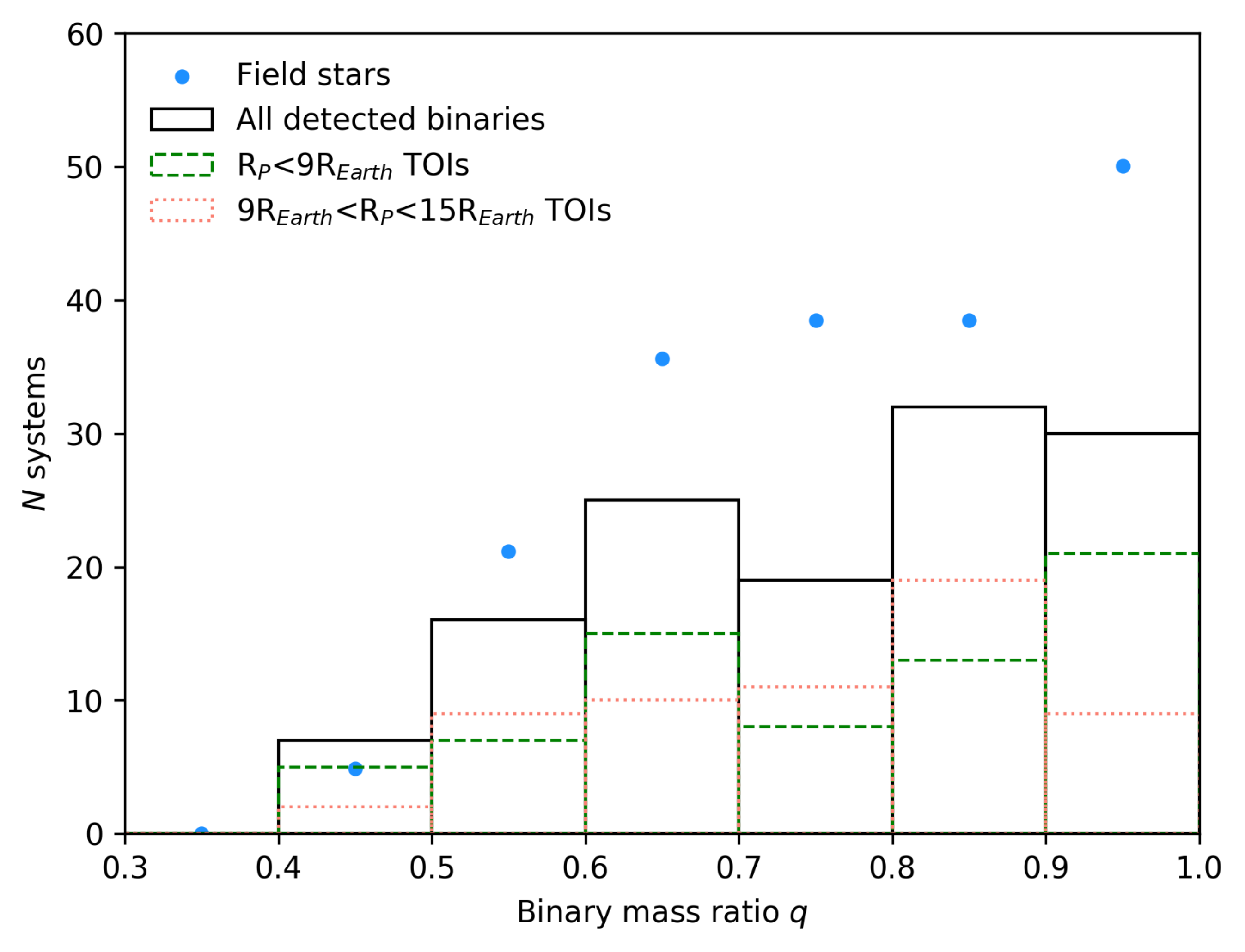}
    \caption{The mass ratio distribution of observed binaries with TESS planet candidates resolved in SOAR speckle imaging, including the individual distributions for large and small planet candidate hosting stars. The expected distribution of observed binaries based on the near-uniform mass ratio distribution of field stars \citep{raghavan10} and the survey sensitivity is included. The observed mass-ratio distribution is consistent with uniform, however, fewer binaries were detected than expected, likely due to the close binary suppression in planet hosts.}
    \label{fig:qs}
\end{figure}

The distribution of mass ratios for binaries in this sample is plotted in Figure \ref{fig:qs}, along with the expected number of binaries based on a near-uniform mass ratio distribution and our survey sensitivity. The observed distribution is consistent with uniform; however, fewer binaries were detected at most mass ratios due to binary suppression. Far fewer low-$q$ companions were detected compared to Paper I, suggesting that a large fraction of the false positives have indeed been removed. The distributions for small and large planet candidates are both similar to each other.

\subsubsection{Mass ratios and binary distribution}\label{sec:high_low_q}

We find significant suppression in transiting planets around close binary systems (Section \ref{sec:closebinarysuppression}). Many theoretical mechanisms employ dynamical interactions to explain the lack of planets in these systems. Presumably then, higher mass companions will result in a larger suppression effect which begins at wider separations.

\begin{figure}
    \centering
    \includegraphics[scale=0.62]{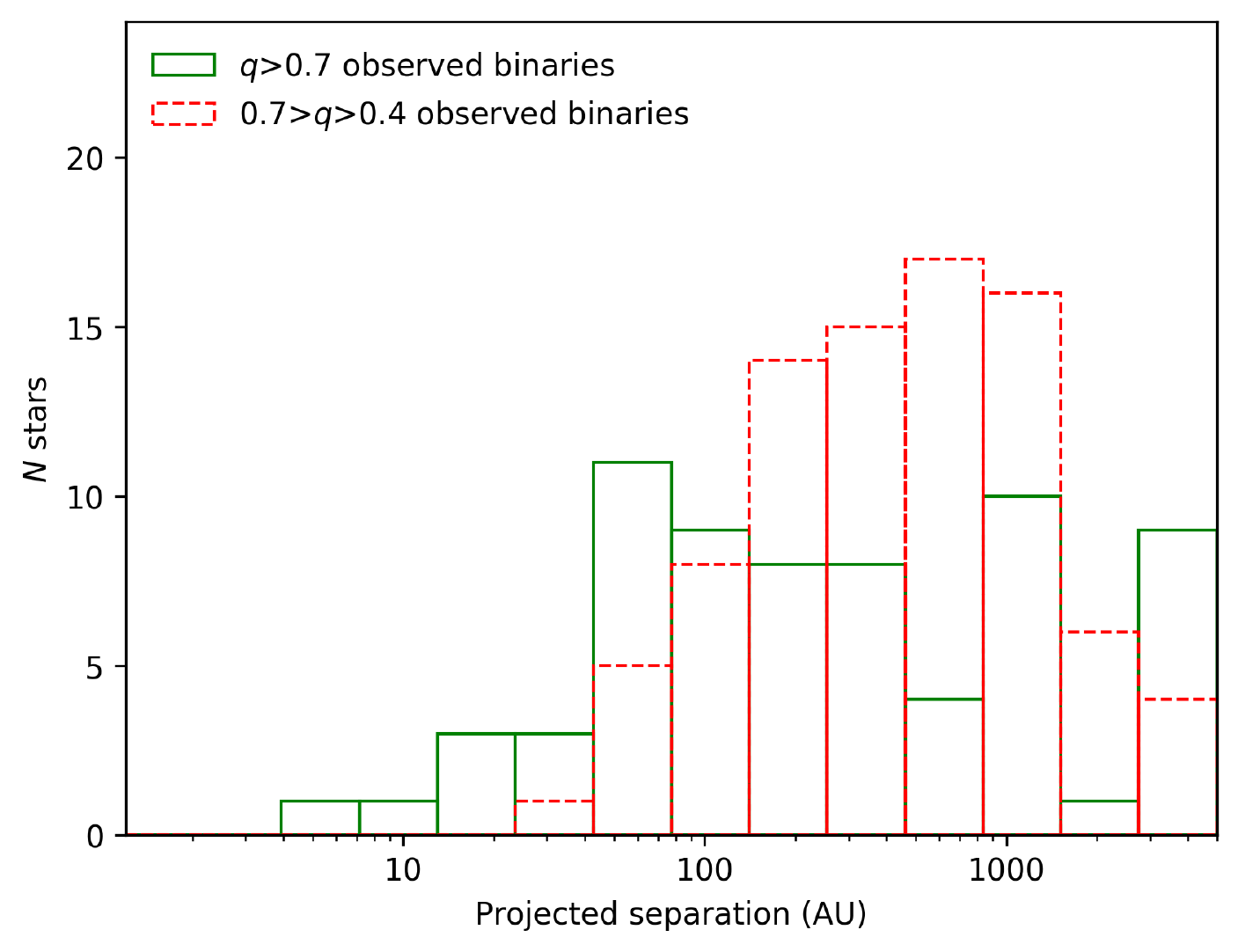}
    \caption{The distribution of the projected separation for high and moderate mass companions to TESS planet candidates. Higher mass companions are seemingly found at closer separations, while moderate mass companions peak at wide separations.}
    \label{fig:qs_high_low}
\end{figure}

We search for evidence of any variation between high and low mass companions in our sample by splitting our sample into two populations: 68 high mass ratio systems, $q>0.7$, and 86 moderate mass ratios systems, $0.7>q>0.4$ (all low mass ratio systems, with $q<0.4$ were excluded from our analysis, as explained in Section \ref{sec:prepseample}). The resulting distribution of the projected separations for binaries in these two populations is shown in Figure \ref{fig:qs_high_low}.

The resulting distributions seemingly run counter to the hypothesis that higher mass companions will have a higher suppression effect. The majority of systems at close separations ($\rho<$50 AU) are high $q$, while most wide separated systems (200$<\rho<$1000 AU) are low $q$. A Kolmogorov-Smirnov test gives a 80$\%$ likelihood that the two populations are in fact different.

This disparity between high and low mass companion separation distributions may be evidence that dynamical interactions are not the primary mechanism for close binary suppression. It may also be a result of observational bias: the lower mass companions are fainter and thus more easily detected at wide separations. Also, as noted in Section \ref{sec:closestbinaries}, the majority of the closest separated systems are likely false positives (disposed as ``anomalous'' planet candidates).

It is likely that how planet suppression depends on the mass of the companion star may provide important insight into the sequence of events that lead to planets either not forming or being destroyed or displaced. The TESS planet candidate list is continuously improving with false positives being identified, and further high-angular resolution observations with large aperture telescope will detect faint, low-mass companions at close separations.

\subsection{Orbital stability of planets in binary systems}

The presence of a binary star makes some planetary orbits dynamically unstable, particularly wide orbits with large semi-major axes. A recent study by \citet{quarles20} used $N$-body simulations to estimate the critical planetary orbital parameters for stability based on binary properties, in particular $\mu$, the dynamical mass (M$_{sec}$/(M$_{prim}$+M$_{sec}$), $e_{bin}$, the eccentricity of the binary orbit, and $a_{p}/a_{bin}$, the ratio of the planetary semi-major axis to the binary semi-major axis.

For each of our resolved systems, we estimate the critical planetary semi-major axis, $a_{crit}$, where $a_{p}>a_{crit}$ results in an unstable planetary orbit, using the relations from \citet{quarles20}. We use the mass ratio from Section \ref{sec:qs} to estimate $\mu$, use the projected separation of the binary for $a_{bin}$, and estimate $a_{p}$ using the host mass and planetary orbital period and assuming a Keplerian orbit. We set $e_{bin}$=0.8 and average binary inclinations of 45$^{\circ}$, reasonably extreme assumptions that will provide a lower limit for $a_{crit}$. In general, \citet{vaneylen19} found no correlation between the presence of a stellar companion and high planetary eccentricity. For each system, we estimate $a_{crit}$ for both the primary and secondary host scenarios using the above orbital assumptions.

\begin{figure}
    \centering
    \includegraphics[scale=0.39]{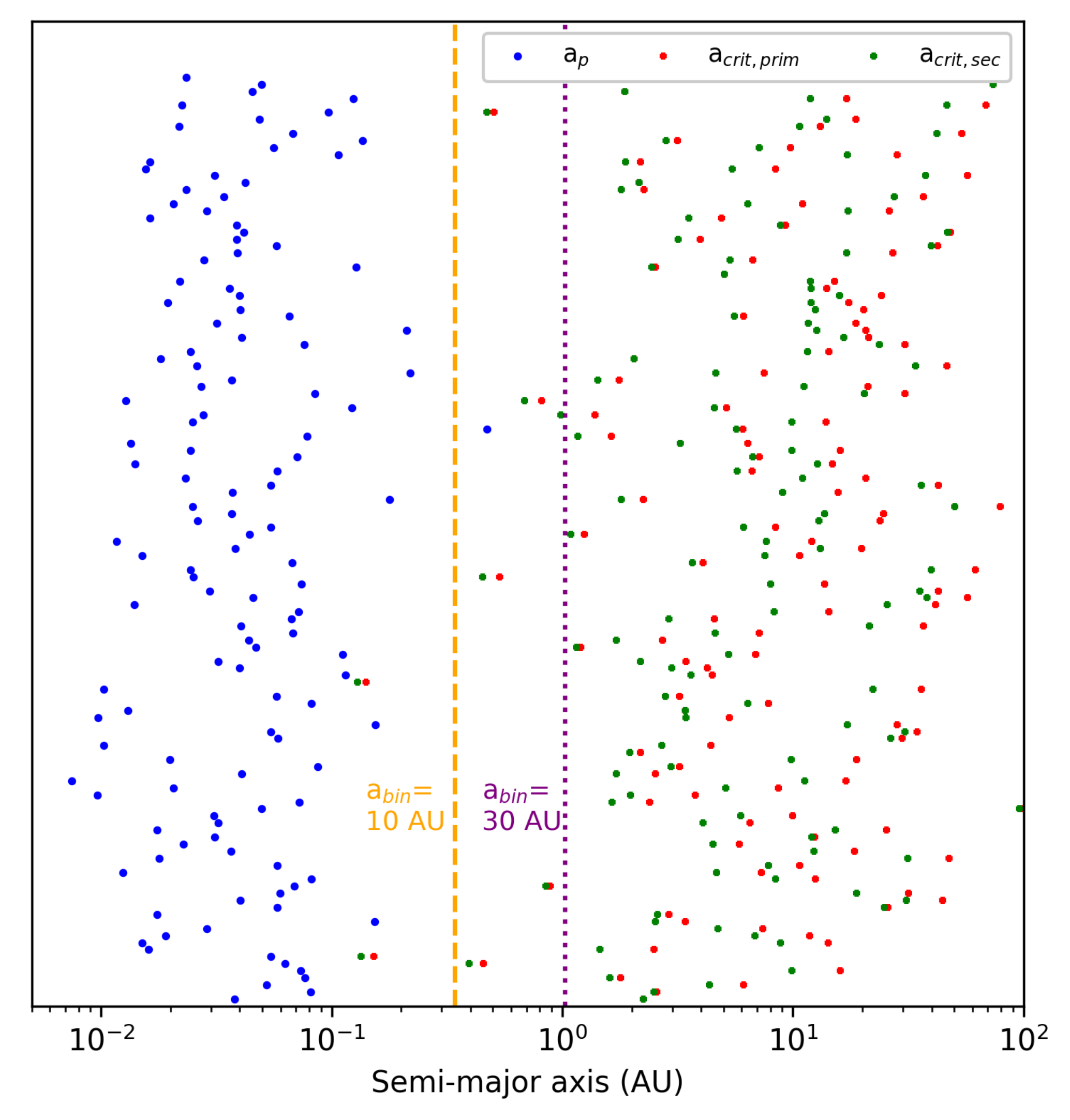}
    \caption{The derived semi-major axis of observed TESS planet candidates, $a_{p}$, compared to the critical stability semi-major axis, $a_{crit}$, for both the primary and secondary host scenario. If $a_{p}>a_{crit}$, the planetary orbit will be unstable due to dynamical interactions with the companion star. All observed planets are well within the stability limit. For reference, the  $a_{crit}$ values for a planet in an equal-mass binary with a highly eccentric ($e=$0.8) $a_{bin}$=10 and 30 AU orbit are also plotted.}
    \label{fig:stability}
\end{figure}

The $a_{p}$ and $a_{crit}$ for the observed TESS planet candidates in binary systems are plotted in Figure \ref{fig:stability}. All of the systems are well below the critical stability limit, by a median factor of 140 in their systems for either host star scenario. For reference, the $a_{crit}$ for a planet in a 10 and 30 AU equal-mass binary system on a highly eccentric orbit ($e_{bin}$= 0.8) is 0.34 and 1.0 AU, respectively. All our planet candidates orbit well within the limits deduced from these reasonable but extreme scenarios, so it is clear that close binary systems hosting TESS planet candidates are not disallowed due to dynamical stability. Binary companions must therefore suppress the formation or early evolution of planets in some manner.

\section{Discussion}\label{sec:discussion}

The suppression seen in our culled sample of TESS planet candidates is significantly stronger than that observed by the \citet{kraus16} survey of \textit{Kepler} planet candidates. While the best fit two-parameter model for the \textit{Kepler} targets had binaries reduced by approximately two-thirds at separations within 47 AU, we found for TESS planet candidates a close binary suppression factor of nearly seven below approximately 58 AU. More careful analysis of the closest pairs, many of which are likely (but not confirmed) false positives, suggests that even this substantial suppression model may be an underestimation of the true effect.

Why is the suppression of planets in close binaries found in this paper using TESS targets nearly twice that of the \textit{Kepler} targets investigated in \citet{kraus16}? Both surveys have similar sensitivity to close-in targets: while the TESS targets are a factor of a few closer on average, the \citeauthor{kraus16} survey used a larger-aperture telescope (10-m Keck telescope compared to the 4-m SOAR telescope) with adaptive optics corrections and non-redundant aperture masking, which provides sub-diffraction limited resolution. These two factors generally cancel each other out such that both surveys are able to probe nearly the same physical separation range around their respective targets.

The explanation may lie in the large differences in the two populations, one of which may be more effected by stellar companions. As shown in Figure 1 in Paper I, the TESS host stars are hotter and have larger planet candidates that orbit at shorter periods than the \textit{Kepler} sample. It maybe, for example, that nearby companion stars result in fewer close-in large planets, perhaps due to reducing the size and lifetime of the protoplanetary disk, as observed by \citet{kraus12}.

A contributing factor is that the TESS targets, bright and nearby, are more amendable to follow-up observations to identify false positives. Nearly a fifth of the observed targets by SOAR have been identified as not hosting transiting planets, a large fraction of which are eclipsing binaries with wide tertiary companions. Removing these systems increased the modeled suppression factor between Paper I and this work by a factor of 60$\%$ (from S$_{bin}$=0.24 previously to 0.15 here).

Close binary suppression also is apparent in M-dwarf planet candidate hosts, although at a low-significance due to the scarcity of systems. Taken together, this suggests that, if not accounted for, binary suppression may result in a factor of two overestimation of the planet occurrence rate in the Galaxy.

The second result in Paper I of this survey, an increase in companion fraction at wide separations around planet candidate hosting stars, appears to be firmly attributable to the false-positive contamination. Unlike the distant, faint stars discovered by \textit{Kepler} which were largely out of reach of ground-based follow up for all but the largest  telescopes, many of the TESS targets are readily accessible to follow-up by sub-meter size telescopes. The timely identification of the false-positive systems has allowed us to remove a large fraction from our sample, clarifying that, as suggested in \citet{moe19},  tertiary companions to close eclipsing binaries likely resulted in the high binary rates at wide separations.

\section{Conclusions}\label{sec:conclusions}

We presented the second year results of the SOAR TESS survey, including 589 newly targeted planet candidates. In the combined survey with false positives removed, we find that 13$\%$ of TESS planet candidates have companions within 1\farcs5. If the planetary candidates orbit the secondary stars, estimates of their radii may rise by a factor of a few.

We also compare the companion fraction of planet-hosting stars to field stars at different orbital separations. In our larger sample, with significantly fewer false positives than in Paper I, we find firm evidence that close binaries disrupt the existence of transiting planets. This effect, if not accounted for, will result in planet occurrence rates being overestimated by a factor of two in magnitude-limited samples. The removal of much of the false positive contamination also explained the anomalously high wide binary rates seen in Paper I.

SOAR observations of TESS planet candidate hosts are on-going. Along with the continuing efforts of the TESS follow-up team to vet planet candidates, this will result in an ever-growing and improving list of candidate planets. Further observations will confirm the association of detected multiple systems and may lead to orbital solutions, providing more details on these systems than our initial discovery snapshot.

\acknowledgments

C.Z. is supported by a Dunlap Fellowship at the Dunlap Institute for Astronomy \& Astrophysics, funded through an endowment established by the Dunlap family and the University of Toronto. A.W.M was supported by NASA grant 80NSSC19K0097 to the University of North Carolina at Chapel Hill.

Based on observations obtained at the Southern Astrophysical Research (SOAR) telescope, which is a joint project of the Minist\'{e}rio da Ci\^{e}ncia, Tecnologia, Inova\c{c}\~{o}es e Comunica\c{c}\~{o}es (MCTIC) do Brasil, the U.S. National Optical Astronomy Observatory (NOAO), the University of North Carolina at Chapel Hill (UNC), and Michigan State University (MSU).

This paper includes data collected by the TESS mission. Funding for the TESS mission is provided by the NASA Explorer Program. This work has made use of data from the European Space Agency (ESA) mission {\it Gaia} (\url{https://www.cosmos.esa.int/gaia}), processed by the {\it Gaia} Data Processing and Analysis Consortium (DPAC, \url{https://www.cosmos.esa.int/web/gaia/dpac/consortium}). Funding for the DPAC has been provided by national institutions, in particular the institutions participating in the {\it Gaia} Multilateral Agreement. This research has made use of the Exoplanet Follow-up Observation Program website, which is operated by the California Institute of Technology, under contract with the National Aeronautics and Space Administration under the Exoplanet Exploration Program. This work made use of the Washington Double Star Catalog maintained at USNO.

\vspace{5mm}
\facilities{SOAR (HRCam), TESS, Gaia}

\software{astropy \citep{astropy:2013, astropy:2018}, emcee \citep{emcee}, corner \citep{corner}}

\bibliographystyle{aasjournal}
\bibliography{refs.bib}

\begin{thebibliography}{}
\expandafter\ifx\csname natexlab\endcsname\relax\def\natexlab#1{#1}\fi
\providecommand{\url}[1]{\href{#1}{#1}}
\providecommand{\dodoi}[1]{doi:~\href{http://doi.org/#1}{\nolinkurl{#1}}}
\providecommand{\doeprint}[1]{\href{http://ascl.net/#1}{\nolinkurl{http://ascl.net/#1}}}
\providecommand{\doarXiv}[1]{\href{https://arxiv.org/abs/#1}{\nolinkurl{https://arxiv.org/abs/#1}}}

\bibitem[{{Arenou} {et~al.}(2017){Arenou}, {Luri}, {Babusiaux}, {Fabricius},
  {Helmi}, {Robin}, {Vallenari}, {Blanco-Cuaresma}, {Cantat-Gaudin},
  {Findeisen}, {Reyl{\'e}}, {Ruiz-Dern}, {Sordo}, {Turon}, {Walton}, {Shih},
  {Antiche}, {Barache}, {Barros}, {Breddels}, {Carrasco}, {Costigan},
  {Diakit{\'e}}, {Eyer}, {Figueras}, {Galluccio}, {Heu}, {Jordi},
  {Krone-Martins}, {Lallement}, {Lambert}, {Leclerc}, {Marrese}, {Moitinho},
  {Mor}, {Romero-G{\'o}mez}, {Sartoretti}, {Soria}, {Soubiran}, {Souchay},
  {Veljanoski}, {Ziaeepour}, {Giuffrida}, {Pancino}, \& {Bragaglia}}]{arenou17}
{Arenou}, F., {Luri}, X., {Babusiaux}, C., {et~al.} 2017, \aap, 599, A50,
  \dodoi{10.1051/0004-6361/201629895}

\bibitem[{{Arenou} {et~al.}(2018){Arenou}, {Luri}, {Babusiaux}, {Fabricius},
  {Helmi}, {Muraveva}, {Robin}, {Spoto}, {Vallenari}, \& {Antoja}}]{arenou18}
---. 2018, \aap, 616, A17, \dodoi{10.1051/0004-6361/201833234}

\bibitem[{{Astropy Collaboration} {et~al.}(2013){Astropy Collaboration},
  {Robitaille}, {Tollerud}, {Greenfield}, {Droettboom}, {Bray}, {Aldcroft},
  {Davis}, {Ginsburg}, {Price-Whelan}, {Kerzendorf}, {Conley}, {Crighton},
  {Barbary}, {Muna}, {Ferguson}, {Grollier}, {Parikh}, {Nair}, {Unther},
  {Deil}, {Woillez}, {Conseil}, {Kramer}, {Turner}, {Singer}, {Fox}, {Weaver},
  {Zabalza}, {Edwards}, {Azalee Bostroem}, {Burke}, {Casey}, {Crawford},
  {Dencheva}, {Ely}, {Jenness}, {Labrie}, {Lim}, {Pierfederici}, {Pontzen},
  {Ptak}, {Refsdal}, {Servillat}, \& {Streicher}}]{astropy:2013}
{Astropy Collaboration}, {Robitaille}, T.~P., {Tollerud}, E.~J., {et~al.} 2013,
  \aap, 558, A33, \dodoi{10.1051/0004-6361/201322068}

\bibitem[{{Bailer-Jones} {et~al.}(2018){Bailer-Jones}, {Rybizki}, {Fouesneau},
  {Mantelet}, \& {Andrae}}]{bailerjones18}
{Bailer-Jones}, C.~A.~L., {Rybizki}, J., {Fouesneau}, M., {Mantelet}, G., \&
  {Andrae}, R. 2018, \aj, 156, 58, \dodoi{10.3847/1538-3881/aacb21}

\bibitem[{{Bakos}(2018)}]{hats}
{Bakos}, G.~{\'A}. 2018, {The HATNet and HATSouth Exoplanet Surveys}, 111,
  \dodoi{10.1007/978-3-319-55333-7_111}

\bibitem[{{Barclay} {et~al.}(2018){Barclay}, {Pepper}, \&
  {Quintana}}]{barclay18}
{Barclay}, T., {Pepper}, J., \& {Quintana}, E.~V. 2018, The Astrophysical
  Journal Supplement Series, 239, 2, \dodoi{10.3847/1538-4365/aae3e9}

\bibitem[{{Ciardi} {et~al.}(2015){Ciardi}, {Beichman}, {Horch}, \&
  {Howell}}]{ciardi15}
{Ciardi}, D.~R., {Beichman}, C.~A., {Horch}, E.~P., \& {Howell}, S.~B. 2015,
  \apj, 805, 16, \dodoi{10.1088/0004-637X/805/1/16}

\bibitem[{{Cooke} {et~al.}(2018){Cooke}, {Pollacco}, {West}, {McCormac}, \&
  {Wheatley}}]{cooke18}
{Cooke}, B.~F., {Pollacco}, D., {West}, R., {McCormac}, J., \& {Wheatley},
  P.~J. 2018, \aap, 619, A175, \dodoi{10.1051/0004-6361/201834014}

\bibitem[{{Dotter} {et~al.}(2008){Dotter}, {Chaboyer}, {Jevremovi{\'c}},
  {Kostov}, {Baron}, \& {Ferguson}}]{dotter08}
{Dotter}, A., {Chaboyer}, B., {Jevremovi{\'c}}, D., {et~al.} 2008, \apjs, 178,
  89, \dodoi{10.1086/589654}

\bibitem[{{Dressing} \& {Charbonneau}(2013)}]{dressing13}
{Dressing}, C.~D., \& {Charbonneau}, D. 2013, \apj, 767, 95,
  \dodoi{10.1088/0004-637X/767/1/95}

\bibitem[{{Eastman} {et~al.}(2013){Eastman}, {Gaudi}, \& {Agol}}]{exofast}
{Eastman}, J., {Gaudi}, B.~S., \& {Agol}, E. 2013, \pasp, 125, 83,
  \dodoi{10.1086/669497}

\bibitem[{{Evans} {et~al.}(2016){Evans}, {Southworth}, {Maxted}, {Skottfelt},
  {Hundertmark}, {J{\o}rgensen}, {Dominik}, {Alsubai}, {Andersen}, {Bozza},
  {Bramich}, {Burgdorf}, {Ciceri}, {D'Ago}, {Figuera Jaimes}, {Gu},
  {Haugb{\o}lle}, {Hinse}, {Juncher}, {Kains}, {Kerins}, {Korhonen},
  {Kuffmeier}, {Mancini}, {Peixinho}, {Popovas}, {Rabus}, {Rahvar}, {Schmidt},
  {Snodgrass}, {Starkey}, {Surdej}, {Tronsgaard}, {von Essen}, {Wang}, \&
  {Wertz}}]{evans16}
{Evans}, D.~F., {Southworth}, J., {Maxted}, P.~F.~L., {et~al.} 2016, \aap, 589,
  A58, \dodoi{10.1051/0004-6361/201527970}

\bibitem[{{Evans} {et~al.}(2018){Evans}, {Southworth}, {Smalley},
  {J{\o}rgensen}, {Dominik}, {Andersen}, {Bozza}, {Bramich}, {Burgdorf},
  {Ciceri}, {D'Ago}, {Figuera Jaimes}, {Gu}, {Hinse}, {Henning}, {Hundertmark},
  {Kains}, {Kerins}, {Korhonen}, {Kokotanekova}, {Kuffmeier}, {Longa-Pe{\~n}a},
  {Mancini}, {MacKenzie}, {Popovas}, {Rabus}, {Rahvar}, {Sajadian},
  {Snodgrass}, {Skottfelt}, {Surdej}, {Tronsgaard}, {Unda-Sanzana}, {von
  Essen}, {Wang}, \& {Wertz}}]{evans18}
{Evans}, D.~F., {Southworth}, J., {Smalley}, B., {et~al.} 2018, \aap, 610, A20,
  \dodoi{10.1051/0004-6361/201731855}

\bibitem[{{Fischer} \& {Marcy}(1992)}]{fischer92}
{Fischer}, D.~A., \& {Marcy}, G.~W. 1992, \apj, 396, 178,
  \dodoi{10.1086/171708}

\bibitem[{{Fontanive} {et~al.}(2019){Fontanive}, {Rice}, {Bonavita}, {Lopez},
  {Mu{\v{z}}i{\'c}}, {}, \& {Biller}}]{fontanive19}
{Fontanive}, C., {Rice}, K., {Bonavita}, M., {et~al.} 2019, \mnras, 485, 4967,
  \dodoi{10.1093/mnras/stz671}

\bibitem[{Foreman-Mackey(2016)}]{corner}
Foreman-Mackey, D. 2016, The Journal of Open Source Software, 1, 24,
  \dodoi{10.21105/joss.00024}

\bibitem[{{Foreman-Mackey} {et~al.}(2013){Foreman-Mackey}, {Hogg}, {Lang}, \&
  {Goodman}}]{emcee}
{Foreman-Mackey}, D., {Hogg}, D.~W., {Lang}, D., \& {Goodman}, J. 2013, \pasp,
  125, 306, \dodoi{10.1086/670067}

\bibitem[{{Fressin} {et~al.}(2013){Fressin}, {Torres}, {Charbonneau}, {Bryson},
  {Christiansen}, {Dressing}, {Jenkins}, {Walkowicz}, \& {Batalha}}]{fressin13}
{Fressin}, F., {Torres}, G., {Charbonneau}, D., {et~al.} 2013, \apj, 766, 81,
  \dodoi{10.1088/0004-637X/766/2/81}

\bibitem[{{Gaia Collaboration} {et~al.}(2016){Gaia Collaboration}, {Prusti},
  {de Bruijne}, {Brown}, {Vallenari}, {Babusiaux}, {Bailer-Jones}, {Bastian},
  {Biermann}, {Evans}, {Eyer}, {Jansen}, {Jordi}, {Klioner}, {Lammers},
  {Lindegren}, {Luri}, {Mignard}, {Milligan}, {Panem}, {Poinsignon},
  {Pourbaix}, {Randich}, {Sarri}, {Sartoretti}, {Siddiqui}, {Soubiran},
  {Valette}, {van Leeuwen}, {Walton}, {Aerts}, {Arenou}, {Cropper}, {Drimmel},
  {H{\o}g}, {Katz}, {Lattanzi}, {O'Mullane}, {Grebel}, {Holland}, {Huc},
  {Passot}, {Bramante}, {Cacciari}, {Casta{\~n}eda}, {Chaoul}, {Cheek}, {De
  Angeli}, {Fabricius}, {Guerra}, {Hern{\'a}ndez}, {Jean-Antoine-Piccolo},
  {Masana}, {Messineo}, {Mowlavi}, {Nienartowicz}, {Ord{\'o}{\~n}ez-Blanco},
  {Panuzzo}, {Portell}, {Richards}, {Riello}, {Seabroke}, {Tanga},
  {Th{\'e}venin}, {Torra}, {Els}, {Gracia-Abril}, {Comoretto},
  {Garcia-Reinaldos}, {Lock}, {Mercier}, {Altmann}, {Andrae}, {Astraatmadja},
  {Bellas-Velidis}, {Benson}, {Berthier}, {Blomme}, {Busso}, {Carry},
  {Cellino}, {Clementini}, {Cowell}, {Creevey}, {Cuypers}, {Davidson}, {De
  Ridder}, {de Torres}, {Delchambre}, {Dell'Oro}, {Ducourant}, {Fr{\'e}mat},
  {Garc{\'\i}a-Torres}, {Gosset}, {Halbwachs}, {Hambly}, {Harrison}, {Hauser},
  {Hestroffer}, {Hodgkin}, {Huckle}, {Hutton}, {Jasniewicz}, {Jordan},
  {Kontizas}, {Korn}, {Lanzafame}, {Manteiga}, {Moitinho}, {Muinonen},
  {Osinde}, {Pancino}, {Pauwels}, {Petit}, {Recio-Blanco}, {Robin}, {Sarro},
  {Siopis}, {Smith}, {Smith}, {Sozzetti}, {Thuillot}, {van Reeven}, {Viala},
  {Abbas}, {Abreu Aramburu}, {Accart}, {Aguado}, {Allan}, {Allasia},
  {Altavilla}, {{\'A}lvarez}, {Alves}, {Anderson}, {Andrei}, {Anglada Varela},
  {Antiche}, {Antoja}, {Ant{\'o}n}, {Arcay}, {Atzei}, {Ayache}, {Bach},
  {Baker}, {Balaguer-N{\'u}{\~n}ez}, {Barache}, {Barata}, {Barbier}, {Barblan},
  {Baroni}, {Barrado y Navascu{\'e}s}, {Barros}, {Barstow}, {Becciani},
  {Bellazzini}, {Bellei}, {Bello Garc{\'\i}a}, {Belokurov}, {Bendjoya},
  {Berihuete}, {Bianchi}, {Bienaym{\'e}}, {Billebaud}, {Blagorodnova},
  {Blanco-Cuaresma}, {Boch}, {Bombrun}, {Borrachero}, {Bouquillon}, {Bourda},
  {Bouy}, {Bragaglia}, {Breddels}, {Brouillet}, {Br{\"u}semeister},
  {Bucciarelli}, {Budnik}, {Burgess}, {Burgon}, {Burlacu}, {Busonero}, {Buzzi},
  {Caffau}, {Cambras}, {Campbell}, {Cancelliere}, {Cantat-Gaudin}, {Carlucci},
  {Carrasco}, {Castellani}, {Charlot}, {Charnas}, {Charvet}, {Chassat},
  {Chiavassa}, {Clotet}, {Cocozza}, {Collins}, {Collins}, {Costigan}, {Crifo},
  {Cross}, {Crosta}, {Crowley}, {Dafonte}, {Damerdji}, {Dapergolas}, {David},
  {David}, {De Cat}, {de Felice}, {de Laverny}, {De Luise}, {De March}, {de
  Martino}, {de Souza}, {Debosscher}, {del Pozo}, {Delbo}, {Delgado},
  {Delgado}, {di Marco}, {Di Matteo}, {Diakite}, {Distefano}, {Dolding}, {Dos
  Anjos}, {Drazinos}, {Dur{\'a}n}, {Dzigan}, {Ecale}, {Edvardsson}, {Enke},
  {Erdmann}, {Escolar}, {Espina}, {Evans}, {Eynard Bontemps}, {Fabre},
  {Fabrizio}, {Faigler}, {Falc{\~a}o}, {Farr{\`a}s Casas}, {Faye}, {Federici},
  {Fedorets}, {Fern{\'a}ndez-Hern{\'a}ndez}, {Fernique}, {Fienga}, {Figueras},
  {Filippi}, {Findeisen}, {Fonti}, {Fouesneau}, {Fraile}, {Fraser}, {Fuchs},
  {Furnell}, {Gai}, {Galleti}, {Galluccio}, {Garabato}, {Garc{\'\i}a-Sedano},
  {Gar{\'e}}, {Garofalo}, {Garralda}, {Gavras}, {Gerssen}, {Geyer}, {Gilmore},
  {Girona}, {Giuffrida}, {Gomes}, {Gonz{\'a}lez-Marcos},
  {Gonz{\'a}lez-N{\'u}{\~n}ez}, {Gonz{\'a}lez-Vidal}, {Granvik}, {Guerrier},
  {Guillout}, {Guiraud}, {G{\'u}rpide}, {Guti{\'e}rrez-S{\'a}nchez}, {Guy},
  {Haigron}, {Hatzidimitriou}, {Haywood}, {Heiter}, {Helmi}, {Hobbs},
  {Hofmann}, {Holl}, {Holland }, {Hunt}, {Hypki}, {Icardi}, {Irwin}, {Jevardat
  de Fombelle}, {Jofr{\'e}}, {Jonker}, {Jorissen}, {Julbe}, {Karampelas},
  {Kochoska}, {Kohley}, {Kolenberg}, {Kontizas}, {Koposov}, {Kordopatis},
  {Koubsky}, {Kowalczyk}, {Krone-Martins}, {Kudryashova}, {Kull}, {Bachchan},
  {Lacoste-Seris}, {Lanza}, {Lavigne}, {Le Poncin-Lafitte}, {Lebreton},
  {Lebzelter}, {Leccia}, {Leclerc}, {Lecoeur-Taibi}, {Lemaitre}, {Lenhardt},
  {Leroux}, {Liao}, {Licata}, {Lindstr{\o}m}, {Lister}, {Livanou}, {Lobel},
  {L{\"o}ffler}, {L{\'o}pez}, {Lopez-Lozano}, {Lorenz}, {Loureiro},
  {MacDonald}, {Magalh{\~a}es Fernandes}, {Managau}, {Mann}, {Mantelet},
  {Marchal}, {Marchant}, {Marconi}, {Marie}, {Marinoni}, {Marrese},
  {Marschalk{\'o}}, {Marshall}, {Mart{\'\i}n-Fleitas}, {Martino}, {Mary},
  {Matijevi{\v{c}}}, {Mazeh}, {McMillan}, {Messina}, {Mestre}, {Michalik},
  {Millar}, {Miranda}, {Molina}, {Molinaro}, {Molinaro}, {Moln{\'a}r},
  {Moniez}, {Montegriffo}, {Monteiro}, {Mor}, {Mora}, {Morbidelli}, {Morel},
  {Morgenthaler}, {Morley}, {Morris}, {Mulone}, {Muraveva}, {Musella},
  {Narbonne}, {Nelemans}, {Nicastro}, {Noval}, {Ord{\'e}novic},
  {Ordieres-Mer{\'e}}, {Osborne}, {Pagani}, {Pagano}, {Pailler}, {Palacin},
  {Palaversa}, {Parsons}, {Paulsen}, {Pecoraro}, {Pedrosa}, {Pentik{\"a}inen},
  {Pereira}, {Pichon}, {Piersimoni}, {Pineau}, {Plachy}, {Plum}, {Poujoulet},
  {Pr{\v{s}}a}, {Pulone}, {Ragaini}, {Rago}, {Rambaux}, {Ramos-Lerate},
  {Ranalli}, {Rauw}, {Read}, {Regibo}, {Renk}, {Reyl{\'e}}, {Ribeiro},
  {Rimoldini}, {Ripepi}, {Riva}, {Rixon}, {Roelens}, {Romero-G{\'o}mez},
  {Rowell}, {Royer}, {Rudolph}, {Ruiz-Dern}, {Sadowski}, {Sagrist{\`a}
  Sell{\'e}s}, {Sahlmann}, {Salgado}, {Salguero}, {Sarasso}, {Savietto},
  {Schnorhk}, {Schultheis}, {Sciacca}, {Segol}, {Segovia}, {Segransan},
  {Serpell}, {Shih}, {Smareglia}, {Smart}, {Smith}, {Solano}, {Solitro},
  {Sordo}, {Soria Nieto}, {Souchay}, {Spagna}, {Spoto}, {Stampa}, {Steele},
  {Steidelm{\"u}ller}, {Stephenson}, {Stoev}, {Suess}, {S{\"u}veges}, {Surdej},
  {Szabados}, {Szegedi-Elek}, {Tapiador}, {Taris}, {Tauran}, {Taylor},
  {Teixeira}, {Terrett}, {Tingley}, {Trager}, {Turon}, {Ulla}, {Utrilla},
  {Valentini}, {van Elteren}, {Van Hemelryck}, {van Leeuwen}, {Varadi},
  {Vecchiato}, {Veljanoski}, {Via}, {Vicente}, {Vogt}, {Voss}, {Votruba},
  {Voutsinas}, {Walmsley}, {Weiler}, {Weingrill}, {Werner}, {Wevers},
  {Whitehead}, {Wyrzykowski}, {Yoldas}, {{\v{Z}}erjal}, {Zucker}, {Zurbach},
  {Zwitter}, {Alecu}, {Allen}, {Allende Prieto}, {Amorim},
  {Anglada-Escud{\'e}}, {Arsenijevic}, {Azaz}, {Balm}, {Beck}, {Bernstein},
  {Bigot}, {Bijaoui}, {Blasco}, {Bonfigli}, {Bono}, {Boudreault}, {Bressan},
  {Brown}, {Brunet}, {Bunclark}, {Buonanno}, {Butkevich}, {Carret}, {Carrion},
  {Chemin}, {Ch{\'e}reau}, {Corcione}, {Darmigny}, {de Boer}, {de Teodoro}, {de
  Zeeuw}, {Delle Luche}, {Domingues}, {Dubath}, {Fodor}, {Fr{\'e}zouls},
  {Fries}, {Fustes}, {Fyfe}, {Gallardo}, {Gallegos}, {Gardiol}, {Gebran},
  {Gomboc}, {G{\'o}mez}, {Grux}, {Gueguen}, {Heyrovsky}, {Hoar}, {Iannicola},
  {Isasi Parache}, {Janotto}, {Joliet}, {Jonckheere}, {Keil}, {Kim},
  {Klagyivik}, {Klar}, {Knude}, {Kochukhov}, {Kolka}, {Kos}, {Kutka}, {Lainey},
  {LeBouquin}, {Liu}, {Loreggia}, {Makarov}, {Marseille}, {Martayan},
  {Martinez-Rubi}, {Massart}, {Meynadier}, {Mignot}, {Munari}, {Nguyen},
  {Nordlander}, {Ocvirk}, {O'Flaherty}, {Olias Sanz}, {Ortiz}, {Osorio},
  {Oszkiewicz}, {Ouzounis}, {Palmer}, {Park}, {Pasquato}, {Peltzer}, {Peralta},
  {P{\'e}turaud}, {Pieniluoma}, {Pigozzi}, {Poels}, {Prat}, {Prod'homme},
  {Raison}, {Rebordao}, {Risquez}, {Rocca-Volmerange}, {Rosen}, {Ruiz-Fuertes},
  {Russo}, {Sembay}, {Serraller Vizcaino}, {Short}, {Siebert}, {Silva},
  {Sinachopoulos}, {Slezak}, {Soffel}, {Sosnowska}, {Strai{\v{z}}ys}, {ter
  Linden}, {Terrell}, {Theil}, {Tiede}, {Troisi}, {Tsalmantza}, {Tur},
  {Vaccari}, {Vachier}, {Valles}, {Van Hamme}, {Veltz}, {Virtanen}, {Wallut},
  {Wichmann}, {Wilkinson}, {Ziaeepour}, \& {Zschocke}}]{gaia2016}
{Gaia Collaboration}, {Prusti}, T., {de Bruijne}, J.~H.~J., {et~al.} 2016,
  Astronomy and Astrophysics, 595, A1, \dodoi{10.1051/0004-6361/201629272}

\bibitem[{{Gaia Collaboration} {et~al.}(2018){Gaia Collaboration}, {Brown},
  {Vallenari}, {Prusti}, {de Bruijne}, {Babusiaux}, {Bailer-Jones}, {Biermann},
  {Evans}, {Eyer}, {Jansen}, {Jordi}, {Klioner}, {Lammers}, {Lindegren},
  {Luri}, {Mignard}, {Panem}, {Pourbaix}, {Randich}, {Sartoretti}, {Siddiqui},
  {Soubiran}, {van Leeuwen}, {Walton}, {Arenou}, {Bastian}, {Cropper},
  {Drimmel}, {Katz}, {Lattanzi}, {Bakker}, {Cacciari}, {Casta{\~n}eda},
  {Chaoul}, {Cheek}, {De Angeli}, {Fabricius}, {Guerra}, {Holl}, {Masana},
  {Messineo}, {Mowlavi}, {Nienartowicz}, {Panuzzo}, {Portell}, {Riello},
  {Seabroke}, {Tanga}, {Th{\'e}venin}, {Gracia-Abril}, {Comoretto},
  {Garcia-Reinaldos}, {Teyssier}, {Altmann}, {Andrae}, {Audard},
  {Bellas-Velidis}, {Benson}, {Berthier}, {Blomme}, {Burgess}, {Busso},
  {Carry}, {Cellino}, {Clementini}, {Clotet}, {Creevey}, {Davidson}, {De
  Ridder}, {Delchambre}, {Dell'Oro}, {Ducourant},
  {Fern{\'a}ndez-Hern{\'a}ndez}, {Fouesneau}, {Fr{\'e}mat}, {Galluccio},
  {Garc{\'\i}a-Torres}, {Gonz{\'a}lez-N{\'u}{\~n}ez}, {Gonz{\'a}lez-Vidal},
  {Gosset}, {Guy}, {Halbwachs}, {Hambly}, {Harrison}, {Hern{\'a}ndez},
  {Hestroffer}, {Hodgkin}, {Hutton}, {Jasniewicz}, {Jean-Antoine-Piccolo},
  {Jordan}, {Korn}, {Krone-Martins}, {Lanzafame}, {Lebzelter}, {L{\"o}ffler},
  {Manteiga}, {Marrese}, {Mart{\'\i}n-Fleitas}, {Moitinho}, {Mora}, {Muinonen},
  {Osinde}, {Pancino}, {Pauwels}, {Petit}, {Recio-Blanco}, {Richards},
  {Rimoldini}, {Robin}, {Sarro}, {Siopis}, {Smith}, {Sozzetti}, {S{\"u}veges},
  {Torra}, {van Reeven}, {Abbas}, {Abreu Aramburu}, {Accart}, {Aerts},
  {Altavilla}, {{\'A}lvarez}, {Alvarez}, {Alves}, {Anderson}, {Andrei},
  {Anglada Varela}, {Antiche}, {Antoja}, {Arcay}, {Astraatmadja}, {Bach},
  {Baker}, {Balaguer-N{\'u}{\~n}ez}, {Balm}, {Barache}, {Barata}, {Barbato},
  {Barblan}, {Barklem}, {Barrado}, {Barros}, {Barstow}, {Bartholom{\'e}
  Mu{\~n}oz}, {Bassilana}, {Becciani}, {Bellazzini}, {Berihuete}, {Bertone},
  {Bianchi}, {Bienaym{\'e}}, {Blanco-Cuaresma}, {Boch}, {Boeche}, {Bombrun},
  {Borrachero}, {Bossini}, {Bouquillon}, {Bourda}, {Bragaglia}, {Bramante},
  {Breddels}, {Bressan}, {Brouillet}, {Br{\"u}semeister}, {Brugaletta},
  {Bucciarelli}, {Burlacu}, {Busonero}, {Butkevich}, {Buzzi}, {Caffau},
  {Cancelliere}, {Cannizzaro}, {Cantat-Gaudin}, {Carballo}, {Carlucci},
  {Carrasco}, {Casamiquela}, {Castellani}, {Castro-Ginard}, {Charlot},
  {Chemin}, {Chiavassa}, {Cocozza}, {Costigan}, {Cowell}, {Crifo}, {Crosta},
  {Crowley}, {Cuypers}, {Dafonte}, {Damerdji}, {Dapergolas}, {David}, {David},
  {de Laverny}, {De Luise}, {De March}, {de Martino}, {de Souza}, {de Torres},
  {Debosscher}, {del Pozo}, {Delbo}, {Delgado}, {Delgado}, {Di Matteo},
  {Diakite}, {Diener}, {Distefano}, {Dolding}, {Drazinos}, {Dur{\'a}n},
  {Edvardsson}, {Enke}, {Eriksson}, {Esquej}, {Eynard Bontemps}, {Fabre},
  {Fabrizio}, {Faigler}, {Falc{\~a}o}, {Farr{\`a}s Casas}, {Federici},
  {Fedorets}, {Fernique}, {Figueras}, {Filippi}, {Findeisen}, {Fonti},
  {Fraile}, {Fraser}, {Fr{\'e}zouls}, {Gai}, {Galleti}, {Garabato},
  {Garc{\'\i}a-Sedano}, {Garofalo}, {Garralda}, {Gavel}, {Gavras}, {Gerssen},
  {Geyer}, {Giacobbe}, {Gilmore}, {Girona}, {Giuffrida}, {Glass}, {Gomes},
  {Granvik}, {Gueguen}, {Guerrier}, {Guiraud}, {Guti{\'e}rrez-S{\'a}nchez},
  {Haigron}, {Hatzidimitriou}, {Hauser}, {Haywood}, {Heiter}, {Helmi}, {Heu},
  {Hilger}, {Hobbs}, {Hofmann}, {Holland}, {Huckle}, {Hypki}, {Icardi},
  {Jan{\ss}en}, {Jevardat de Fombelle}, {Jonker}, {Juh{\'a}sz}, {Julbe},
  {Karampelas}, {Kewley}, {Klar}, {Kochoska}, {Kohley}, {Kolenberg},
  {Kontizas}, {Kontizas}, {Koposov}, {Kordopatis}, {Kostrzewa-Rutkowska},
  {Koubsky}, {Lambert}, {Lanza}, {Lasne}, {Lavigne}, {Le Fustec}, {Le
  Poncin-Lafitte}, {Lebreton}, {Leccia}, {Leclerc}, {Lecoeur-Taibi},
  {Lenhardt}, {Leroux}, {Liao}, {Licata}, {Lindstr{\o}m}, {Lister}, {Livanou},
  {Lobel}, {L{\'o}pez}, {Managau}, {Mann}, {Mantelet}, {Marchal}, {Marchant},
  {Marconi}, {Marinoni}, {Marschalk{\'o}}, {Marshall}, {Martino}, {Marton},
  {Mary}, {Massari}, {Matijevi{\v{c}}}, {Mazeh}, {McMillan}, {Messina},
  {Michalik}, {Millar}, {Molina}, {Molinaro}, {Moln{\'a}r}, {Montegriffo},
  {Mor}, {Morbidelli}, {Morel}, {Morris}, {Mulone}, {Muraveva}, {Musella},
  {Nelemans}, {Nicastro}, {Noval}, {O'Mullane}, {Ord{\'e}novic},
  {Ord{\'o}{\~n}ez-Blanco}, {Osborne}, {Pagani}, {Pagano}, {Pailler},
  {Palacin}, {Palaversa}, {Panahi}, {Pawlak}, {Piersimoni}, {Pineau}, {Plachy},
  {Plum}, {Poggio}, {Poujoulet}, {Pr{\v{s}}a}, {Pulone}, {Racero}, {Ragaini},
  {Rambaux}, {Ramos-Lerate}, {Regibo}, {Reyl{\'e}}, {Riclet}, {Ripepi}, {Riva},
  {Rivard}, {Rixon}, {Roegiers}, {Roelens}, {Romero-G{\'o}mez}, {Rowell},
  {Royer}, {Ruiz-Dern}, {Sadowski}, {Sagrist{\`a} Sell{\'e}s}, {Sahlmann},
  {Salgado}, {Salguero}, {Sanna}, {Santana-Ros}, {Sarasso}, {Savietto},
  {Schultheis}, {Sciacca}, {Segol}, {Segovia}, {S{\'e}gransan}, {Shih},
  {Siltala}, {Silva}, {Smart}, {Smith}, {Solano}, {Solitro}, {Sordo}, {Soria
  Nieto}, {Souchay}, {Spagna}, {Spoto}, {Stampa}, {Steele},
  {Steidelm{\"u}ller}, {Stephenson}, {Stoev}, {Suess}, {Surdej}, {Szabados},
  {Szegedi-Elek}, {Tapiador}, {Taris}, {Tauran}, {Taylor}, {Teixeira},
  {Terrett}, {Teyssand ier}, {Thuillot}, {Titarenko}, {Torra Clotet}, {Turon},
  {Ulla}, {Utrilla}, {Uzzi}, {Vaillant}, {Valentini}, {Valette}, {van Elteren},
  {Van Hemelryck}, {van Leeuwen}, {Vaschetto}, {Vecchiato}, {Veljanoski},
  {Viala}, {Vicente}, {Vogt}, {von Essen}, {Voss}, {Votruba}, {Voutsinas},
  {Walmsley}, {Weiler}, {Wertz}, {Wevers}, {Wyrzykowski}, {Yoldas},
  {{\v{Z}}erjal}, {Ziaeepour}, {Zorec}, {Zschocke}, {Zucker}, {Zurbach}, \&
  {Zwitter}}]{gaia}
{Gaia Collaboration}, {Brown}, A.~G.~A., {Vallenari}, A., {et~al.} 2018, \aap,
  616, A1, \dodoi{10.1051/0004-6361/201833051}

\bibitem[{{Giacalone} \& {Dressing}(2020)}]{triceratops}
{Giacalone}, S., \& {Dressing}, C.~D. 2020, arXiv e-prints, arXiv:2002.00691.
\newblock \doarXiv{2002.00691}

\bibitem[{{Henden} {et~al.}(2009){Henden}, {Welch}, {Terrell}, \&
  {Levine}}]{Henden09}
{Henden}, A.~A., {Welch}, D.~L., {Terrell}, D., \& {Levine}, S.~E. 2009, in
  American Astronomical Society Meeting Abstracts, Vol. 214, American
  Astronomical Society Meeting Abstracts \#214, 669

\bibitem[{{Hirsch} {et~al.}(2017){Hirsch}, {Ciardi}, {Howard}, {Everett},
  {Furlan}, {Saylors}, {Horch}, {Howell}, {Teske}, \& {Marcy}}]{hirsch17}
{Hirsch}, L.~A., {Ciardi}, D.~R., {Howard}, A.~W., {et~al.} 2017, \aj, 153,
  117, \dodoi{10.3847/1538-3881/153/3/117}

\bibitem[{{Horch} {et~al.}(2014){Horch}, {Howell}, {Everett}, \&
  {Ciardi}}]{horch14}
{Horch}, E.~P., {Howell}, S.~B., {Everett}, M.~E., \& {Ciardi}, D.~R. 2014,
  \apj, 795, 60, \dodoi{10.1088/0004-637X/795/1/60}

\bibitem[{{Howell} {et~al.}(2021){Howell}, {Matson}, {Ciardi}, {Everett},
  {Livingston}, {Scott}, {Horch}, \& {Winn}}]{howell21}
{Howell}, S.~B., {Matson}, R.~A., {Ciardi}, D.~R., {et~al.} 2021, arXiv
  e-prints, arXiv:2101.08671.
\newblock \doarXiv{2101.08671}

\bibitem[{{Huang} {et~al.}(2020{\natexlab{a}}){Huang}, {Vanderburg}, {P{\'a}l},
  {Sha}, {Yu}, {Fong}, {Fausnaugh}, {Shporer}, {Guerrero}, {Vanderspek}, \&
  {Ricker}}]{qlp}
{Huang}, C.~X., {Vanderburg}, A., {P{\'a}l}, A., {et~al.} 2020{\natexlab{a}},
  Research Notes of the American Astronomical Society, 4, 204,
  \dodoi{10.3847/2515-5172/abca2e}

\bibitem[{{Huang} {et~al.}(2020{\natexlab{b}}){Huang}, {Vanderburg}, {P{\'a}l},
  {Sha}, {Yu}, {Fong}, {Fausnaugh}, {Shporer}, {Guerrero}, {Vanderspek}, \&
  {Ricker}}]{huang20}
---. 2020{\natexlab{b}}, Research Notes of the American Astronomical Society,
  4, 204, \dodoi{10.3847/2515-5172/abca2e}

\bibitem[{{Jenkins} {et~al.}(2016){Jenkins}, {Twicken}, {McCauliff},
  {Campbell}, {Sanderfer}, {Lung}, {Mansouri-Samani}, {Girouard}, {Tenenbaum},
  {Klaus}, {Smith}, {Caldwell}, {Chacon}, {Henze}, {Heiges}, {Latham},
  {Morgan}, {Swade}, {Rinehart}, \& {Vanderspek}}]{jenkins16}
{Jenkins}, J.~M., {Twicken}, J.~D., {McCauliff}, S., {et~al.} 2016, in
  \procspie, Vol. 9913, Software and Cyberinfrastructure for Astronomy IV,
  99133E, \dodoi{10.1117/12.2233418}

\bibitem[{{Knutson} {et~al.}(2014){Knutson}, {Fulton}, {Montet}, {Kao}, {Ngo},
  {Howard}, {Crepp}, {Hinkley}, {Bakos}, \& {Batygin}}]{knutson13}
{Knutson}, H.~A., {Fulton}, B.~J., {Montet}, B.~T., {et~al.} 2014, \apj, 785,
  126, \dodoi{10.1088/0004-637X/785/2/126}

\bibitem[{{Kraus} \& {Hillenbrand}(2007)}]{kraus07}
{Kraus}, A.~L., \& {Hillenbrand}, L.~A. 2007, \aj, 134, 2340,
  \dodoi{10.1086/522831}

\bibitem[{{Kraus} {et~al.}(2012){Kraus}, {Ireland}, {Hillenbrand}, \&
  {Martinache}}]{kraus12}
{Kraus}, A.~L., {Ireland}, M.~J., {Hillenbrand}, L.~A., \& {Martinache}, F.
  2012, \apj, 745, 19, \dodoi{10.1088/0004-637X/745/1/19}

\bibitem[{{Kraus} {et~al.}(2016){Kraus}, {Ireland}, {Huber}, {Mann}, \&
  {Dupuy}}]{kraus16}
{Kraus}, A.~L., {Ireland}, M.~J., {Huber}, D., {Mann}, A.~W., \& {Dupuy}, T.~J.
  2016, \aj, 152, 8, \dodoi{10.3847/0004-6256/152/1/8}

\bibitem[{{Law} {et~al.}(2014){Law}, {Morton}, {Baranec}, {Riddle},
  {Ravichandran}, {Ziegler}, {Johnson}, {Tendulkar}, {Bui}, {Burse}, {Das},
  {Dekany}, {Kulkarni}, {Punnadi}, \& {Ramaprakash}}]{law14}
{Law}, N.~M., {Morton}, T., {Baranec}, C., {et~al.} 2014, \apj, 791, 35,
  \dodoi{10.1088/0004-637X/791/1/35}

\bibitem[{{Mason} {et~al.}(2009){Mason}, {Wycoff}, {Hartkopf}, {Douglass}, \&
  {Worley}}]{mason09}
{Mason}, B.~D., {Wycoff}, G.~L., {Hartkopf}, W.~I., {Douglass}, G.~G., \&
  {Worley}, C.~E. 2009, VizieR Online Data Catalog, B/wds

\bibitem[{{Matson} {et~al.}(2019){Matson}, {Howell}, \& {Ciardi}}]{matson19}
{Matson}, R.~A., {Howell}, S.~B., \& {Ciardi}, D.~R. 2019, \aj, 157, 211,
  \dodoi{10.3847/1538-3881/ab1755}

\bibitem[{{Moe} \& {Kratter}(2019)}]{moe19}
{Moe}, M., \& {Kratter}, K.~M. 2019, arXiv e-prints, arXiv:1912.01699.
\newblock \doarXiv{1912.01699}

\bibitem[{{Naoz} {et~al.}(2012){Naoz}, {Farr}, \& {Rasio}}]{naoz12}
{Naoz}, S., {Farr}, W.~M., \& {Rasio}, F.~A. 2012, \apjl, 754, L36,
  \dodoi{10.1088/2041-8205/754/2/L36}

\bibitem[{{Ngo} {et~al.}(2015){Ngo}, {Knutson}, {Hinkley}, {Crepp}, {Bechter},
  {Batygin}, {Howard}, {Johnson}, {Morton}, \& {Muirhead}}]{ngo15}
{Ngo}, H., {Knutson}, H.~A., {Hinkley}, S., {et~al.} 2015, \apj, 800, 138,
  \dodoi{10.1088/0004-637X/800/2/138}

\bibitem[{{Ngo} {et~al.}(2016){Ngo}, {Knutson}, {Hinkley}, {Bryan}, {Crepp},
  {Batygin}, {Crossfield}, {Hansen}, {Howard}, {Johnson}, {Mawet}, {Morton},
  {Muirhead}, \& {Wang}}]{ngo16}
---. 2016, \apj, 827, 8, \dodoi{10.3847/0004-637X/827/1/8}

\bibitem[{{Pecaut} \& {Mamajek}(2013)}]{pecaut}
{Pecaut}, M.~J., \& {Mamajek}, E.~E. 2013, The Astrophysical Journal Supplement
  Series, 208, 9, \dodoi{10.1088/0067-0049/208/1/9}

\bibitem[{{Pecaut} {et~al.}(2012){Pecaut}, {Mamajek}, \& {Bubar}}]{pecaut12}
{Pecaut}, M.~J., {Mamajek}, E.~E., \& {Bubar}, E.~J. 2012, \apj, 746, 154,
  \dodoi{10.1088/0004-637X/746/2/154}

\bibitem[{{Price-Whelan} {et~al.}(2018){Price-Whelan}, {Sip{\H{o}}cz},
  {G{\"u}nther}, {Lim}, {Crawford}, {Conseil}, {Shupe}, {Craig}, {Dencheva},
  {Ginsburg}, {VanderPlas}, {Bradley}, {P{\'e}rez-Su{\'a}rez}, {de Val-Borro},
  {Paper Contributors}, {Aldcroft}, {Cruz}, {Robitaille}, {Tollerud},
  {Coordination Committee}, {Ardelean}, {Babej}, {Bach}, {Bachetti}, {Bakanov},
  {Bamford}, {Barentsen}, {Barmby}, {Baumbach}, {Berry}, {Biscani}, {Boquien},
  {Bostroem}, {Bouma}, {Brammer}, {Bray}, {Breytenbach}, {Buddelmeijer},
  {Burke}, {Calderone}, {Cano Rodr{\'\i}guez}, {Cara}, {Cardoso}, {Cheedella},
  {Copin}, {Corrales}, {Crichton}, {D{\textquoteright}Avella}, {Deil},
  {Depagne}, {Dietrich}, {Donath}, {Droettboom}, {Earl}, {Erben}, {Fabbro},
  {Ferreira}, {Finethy}, {Fox}, {Garrison}, {Gibbons}, {Goldstein}, {Gommers},
  {Greco}, {Greenfield}, {Groener}, {Grollier}, {Hagen}, {Hirst}, {Homeier},
  {Horton}, {Hosseinzadeh}, {Hu}, {Hunkeler}, {Ivezi{\'c}}, {Jain}, {Jenness},
  {Kanarek}, {Kendrew}, {Kern}, {Kerzendorf}, {Khvalko}, {King}, {Kirkby},
  {Kulkarni}, {Kumar}, {Lee}, {Lenz}, {Littlefair}, {Ma}, {Macleod},
  {Mastropietro}, {McCully}, {Montagnac}, {Morris}, {Mueller}, {Mumford},
  {Muna}, {Murphy}, {Nelson}, {Nguyen}, {Ninan}, {N{\"o}the}, {Ogaz}, {Oh},
  {Parejko}, {Parley}, {Pascual}, {Patil}, {Patil}, {Plunkett}, {Prochaska},
  {Rastogi}, {Reddy Janga}, {Sabater}, {Sakurikar}, {Seifert}, {Sherbert},
  {Sherwood-Taylor}, {Shih}, {Sick}, {Silbiger}, {Singanamalla}, {Singer},
  {Sladen}, {Sooley}, {Sornarajah}, {Streicher}, {Teuben}, {Thomas},
  {Tremblay}, {Turner}, {Terr{\'o}n}, {van Kerkwijk}, {de la Vega}, {Watkins},
  {Weaver}, {Whitmore}, {Woillez}, {Zabalza}, \& {Contributors}}]{astropy:2018}
{Price-Whelan}, A.~M., {Sip{\H{o}}cz}, B.~M., {G{\"u}nther}, H.~M., {et~al.}
  2018, \aj, 156, 123, \dodoi{10.3847/1538-3881/aabc4f}

\bibitem[{{Quarles} {et~al.}(2019){Quarles}, {Li}, {Kostov}, \&
  {Haghighipour}}]{quarles20}
{Quarles}, B., {Li}, G., {Kostov}, V., \& {Haghighipour}, N. 2019, arXiv
  e-prints, arXiv:1912.11019.
\newblock \doarXiv{1912.11019}

\bibitem[{{Quintana} {et~al.}(2007){Quintana}, {Adams}, {Lissauer}, \&
  {Chambers}}]{quintana07}
{Quintana}, E.~V., {Adams}, F.~C., {Lissauer}, J.~J., \& {Chambers}, J.~E.
  2007, \apj, 660, 807, \dodoi{10.1086/512542}

\bibitem[{{Raghavan} {et~al.}(2010){Raghavan}, {McAlister}, {Henry}, {Latham},
  {Marcy}, {Mason}, {Gies}, {White}, \& {ten Brummelaar}}]{raghavan10}
{Raghavan}, D., {McAlister}, H.~A., {Henry}, T.~J., {et~al.} 2010, \apjs, 190,
  1, \dodoi{10.1088/0067-0049/190/1/1}

\bibitem[{{Ricker} {et~al.}(2014){Ricker}, {Winn}, {Vanderspek}, {Latham},
  {Bakos}, {Bean}, {Berta-Thompson}, {Brown}, {Buchhave}, {Butler}, {Butler},
  {Chaplin}, {Charbonneau}, {Christensen-Dalsgaard}, {Clampin}, {Deming},
  {Doty}, {De Lee}, {Dressing}, {Dunham}, {Endl}, {Fressin}, {Ge}, {Henning},
  {Holman}, {Howard}, {Ida}, {Jenkins}, {Jernigan}, {Johnson}, {Kaltenegger},
  {Kawai}, {Kjeldsen}, {Laughlin}, {Levine}, {Lin}, {Lissauer}, {MacQueen},
  {Marcy}, {McCullough}, {Morton}, {Narita}, {Paegert}, {Palle}, {Pepe},
  {Pepper}, {Quirrenbach}, {Rinehart}, {Sasselov}, {Sato}, {Seager},
  {Sozzetti}, {Stassun}, {Sullivan}, {Szentgyorgyi}, {Torres}, {Udry}, \&
  {Villasenor}}]{tess}
{Ricker}, G.~R., {Winn}, J.~N., {Vanderspek}, R., {et~al.} 2014, in Society of
  Photo-Optical Instrumentation Engineers (SPIE) Conference Series, Vol. 9143,
  Space Telescopes and Instrumentation 2014: Optical, Infrared, and Millimeter
  Wave, 914320, \dodoi{10.1117/12.2063489}

\bibitem[{{Schmidt}(1968)}]{schmidt68}
{Schmidt}, M. 1968, \apj, 151, 393, \dodoi{10.1086/149446}

\bibitem[{{Skrutskie} {et~al.}(2006){Skrutskie}, {Cutri}, {Stiening},
  {Weinberg}, {Schneider}, {Carpenter}, {Beichman}, {Capps}, {Chester},
  {Elias}, {Huchra}, {Liebert}, {Lonsdale}, {Monet}, {Price}, {Seitzer},
  {Jarrett}, {Kirkpatrick}, {Gizis}, {Howard}, {Evans}, {Fowler}, {Fullmer},
  {Hurt}, {Light}, {Kopan}, {Marsh}, {McCallon}, {Tam}, {Van Dyk}, \&
  {Wheelock}}]{skrutskie06}
{Skrutskie}, M.~F., {Cutri}, R.~M., {Stiening}, R., {et~al.} 2006, \aj, 131,
  1163, \dodoi{10.1086/498708}

\bibitem[{{Stassun} {et~al.}(2019){Stassun}, {Oelkers}, {Paegert}, {Torres},
  {Pepper}, {De Lee}, {Collins}, {Latham}, {Muirhead}, {Chittidi},
  {Rojas-Ayala}, {Fleming}, {Rose}, {Tenenbaum}, {Ting}, {Kane}, {Barclay},
  {Bean}, {Brassuer}, {Charbonneau}, {Lissauer}, {Mann}, {McLean}, {Mulally},
  {Narita}, {Plavchan}, {Ricker}, {Sasselov}, {Seager}, {Sharma}, {Shiao},
  {Sozzetti}, {Stello}, {Vand erspek}, {Wallace}, \& {Winn}}]{tic}
{Stassun}, K.~G., {Oelkers}, R.~J., {Paegert}, M., {et~al.} 2019, arXiv
  e-prints, arXiv:1905.10694.
\newblock \doarXiv{1905.10694}

\bibitem[{{Street} {et~al.}(2003){Street}, {Pollaco}, {Fitzsimmons}, {Keenan},
  {Horne}, {Kane}, {Collier Cameron}, {Lister}, {Haswell}, \& {Norton}}]{wasp}
{Street}, R.~A., {Pollaco}, D.~L., {Fitzsimmons}, A., {et~al.} 2003, in
  Astronomical Society of the Pacific Conference Series, Vol. 294, Scientific
  Frontiers in Research on Extrasolar Planets, ed. D.~{Deming} \& S.~{Seager},
  405--408.
\newblock \doarXiv{astro-ph/0208233}

\bibitem[{{Sullivan} {et~al.}(2015){Sullivan}, {Winn}, {Berta-Thompson},
  {Charbonneau}, {Deming}, {Dressing}, {Latham}, {Levine}, {McCullough},
  {Morton}, {Ricker}, {Vanderspek}, \& {Woods}}]{sullivan15}
{Sullivan}, P.~W., {Winn}, J.~N., {Berta-Thompson}, Z.~K., {et~al.} 2015, \apj,
  809, 77, \dodoi{10.1088/0004-637X/809/1/77}

\bibitem[{{Tokovinin}(2018)}]{tokovinin18}
{Tokovinin}, A. 2018, \pasp, 130, 035002, \dodoi{10.1088/1538-3873/aaa7d9}

\bibitem[{{Tokovinin} {et~al.}(2010){Tokovinin}, {Mason}, \&
  {Hartkopf}}]{tokovinin10}
{Tokovinin}, A., {Mason}, B.~D., \& {Hartkopf}, W.~I. 2010, \aj, 139, 743,
  \dodoi{10.1088/0004-6256/139/2/743}

\bibitem[{{Tokovinin} {et~al.}(2006){Tokovinin}, {Thomas}, {Sterzik}, \&
  {Udry}}]{tokovinin06}
{Tokovinin}, A., {Thomas}, S., {Sterzik}, M., \& {Udry}, S. 2006, \aap, 450,
  681, \dodoi{10.1051/0004-6361:20054427}

\bibitem[{{Van Eylen} \& {Albrecht}(2015)}]{vaneylen15}
{Van Eylen}, V., \& {Albrecht}, S. 2015, \apj, 808, 126,
  \dodoi{10.1088/0004-637X/808/2/126}

\bibitem[{{Van Eylen} {et~al.}(2019){Van Eylen}, {Albrecht}, {Huang},
  {MacDonald}, {Dawson}, {Cai}, {Foreman-Mackey}, {Lundkvist}, {Silva Aguirre},
  {Snellen}, \& {Winn}}]{vaneylen19}
{Van Eylen}, V., {Albrecht}, S., {Huang}, X., {et~al.} 2019, \aj, 157, 61,
  \dodoi{10.3847/1538-3881/aaf22f}

\bibitem[{{Wang} {et~al.}(2015){Wang}, {Fischer}, {Xie}, \& {Ciardi}}]{wang15}
{Wang}, J., {Fischer}, D.~A., {Xie}, J.-W., \& {Ciardi}, D.~R. 2015, \apj, 813,
  130, \dodoi{10.1088/0004-637X/813/2/130}

\bibitem[{{Wang} {et~al.}(2014){Wang}, {Xie}, {Barclay}, \& {Fischer}}]{wang14}
{Wang}, J., {Xie}, J.-W., {Barclay}, T., \& {Fischer}, D.~A. 2014, \apj, 783,
  4, \dodoi{10.1088/0004-637X/783/1/4}

\bibitem[{{Winn}(2010)}]{winn10}
{Winn}, J.~N. 2010, arXiv e-prints, arXiv:1001.2010.
\newblock \doarXiv{1001.2010}

\bibitem[{{Winters} {et~al.}(2019{\natexlab{a}}){Winters}, {Henry}, {Jao},
  {Subasavage}, {Chatelain}, {Slatten}, {Riedel}, {Silverstein}, \&
  {Payne}}]{winters19}
{Winters}, J.~G., {Henry}, T.~J., {Jao}, W.-C., {et~al.} 2019{\natexlab{a}},
  \aj, 157, 216, \dodoi{10.3847/1538-3881/ab05dc}

\bibitem[{{Winters} {et~al.}(2019{\natexlab{b}}){Winters}, {Medina}, {Irwin},
  {Charbonneau}, {Astudillo-Defru}, {Horch}, {Eastman}, {Halley Vrijmoet},
  {Henry}, {Diamond-Lowe}, {Winston}, {Barclay}, {Bonfils}, {Ricker},
  {Vanderspek}, {Latham}, {Seager}, {Winn}, {Jenkins}, {Udry}, {Twicken},
  {Teske}, {Tenenbaum}, {Pepe}, {Murgas}, {Muirhead}, {Mink}, {Lovis},
  {Levine}, {L{\'e}pine}, {Jao}, {Henze}, {Fur{\'e}sz}, {Forveille},
  {Figueira}, {Esquerdo}, {Dressing}, {D{\'\i}az}, {Delfosse}, {Burke},
  {Bouchy}, {Berlind}, \& {Almenara}}]{winters19b}
{Winters}, J.~G., {Medina}, A.~A., {Irwin}, J.~M., {et~al.} 2019{\natexlab{b}},
  \aj, 158, 152, \dodoi{10.3847/1538-3881/ab364d}

\bibitem[{{Ziegler} {et~al.}(2020){Ziegler}, {Tokovinin}, {Brice{\~n}o},
  {Mang}, {Law}, \& {Mann}}]{ziegler20}
{Ziegler}, C., {Tokovinin}, A., {Brice{\~n}o}, C., {et~al.} 2020, \aj, 159, 19,
  \dodoi{10.3847/1538-3881/ab55e9}

\bibitem[{{Ziegler} {et~al.}(2018{\natexlab{a}}){Ziegler}, {Law}, {Baranec},
  {Howard}, {Morton}, {Riddle}, {Duev}, {Salama}, {Jensen-Clem}, \&
  {Kulkarni}}]{ziegler18b}
{Ziegler}, C., {Law}, N.~M., {Baranec}, C., {et~al.} 2018{\natexlab{a}}, \aj,
  156, 83, \dodoi{10.3847/1538-3881/aace59}

\bibitem[{{Ziegler} {et~al.}(2018{\natexlab{b}}){Ziegler}, {Law}, {Baranec},
  {Riddle}, {Duev}, {Howard}, {Jensen-Clem}, {Kulkarni}, {Morton}, \&
  {Salama}}]{ziegler18a}
---. 2018{\natexlab{b}}, \aj, 155, 161, \dodoi{10.3847/1538-3881/aab042}

\bibitem[{{Ziegler} {et~al.}(2018{\natexlab{c}}){Ziegler}, {Law}, {Baranec},
  {Morton}, {Riddle}, {De Lee}, {Huber}, {Mahadevan}, \& {Pepper}}]{roboaogaia}
---. 2018{\natexlab{c}}, \aj, 156, 259, \dodoi{10.3847/1538-3881/aad80a}

\end{thebibliography}

\clearpage

\appendix

\begin{figure*}
    \centering
    \includegraphics[scale=.65]{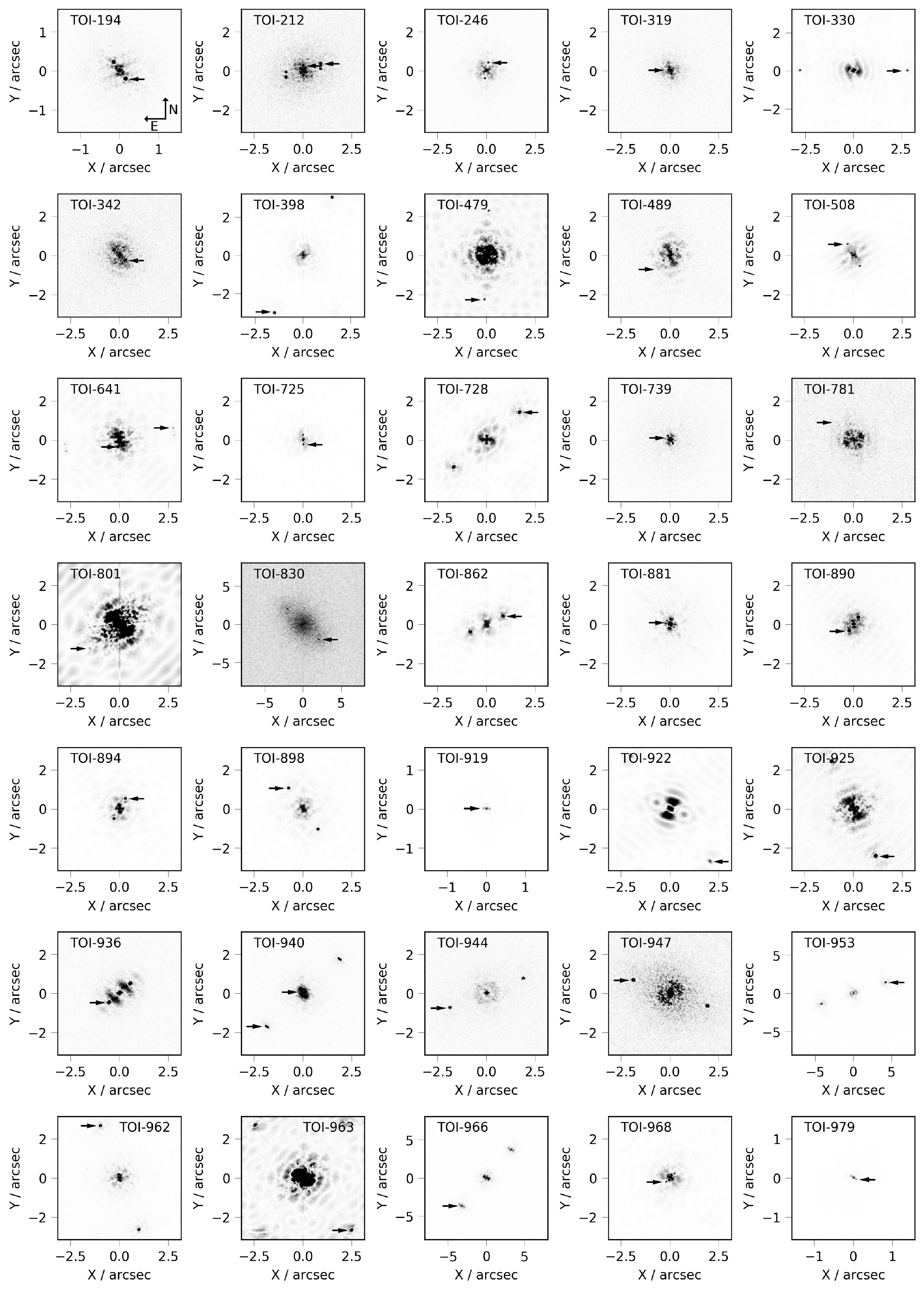}
    \caption{Speckle auto-correlation functions from SOAR speckle observing of TESS planet candidate hosts stars with resolved nearby stars. Each nearby star is mirrored in the images, with the true location marked by an arrow. Images are presented with an inverse linear scale for clarity. The orientation is similar in all images, with North pointed up and East to the left. A compass is shown in the top left image for reference.}
    \label{fig:grid1}
\end{figure*}

\begin{figure*}
    \centering
    \includegraphics[scale=.65]{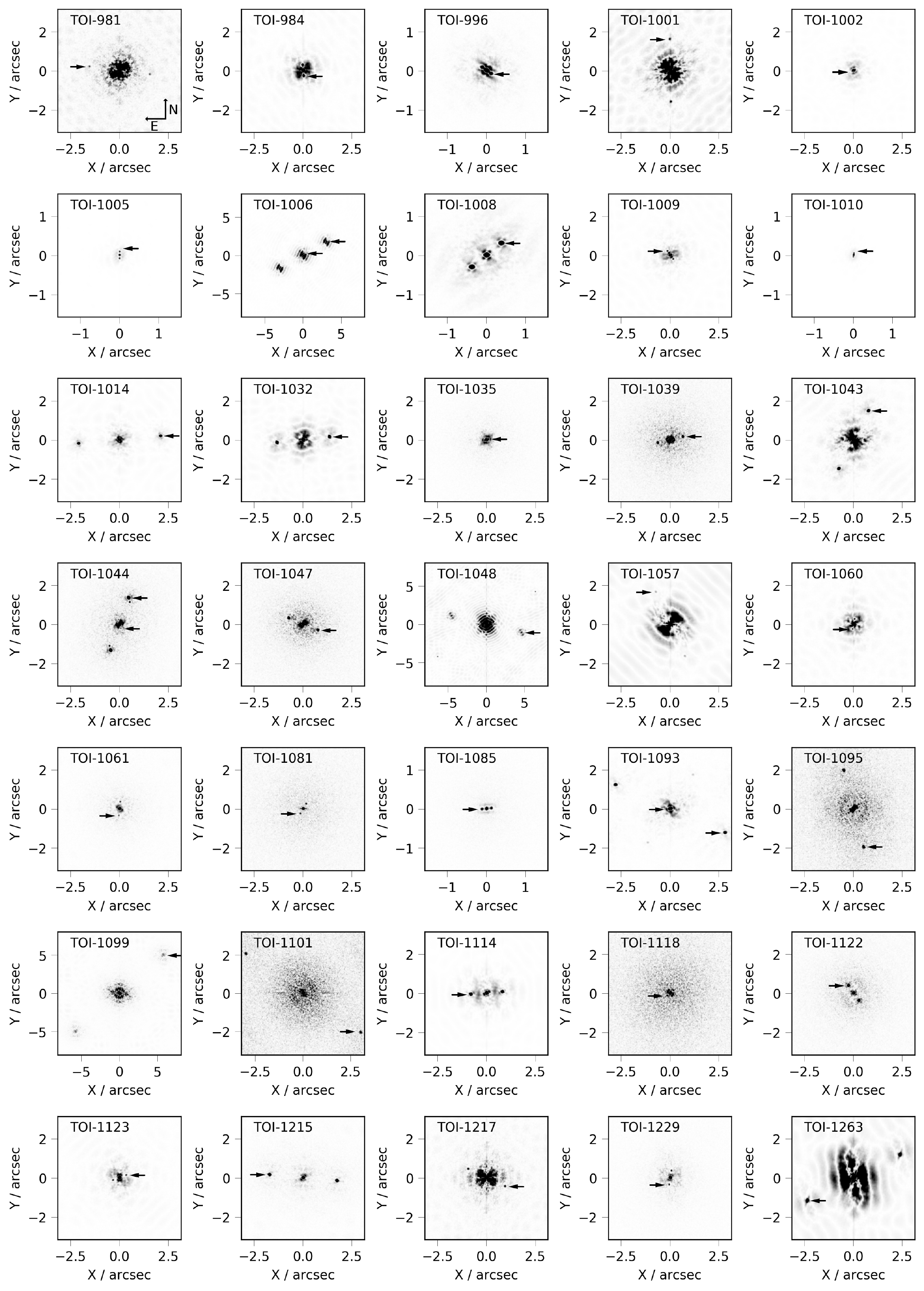}
    \caption{Similar to Figure \ref{fig:grid1}.}
    \label{fig:grid2}
\end{figure*}

\begin{figure*}
    \centering
    \includegraphics[scale=.65]{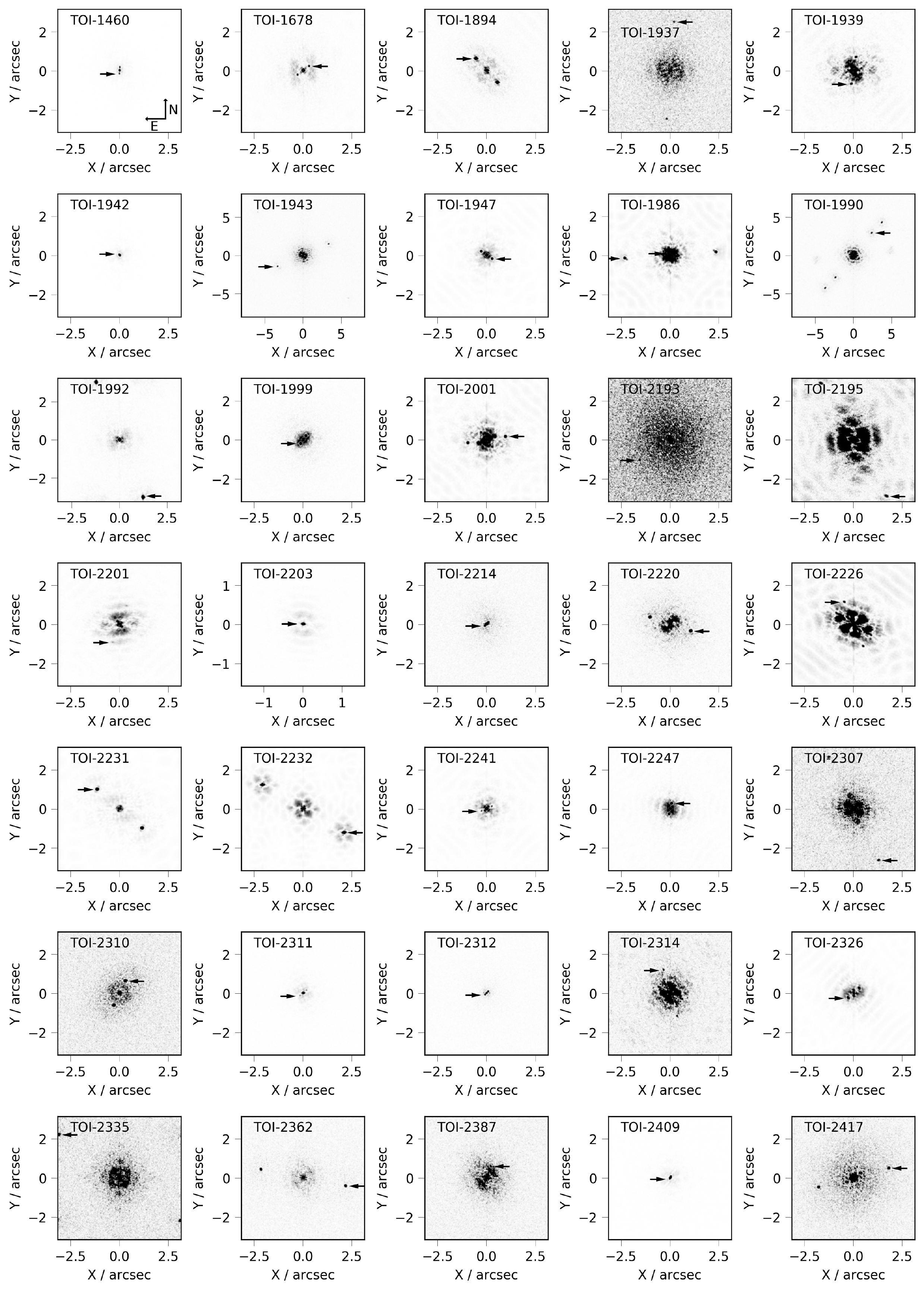}
    \caption{Similar to Figure \ref{fig:grid1}.}
    \label{fig:grid3}
\end{figure*}

\begin{figure*}
    \centering
    \includegraphics[scale=.65]{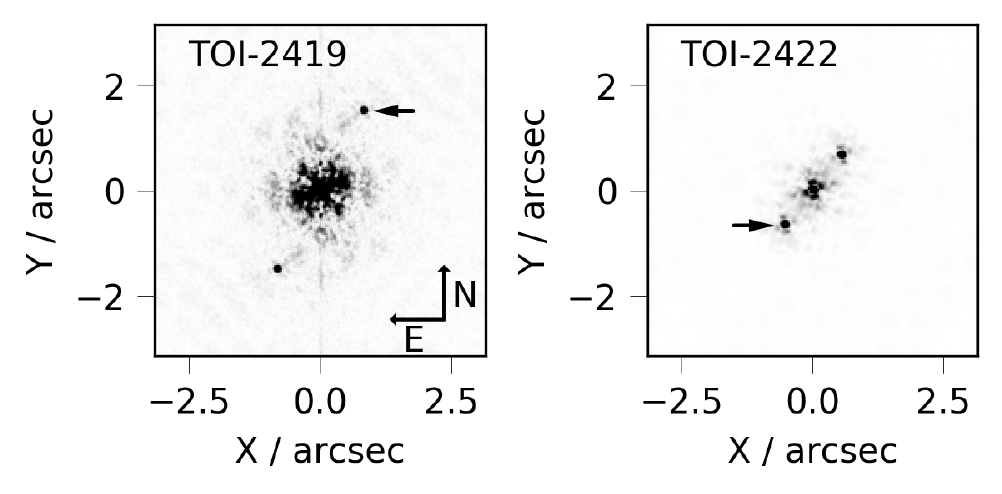}
    \caption{Similar to Figure \ref{fig:grid1}.}
    \label{fig:grid4}
\end{figure*}

\clearpage

\footnotesize

\textbf{Notes for Table \ref{tab:ticbinaries}. -- }Column (1) is the TOI number. Columns (2) and (3) is the TIC number for the primary and secondary stars. Columns (4) to (6) gives the measured separation, position angle, and $I$-band contrast from the SOAR observations. Columns (7) to (9) give the separation and position angle for the system derived from the TIC coordinates, and the TESS band contrast for each pair of stars.

\tabcolsep=0.12cm
\begin{longtable*}{ccc|ccc|ccc}
\caption{TIC matches to resolved binaries detected by SOAR \label{tab:ticbinaries}}\\
\hline
\hline
\noalign{\vskip 3pt}  
\text{TOI} & \text{TIC} & \text{TIC} & \multicolumn{3}{c}{SOAR}  & \multicolumn{3}{c}{TIC} \\ [0.1ex]
 &  Primary & Secondary & Sep. & P.A. & Contrast & Sep. & P.A. & Contrast  \\ [0.1ex]
  &  &  & (\arcsec) & ($^{\circ}$) & (mag)  & (\arcsec) & ($^{\circ}$) & (mag) \\ [0.1ex]
\hline
\noalign{\vskip 3pt}  
\endfirsthead
\multicolumn{9}{c}
{\tablename\ \thetable\ -- \textit{Continued}} \\
\hline \hline
\noalign{\vskip 3pt} 
\text{TOI} & \text{TIC} & \text{TIC} & \multicolumn{3}{c}{SOAR}  & \multicolumn{3}{c}{TIC} \\ [0.1ex]
 &  Primary & Secondary & Sep. & P.A. & Contrast & Sep. & P.A. & Contrast  \\ [0.1ex]
  &  &  & (\arcsec) & ($^{\circ}$) & (mag)  & (\arcsec) & ($^{\circ}$) & (mag) \\ [0.1ex]
\hline
\noalign{\vskip 3pt}  
\endhead
\endfoot
\hline
\endlastfoot
330 & 27966179 & 2051910904 & 2.7559 & 269.9 & 2.8 & 2.71 & 271.8 & 2.89 \\ 
398 & 1129033 & 632613066 & 3.3061 & 153.6 & 1.6 & 3.25 & 153.6 & 1.78 \\ 
479 & 306362738 & 708826792 & 2.2772 & 177.2 & 5.0 & 2.28 & 178.0 & 5.44 \\
830 & 281924357 & 804936024 & 2.8678 & 225.8 & 2.4 & 2.83 & 226.4 & 2.11 \\ 
862 & 309254930 & 806266868 & 0.9341 & 296.2 & 0.5 & 1.25 & 304.2 & 0.34 \\ 
898 & 124543547 & 752617891 & 1.2946 & 35.7 & 2.0 & 1.29 & 36.3 & 2.06 \\ 
922 & 278199349 & 764479189 & 3.3804 & 217.0 & 4.4 & 3.34 & 217.2 & 2.6 \\ 
925 & 300599466 & 764987221 & 2.682 & 205.0 & 2.1 & 2.67 & 205.2 & 1.89 \\ 
940 & 248434716 & 672067975 & 2.5473 & 131.7 & 1.8 & 2.52 & 135.7 & 2.53 \\ 
944 & 434234955 & 673618175 & 2.0189 & 111.9 & 2.0 & 2.0 & 112.1 & 1.9 \\ 
953 & 449050248 & 449050247 & 4.4222 & 288.7 & 0.2 & 4.38 & 288.8 & 0.22 \\ 
966 & 178367144 & 178367145 & 4.9007 & 138.7 & 0.8 & 4.85 & 139.2 & 0.77 \\ 
1032 & 146589986 & 865334360 & 1.3517 & 276.5 & 1.4 & 1.36 & 275.4 & 1.18 \\ 
1043 & 90448944 & 813212079 & 1.6637 & 332.8 & 2.5 & 1.66 & 332.6 & 2.48 \\ 
1048 & 384549882 & 384549880 & 4.706 & 256.5 & 3.5 & 4.7 & 256.9 & 3.42 \\ 
1057 & 323132914 & 804696171 & 1.8025 & 24.5 & 5.6 & 1.83 & 24.0 & 4.95 \\
1095 & 375223080 & 804852605 & 2.014 & 194.7 & 2.2 & 1.99 & 195.3 & 1.99 \\ 
1099 & 290348383 & 290348382 & 7.6379 & 310.5 & 2.3 & 7.6 & 306.1 & 2.3 \\ 
1101 & 271581073 & 271581074 & 3.5573 & 235.7 & 2.5 & 3.56 & 235.6 & 2.42 \\ 
1263 & 406672232 & 1943945558 & 2.6501 & 116.2 & 3.8 & 2.83 & 120.9 & 3.56 \\ 
1894 & 280865159 & 764547428 & 0.8258 & 42.0 & 0.0 & 1.25 & 37.7 & 0.26 \\ 
1937 & 268301217 & 766593811 & 2.4956 & 355.7 & 4.3 & 2.49 & 356.0 & 4.36 \\ 
1943 & 382980571 & 976026865 & 3.6514 & 114.3 & 4.1 & 3.64 & 114.2 & 3.55 \\
1990 & 457939414 & 851216237 & 3.7585 & 320.9 & 4.8 & 3.94 & 319.9 & 3.29 \\ 
1992 & 147340931 & 941340744 & 3.184 & 202.2 & 0.5 & 3.15 & 202.2 & 0.46 \\ 
2193 & 401604346 & 1988059412 & 1.885 & 124.0 & 3.8 & 1.87 & 124.1 & 4.22 \\ 
2195 & 24695044 & 630977959 & 3.3534 & 210.2 & 4.8 & 3.36 & 210.4 & 3.76 \\ 
2307 & 270219048 & 2028202009 & 2.9343 & 205.8 & 4.4 & 2.92 & 206.1 & 4.44 \\ 
2362 & 302924206 & 631367210 & 2.2124 & 259.0 & 2.0 & 2.19 & 259.2 & 1.84 \\ 
2417 & 49617263 & 631878531 & 1.8554 & 285.3 & 2.9 & 1.91 & 284.9 & 3.14 \\ 
2419 & 358248442 & 650660571 & 1.72 & 331.4 & 3.4 & 1.73 & 332.5 & 3.57 \\

\end{longtable*}

\textbf{Notes for Table \ref{tab:gaiamatches}. -- }Columns (1) is the TOI number and Column (2) is the TIC number. Column (3) and (4) is the Gaia DR2 source ID for the primary and secondary stars. Columns (5) to (7) gives the measured separation, position angle, and $I$-band contrast from the SOAR observations. Columns (8) to (10) give the separation and position angle for the system derived from the Gaia DR2 coordinates, and the Gaia $G-$band contrast for each pair of stars.

\footnotesize
\tabcolsep=0.12cm
\begin{longtable*}{cccc|ccc|ccc}
\caption{Gaia DR2 matches to resolved binaries detected by SOAR \label{tab:gaiamatches}}\\
\hline
\hline
\noalign{\vskip 3pt}  
\text{TOI} & \text{TIC} & \multicolumn{2}{c}{Gaia DR2 IDs}& \multicolumn{3}{c}{SOAR}  & \multicolumn{3}{c}{Gaia DR2} \\ [0.1ex]
 & &  Primary & Secondary & Sep. & P.A. & Contrast & Sep. & P.A. & Contrast  \\ [0.1ex]
  &  & &  & (\arcsec) & ($^{\circ}$) & (mag)  & (\arcsec) & ($^{\circ}$) & (mag) \\ [0.1ex]
\hline
\noalign{\vskip 3pt}  
\endfirsthead
\multicolumn{10}{c}
{\tablename\ \thetable\ -- \textit{Continued}} \\
\hline \hline
\noalign{\vskip 3pt} 
\text{TOI} & \text{TIC} & \multicolumn{2}{c}{Gaia DR2 IDs}& \multicolumn{3}{c}{SOAR}  & \multicolumn{3}{c}{Gaia DR2} \\ [0.1ex]
 & &  Primary & Secondary & Sep. & P.A. & Contrast & Sep. & P.A. & Contrast  \\ [0.1ex]
  &  & &  & (\arcsec) & ($^{\circ}$) & (mag)  & (\arcsec) & ($^{\circ}$) & (mag) \\ [0.1ex]
\hline
\noalign{\vskip 3pt}  
\endhead
\endfoot
\hline
\endlastfoot

330 & 27966179 & 2340807311474086272 & 2340807311474086016 & 2.7559 & 269.9 & 2.8 & 2.728 & 270.21 & 3.15 \\ 
398 & 1129033 & 5178405479960844160 & 5178405479961475712 & 3.3061 & 153.6 & 1.6 & 3.276 & 153.81 & 1.74 \\ 
479 & 306362738 & 2991284369063612928 & 2991284162905572992 & 2.2772 & 177.2 & 5.0 & 2.264 & 177.53 & 5.48 \\ 
830 & 281924357 & 5216790594823376640 & 5216790599118165504 & 2.8678 & 225.8 & 2.4 & 2.841 & 226.04 & 2.46 \\ 
862 & 309254930 & 5276025997794611968 & 5276025993495435392 & 0.9341 & 296.2 & 0.5 & 0.916 & 295.51 & 0.39 \\ 
898 & 124543547 & 3049803745151137536 & 3049803745147247616 & 1.2946 & 35.7 & 2.0 & 1.287 & 35.85 & 2.03 \\ 
922 & 278199349 & 5208281616071318400 & 5208281581710118912 & 3.3804 & 217.0 & 4.4 & 3.346 & 217.14 & 2.43 \\ 
925 & 300599466 & 5268832610469785856 & 5268832614766708992 & 2.682 & 205.0 & 2.1 & 2.664 & 205.22 & 1.87 \\ 
940 & 248434716 & 3211644018438077568 & 3211644018440408832 & 2.5473 & 131.7 & 1.8 & 2.518 & 133.47 & 2.84 \\ 
944 & 434234955 & 3311737823250737152 & 3311737823250737024 & 2.0189 & 111.9 & 2.0 & 2.0 & 112.14 & 1.72 \\ 
953 & 449050248 & 3285409673726608768 & 3285409669430237824 & 4.4222 & 288.7 & 0.2 & 4.387 & 288.85 & 0.23 \\ 
966 & 178367144 & 3070964117005132416 & 3070964117005860736 & 4.9007 & 138.7 & 0.8 & 4.862 & 138.92 & 0.86 \\ 
1032 & 146589986 & 5392041345156957696 & 5392041345153233920 & 1.3517 & 276.5 & 1.4 & 1.34 & 276.8 & 1.63 \\ 
1043 & 90448944 & 5321241283189288704 & 5321241283179461120 & 1.6637 & 332.8 & 2.5 & 1.652 & 332.97 & 2.57 \\ 
1048 & 384549882 & 5310970160975211008 & 5310970160975209728 & 4.706 & 256.5 & 3.5 & 4.677 & 256.7 & 3.88 \\ 
1057 & 323132914 & 5209299420242141952 & 5209299420242142080 & 1.8025 & 24.5 & 5.6 & 1.791 & 24.72 & 4.93 \\ 
1095 & 375223080 & 5215690640816366080 & 5215690640816365952 & 2.014 & 194.7 & 2.2 & 1.998 & 195.05 & 2.09 \\ 
1099 & 290348383 & 6356417496318028800 & 6356417496318028928 & 7.6379 & 310.5 & 2.3 & 7.601 & 310.43 & 2.86 \\ 
1101 & 271581073 & 6356211544046121728 & 6356211539750839552 & 3.5573 & 235.7 & 2.5 & 3.548 & 235.77 & 2.7 \\ 
1263 & 406672232 & 1818973354862632192 & 1818973354857904128 & 2.6501 & 116.2 & 3.8 & 2.685 & 117.66 & 3.44 \\ 
1894 & 280865159 & 5211978895716614400 & 5211978895718343168 & 0.8258 & 42.0 & 0.0 & 0.813 & 42.14 & 0.24 \\ 
1943 & 382980571 & 6054233458679730432 & 6054233458662893824 & 3.6514 & 114.3 & 4.1 & 3.626 & 114.43 & 3.96 \\ 
1990 & 457939414 & 5254512781523942912 & 5254512781505812736 & 3.7585 & 320.9 & 4.8 & 3.779 & 320.91 & 3.03 \\ 
1992 & 147340931 & 5387770498397701760 & 5387770498397701504 & 3.184 & 202.2 & 0.5 & 3.153 & 202.28 & 0.46 \\ 
2193 & 401604346 & 6373308503181838592 & 6373308503181838080 & 1.885 & 124.0 & 3.8 & 1.869 & 124.24 & 4.23 \\ 
2195 & 24695044 & 4644501668809042560 & 4644501668809042688 & 3.3534 & 210.2 & 4.8 & 3.344 & 210.34 & 4.19 \\ 
2231 & 100504381 & 6476174516109022464 & 6476174516107691776 & 1.5088 & 49.4 & 0.3 & 1.502 & 49.24 & 0.36 \\ 
2232 & 439444938 & 6698820294975508352 & 6698820294975508736 & 2.4207 & 239.8 & 0.1 & 2.401 & 239.92 & 0.1 \\ 
2307 & 270219048 & 6616873345462433152 & 6616873349756963456 & 2.9343 & 205.8 & 4.4 & 2.917 & 205.83 & 4.87 \\ 
2362 & 302924206 & 4701294983436539776 & 4701294983435599360 & 2.2124 & 259.0 & 2.0 & 2.194 & 259.18 & 1.98 \\ 
2417 & 49617263 & 4952839741811167616 & 4952839741810218880 & 1.8554 & 285.3 & 2.9 & 1.855 & 285.15 & 3.13 \\ 
2419 & 358248442 & 4730528077041899136 & 4730528077040687616 & 1.72 & 331.4 & 3.4 & 1.714 & 331.77 & 3.59 \\

\end{longtable*}

\tabcolsep=0.12cm
\begin{tiny}

\begin{ThreePartTable}
\begin{TableNotes}
\footnotesize

\end{TableNotes}

\begin{longtable*}{ccccccc}
\caption{Nearby stars detected by SOAR to false positive planet candidate hosts \label{tab:fp_binaries}}\\
\hline
\hline
\noalign{\vskip 3pt}  
\text{TOI} & \text{Separation} & \text{P.A.} & \text{Contrast} & \text{T$_{eff}$} & \text{Distance} & \text{Proj. Sep.} \\ 
 & \text{(\arcsec)} & (deg) & \text{($I$-band)} & \text{(K)} &  \text{(pc)} & \text{(AU)} \\ [0.1ex]
\hline
\noalign{\vskip 3pt} 
\endfirsthead

\multicolumn{7}{c}
{\tablename\ \thetable\ -- \textit{Continued}} \\
\hline \hline
\noalign{\vskip 3pt} 
\text{TOI} & \text{Separation} & \text{P.A.} & \text{Contrast} & \text{T$_{eff}$} & \text{Distance} & \text{Proj. Sep.} \\ 
 & \text{(\arcsec)} & (deg) & \text{($I$-band)} & \text{(K)} &  \text{(pc)} & \text{(AU)} \\ [0.1ex]
\hline
\noalign{\vskip 3pt}  
\endhead
\endfoot
\hline
\endlastfoot

342 & 0.3931 & 226.7 & 0.2 & 5800 & 508 & 199 \\ 
725 & 0.2479 & 185.0 & 1.4 & 6547 & 272 & 67 \\ 
881 & 0.1229 & 49.5 & 2.5 & 5274 & 994 & 122 \\ 
898 & 1.2946 & 35.7 & 2.0 & 5652 & 479 & 620 \\ 
922 & 3.3804 & 217.0 & 4.4 & 4120 & 1220 & 4124 \\ 
925 & 2.682 & 205.0 & 2.1 & 5027 & 1811 & 4857 \\ 
944 & 2.0189 & 111.9 & 2.0 & 7011 & 938 & 1893 \\ 
953 & 4.4222 & 288.7 & 0.2 & 5772 & 204 & 902 \\ 
962 & 2.8419 & 20.4 & 1.6 & 6151 & 447 & 1270 \\ 
968 & 0.2984 & 136.2 & 3.1 & 4968 & 1120 & 334 \\ 
979 & 0.0834 & 238.4 & 0.5 & 5806 & 414 & 34 \\ 
984 & 0.286 & 186.1 & 2.2 & 7773 & 442 & 126 \\ 
996 & 0.1391 & 230.1 & 1.6 & 8007 & 379 & 52 \\ 
1005 & 0.1763 & 358.1 & 0.3 & 5613 & 100 & 17 \\ 
1008 & 0.4882 & 309.1 & 0.3 & 6699 & 144 & 70 \\ 
1047 & 0.806 & 247.3 & 2.1 & 5622 & 134 & 108 \\ 
1048 & 4.706 & 256.5 & 3.5 & 11892 & 440 & 2070 \\ 
1061 & 0.3583 & 173.8 & 3.1 & 5525 & 133 & 47 \\ 
1093 & 0.0957 & 111.8 & 2.2 &  & 1710 & 163 \\ 
1093 & 3.0568 & 246.2 & 1.9 &  & 1710 & 5227 \\ 
1118 & 0.2003 & 132.0 & 2.4 & 4898 & 871 & 174 \\ 
1122 & 0.4746 & 35.2 & 0.1 & 6317 & 249 & 118 \\ 
1942 & 0.0741 & 32.3 & 2.2 & 5878 & 130 & 9 \\ 
1947 & 0.3454 & 235.8 & 1.8 & 4990 &  &  \\

\end{longtable*}
\end{ThreePartTable}
\end{tiny}

\tiny
\tabcolsep=0.2cm
\begin{longtable}{cccccccc}
\caption{SOAR speckle observations and resolved binary properties of known eclipsing binaries \label{tab:ebs}}\\
\hline
\hline
\noalign{\vskip 3pt}  
\text{TIC} & \text{Obs.} & \text{$\rho$} & $\theta$ & \text{$\Delta m$ } & Min. $\rho$ & \multicolumn{2}{c}{Limiting $\Delta m$} \\ [0.1ex]
 \hline
  & Year & (\arcsec) & (deg) & (mag) & (\arcsec) & 0\farcs15 & 1\arcsec \\ [0.1ex]
\hline
\noalign{\vskip 3pt}  
\endfirsthead
\multicolumn{8}{c}
{\tablename\ \thetable\ -- \textit{Continued}} \\
\hline \hline
\noalign{\vskip 3pt} 
\text{TIC} & \text{Obs.} & \text{$\rho$} & $\theta$ & \text{$\Delta m$ } & Min. $\rho$ & \multicolumn{2}{c}{Limiting $\Delta m$} \\ [0.1ex]
 \hline
  & Year & (\arcsec) & (deg) & (mag) & (\arcsec) & 0\farcs15 & 1\arcsec \\ [0.1ex]
\hline
\hline
\noalign{\vskip 3pt}  
\endhead
\endfoot
\hline
\endlastfoot

219406747 & 19.2063 &  &  &  & 0.064 & 3.10 & 5.47 \\ 
389669796 & 19.2063 &  &  &  & 0.109 & 2.05 & 2.93 \\ 
149975922 & 19.2064 &  &  &  & 0.085 & 2.47 & 3.95 \\ 
350862394 & 19.2091 &  &  &  & 0.064 & 2.92 & 5.57 \\ 
149989242 & 19.2065 &  &  &  & 0.064 & 2.80 & 5.24 \\ 
350954272 & 19.2090 &  &  &  & 0.064 & 2.81 & 5.07 \\ 
141712167 & 19.2013 & 1.4175 & 296.5 & 1.6 & 0.068 & 2.62 & 4.64 \\ 
141757920 & 19.2063 &  &  &  & 0.097 & 2.12 & 3.45 \\ 
150100627 & 19.2065 & 0.1077 & 108.8 & 1.9 & 0.064 & 2.07 & 5.03 \\ 
41232835 & 19.2064 &  &  &  & 0.075 & 2.32 & 4.02 \\ 
150101472 & 19.2065 & 0.1289 & 84.4 & 0.0 & 0.100 & 1.70 & 2.85 \\ 
260128333 & 19.2091 &  &  &  & 0.081 & 2.38 & 4.37 \\ 
41259805 & 19.2064 &  &  &  & 0.067 & 2.83 & 4.51 \\ 
141806292 & 19.2063 & 0.0456 & 10.0 & 0.0 & 0.102 & 1.76 & 3.15 \\ 
260131665 & 19.2090 &  &  &  & 0.039 & 3.41 & 6.14 \\ 
260160453 & 19.2091 &  &  &  & 0.064 & 2.96 & 5.64 \\ 
260161144 & 19.2091 & 1.9519 & 41.0 & 1.5 & 0.064 & 3.04 & 5.28 \\ 
260162387 & 19.2090 &  &  &  & 0.066 & 2.82 & 4.92 \\ 
41364284 & 19.2064 &  &  &  & 0.064 & 2.78 & 5.47 \\ 
150145456 & 19.2091 & 0.4125 & 87.4 & 2.1 & 0.054 & 2.09 & 4.80 \\ 
150145456 & 19.2091 & 0.4125 & 87.4 & 2.1 & 0.054 & 2.09 & 4.80 \\ 
150145456 & 19.2091 & 0.1119 & 124.6 & 1.6 & 0.054 & 2.09 & 4.80 \\ 
150145456 & 19.2091 & 0.1119 & 124.6 & 1.6 & 0.054 & 2.09 & 4.80 \\ 
260188537 & 19.2091 & 0.1973 & 217.4 & 2.7 & 0.064 & 2.36 & 5.47 \\ 
150162739 & 19.2064 &  &  &  & 0.071 & 2.21 & 3.78 \\ 
150165657 & 19.2091 &  &  &  & 0.064 & 3.14 & 5.46 \\ 
150166721 & 19.2065 &  &  &  & 0.064 & 2.80 & 5.27 \\ 
41483281 & 19.2064 &  &  &  & 0.090 & 2.06 & 3.54 \\ 
150187916 & 19.2065 &  &  &  & 0.064 & 2.82 & 5.58 \\ 
260305166 & 19.2091 &  &  &  & 0.064 & 3.07 & 5.54 \\ 
260304277 & 19.2090 &  &  &  & 0.064 & 2.77 & 5.53 \\ 
166969346 & 19.2064 &  &  &  & 0.067 & 2.49 & 4.32 \\ 
166969516 & 19.2064 &  &  &  & 0.105 & 1.80 & 3.04 \\ 
260353507 & 19.2090 &  &  &  & 0.064 & 2.94 & 5.58 \\ 
167007869 & 19.2064 &  &  &  & 0.064 & 2.82 & 5.14 \\ 
260352274 & 19.2090 & 2.3243 & 209.2 & 2.0 & 0.064 & 2.82 & 5.15 \\ 
167008868 & 19.2064 & 2.2949 & 278.3 & 2.0 & 0.063 & 2.64 & 5.17 \\ 
167009792 & 19.2064 & 1.9533 & 284.8 & 2.8 & 0.068 & 2.79 & 4.63 \\ 
150320620 & 19.2065 &  &  &  & 0.064 & 3.18 & 5.71 \\ 
150357290 & 19.2064 &  &  &  & 0.095 & 1.95 & 3.25 \\ 
260474813 & 19.2090 &  &  &  & 0.064 & 2.85 & 5.48 \\ 
167163906 & 19.2063 &  &  &  & 0.162 & 1.29 & 2.01 \\ 
150360766 & 19.2065 &  &  &  & 0.064 & 2.82 & 5.49 \\ 
150361803 & 19.2064 & 0.6299 & 123.8 & 1.8 & 0.121 & 1.62 & 2.58 \\ 
167163980 & 19.2063 &  &  &  & 0.097 & 2.00 & 3.24 \\ 
260502102 & 19.2091 &  &  &  & 0.072 & 2.69 & 5.05 \\ 
150429807 & 19.2065 &  &  &  & 0.070 & 2.71 & 4.49 \\ 
167205241 & 19.2063 &  &  &  & 0.064 & 2.70 & 5.04 \\ 
150442264 & 19.2065 &  &  &  & 0.064 & 2.87 & 5.72 \\ 
167304218 & 19.2064 &  &  &  & 0.064 & 2.76 & 5.34 \\ 
260640910 & 19.2091 & 2.4521 & 228.7 & 3.3 & 0.064 & 2.92 & 5.55 \\ 
293224065 & 19.2091 &  &  &  & 0.064 & 2.92 & 5.37 \\ 
142082796 & 19.2063 &  &  &  & 0.085 & 2.27 & 3.74 \\ 
293268667 & 19.2065 &  &  &  & 0.064 & 3.03 & 5.07 \\ 
167339584 & 19.2064 &  &  &  & 0.064 & 2.69 & 5.35 \\ 
167344197 & 19.2064 & 0.1384 & 348.5 & 2.0 & 0.064 & 1.99 & 5.08 \\ 
167344197 & 19.2064 & 0.1384 & 348.5 & 2.0 & 0.064 & 1.99 & 5.08 \\ 
167344197 & 19.2064 & 4.4322 & 174.0 & 7.4 & 0.064 & 1.99 & 5.08 \\ 
167344197 & 19.2064 & 4.4322 & 174.0 & 7.4 & 0.064 & 1.99 & 5.08 \\ 
293345927 & 19.2091 &  &  &  & 0.064 & 2.80 & 5.26 \\ 
278683641 & 19.2091 &  &  &  & 0.064 & 3.05 & 5.29 \\ 
176931266 & 19.2064 &  &  &  & 0.064 & 3.02 & 5.27 \\ 
375088647 & 19.2065 &  &  &  & 0.134 & 1.66 & 2.44 \\ 
167526485 & 19.2091 & 0.7385 & 192.6 & 4.8 & 0.064 & 2.88 & 5.40 \\ 
176934407 & 19.2064 &  &  &  & 0.064 & 2.85 & 5.07 \\ 
176958076 & 19.2064 &  &  &  & 0.069 & 2.56 & 4.25 \\ 
167554898 & 19.2091 & 0.7928 & 112.3 & 3.8 & 0.064 & 2.73 & 5.35 \\ 
167574282 & 19.2091 &  &  &  & 0.064 & 2.70 & 5.14 \\ 
167602738 & 19.2091 &  &  &  & 0.064 & 2.67 & 5.53 \\ 
167602025 & 19.2065 & 0.7377 & 132.2 & 0.0 & 0.134 & 1.69 & 2.40 \\ 
279053000 & 19.2091 &  &  &  & 0.064 & 2.83 & 5.48 \\ 
177022232 & 19.2064 &  &  &  & 0.064 & 2.66 & 5.35 \\ 
279088163 & 19.2090 &  &  &  & 0.064 & 2.74 & 5.25 \\ 
167692429 & 19.2065 &  &  &  & 0.147 & 1.56 & 2.15 \\ 
167691903 & 19.2065 &  &  &  & 0.079 & 2.57 & 4.03 \\ 
177035603 & 19.2064 & 0.0618 & 23.9 & 2.0 & 0.064 & 2.73 & 5.05 \\ 
177077475 & 19.2064 & 0.2324 & 94.4 & 2.6 & 0.107 & 1.86 & 3.06 \\ 
167722437 & 19.2065 &  &  &  & 0.064 & 2.71 & 5.44 \\ 
177082055 & 19.2064 &  &  &  & 0.064 & 2.69 & 5.38 \\ 
279245231 & 19.2090 & 0.5767 & 243.7 & 0.2 & 0.064 & 2.71 & 4.86 \\ 
177313167 & 19.2013 &  &  &  & 0.134 & 1.64 & 2.35 \\ 
177313167 & 19.2063 &  &  &  & 0.134 & 1.64 & 2.35 \\ 
177313167 & 19.2013 &  &  &  & 0.134 & 1.64 & 2.35 \\ 
177313167 & 19.2063 &  &  &  & 0.134 & 1.64 & 2.35 \\ 
177118923 & 19.2064 &  &  &  & 0.124 & 1.54 & 2.36 \\ 
167793961 & 19.2064 & 0.0772 & 102.8 & 1.1 & 0.084 & 1.99 & 3.66 \\ 
177175199 & 19.2091 &  &  &  & 0.064 & 2.63 & 5.15 \\ 
279431011 & 19.2091 &  &  &  & 0.064 & 2.67 & 5.60 \\ 
370236000 & 19.2091 &  &  &  & 0.064 & 2.75 & 5.40 \\ 
284196481 & 19.2091 &  &  &  & 0.064 & 2.88 & 5.30 \\ 
271554516 & 19.2091 &  &  &  & 0.064 & 2.81 & 5.41 \\ 
279569731 & 19.2091 &  &  &  & 0.064 & 2.80 & 5.11 \\ 
299897992 & 19.2091 & 0.1231 & 307.3 & 2.1 & 0.072 & 2.07 & 4.85 \\ 
279569718 & 19.2091 &  &  &  & 0.064 & 3.02 & 5.29 \\ 
299899924 & 19.2091 & 0.115 & 72.2 & 1.3 & 0.064 & 2.11 & 5.06 \\ 
348897766 & 19.2091 &  &  &  & 0.064 & 2.82 & 5.40 \\ 
348898049 & 19.2091 &  &  &  & 0.064 & 2.69 & 5.27 \\ 
279615956 & 19.2091 &  &  &  & 0.064 & 2.83 & 5.25 \\ 
348900258 & 19.2091 &  &  &  & 0.064 & 2.74 & 5.50 \\ 
300010961 & 19.2092 & 2.3703 & 21.3 & 2.6 & 0.064 & 2.89 & 5.18 \\ 
349055189 & 19.2091 &  &  &  & 0.064 & 2.73 & 5.29 \\ 
294092960 & 19.2090 &  &  &  & 0.071 & 2.59 & 4.82 \\ 
294092966 & 19.2091 &  &  &  & 0.064 & 2.54 & 5.39 \\ 
300033857 & 19.2091 &  &  &  & 0.064 & 2.77 & 5.40 \\ 
349059448 & 19.2091 &  &  &  & 0.064 & 2.76 & 5.38 \\ 
300039099 & 19.2092 &  &  &  & 0.064 & 2.81 & 5.16 \\ 
300039094 & 19.2092 & 0.6149 & 24.9 & 4.2 & 0.064 & 2.47 & 5.30 \\ 
300038935 & 19.2092 &  &  &  & 0.064 & 2.88 & 5.40 \\ 
349156098 & 19.2091 &  &  &  & 0.064 & 2.78 & 5.21 \\ 
349153143 & 19.2091 &  &  &  & 0.064 & 2.83 & 5.38 \\ 
294273900 & 19.2091 &  &  &  & 0.064 & 2.74 & 5.45 \\ 
271640350 & 19.2091 & 2.1325 & 197.1 & 2.1 & 0.064 & 2.79 & 5.24 \\ 
349270369 & 19.2091 &  &  &  & 0.064 & 2.94 & 5.27 \\ 
300161962 & 19.2092 & 0.6047 & 170.3 & 0.7 & 0.064 & 2.56 & 4.73 \\ 
339633702 & 19.2091 &  &  &  & 0.064 & 2.98 & 5.35 \\ 
349375972 & 19.2092 & 1.051 & 133.5 & 2.7 & 0.064 & 2.64 & 5.03 \\ 
349409844 & 19.2092 &  &  &  & 0.064 & 2.68 & 5.20 \\ 
300327482 & 19.2092 &  &  &  & 0.064 & 2.71 & 5.31 \\ 
349480507 & 19.2092 &  &  &  & 0.064 & 2.65 & 5.21 \\ 
300382665 & 19.2092 &  &  &  & 0.064 & 2.60 & 5.28 \\ 
300448625 & 19.2092 &  &  &  & 0.064 & 2.75 & 5.55 \\ 
300447314 & 19.2092 &  &  &  & 0.064 & 2.74 & 5.10 \\ 
339890862 & 19.2092 &  &  &  & 0.064 & 2.81 & 5.60 \\ 
349644606 & 19.2092 &  &  &  & 0.064 & 2.88 & 5.45 \\ 
300560295 & 19.2092 &  &  &  & 0.064 & 2.91 & 4.98 \\ 
349790953 & 19.2091 &  &  &  & 0.064 & 2.66 & 5.23 \\ 
349832824 & 19.2092 & 2.6525 & 114.5 & 5.0 & 0.064 & 2.64 & 4.59 \\ 
349832824 & 19.2092 & 0.3116 & 30.7 & 1.1 & 0.064 & 2.64 & 4.59 \\ 
349902873 & 19.2092 &  &  &  & 0.064 & 2.89 & 5.25 \\ 
349907707 & 19.2091 & 0.9013 & 196.8 & 2.3 & 0.064 & 2.82 & 5.37 \\ 
300654002 & 19.2092 &  &  &  & 0.064 & 2.56 & 5.22 \\ 
349972600 & 19.2091 &  &  &  & 0.064 & 2.71 & 5.37 \\ 
350027507 & 19.2092 & 3.99 & 283.9 & 2.2 & 0.063 & 2.28 & 5.54 \\ 
350094542 & 19.2092 &  &  &  & 0.064 & 2.58 & 5.00 \\ 
391892842 & 19.2092 & 0.707 & 195.7 & 3.5 & 0.064 & 2.33 & 4.94 \\ 
350144298 & 19.2092 &  &  &  & 0.064 & 2.74 & 5.03 \\ 
391891749 & 19.2092 & 0.7003 & 211.8 & 2.0 & 0.064 & 2.79 & 5.27 \\ 
391894459 & 19.2092 &  &  &  & 0.064 & 2.54 & 5.23 \\ 
350146296 & 19.2092 &  &  &  & 0.064 & 2.85 & 5.00 \\ 
262609754 & 19.2091 &  &  &  & 0.064 & 2.74 & 5.31 \\ 
382437243 & 19.2092 &  &  &  & 0.066 & 2.43 & 4.84 \\ 
300867734 & 19.2092 &  &  &  & 0.064 & 2.57 & 4.92 \\ 
382517745 & 19.2092 &  &  &  & 0.064 & 2.75 & 5.03 \\ 
300871376 & 19.2091 & 1.9442 & 206.8 & 0.3 & 0.064 & 2.69 & 4.94 \\ 
272286042 & 19.2091 &  &  &  & 0.064 & 2.62 & 5.16 \\ 
382575967 & 19.2092 &  &  &  & 0.064 & 2.76 & 5.27 \\ 
300971133 & 19.2092 &  &  &  & 0.064 & 2.63 & 5.02 \\ 
382577618 & 19.2092 &  &  &  & 0.064 & 2.73 & 5.10 \\ 
306470921 & 19.2092 &  &  &  & 0.064 & 2.58 & 5.11 \\ 
306508587 & 19.2092 &  &  &  & 0.066 & 2.59 & 4.78 \\ 
358459933 & 19.2094 &  &  &  & 0.064 & 2.62 & 5.20 \\ 
358511856 & 19.2092 &  &  &  & 0.064 & 2.57 & 5.00 \\ 
272357134 & 19.2091 &  &  &  & 0.064 & 2.73 & 5.29 \\ 
364398097 & 19.2092 &  &  &  & 0.064 & 2.69 & 5.09 \\ 
306580215 & 19.2092 & 3.0195 & 12.7 & 0.3 & 0.063 & 2.31 & 5.15 \\ 
410487677 & 19.2092 & 0.4721 & 358.5 & 3.8 & 0.064 & 2.54 & 5.02 \\ 
308397121 & 19.2092 &  &  &  & 0.064 & 2.80 & 5.03 \\ 
306669607 & 19.2091 &  &  &  & 0.064 & 2.78 & 5.15 \\ 
306740183 & 19.2091 &  &  &  & 0.072 & 2.72 & 4.95 \\ 
308537791 & 19.2092 &  &  &  & 0.064 & 2.90 & 5.25 \\ 
308454245 & 19.2092 &  &  &  & 0.064 & 2.71 & 5.07 \\ 
306773020 & 19.2092 &  &  &  & 0.066 & 2.49 & 4.77 \\ 
306742226 & 19.2092 & 2.123 & 51.1 & 3.3 & 0.056 & 2.57 & 4.86 \\ 
306742226 & 19.2092 & 0.2247 & 45.6 & 1.9 & 0.056 & 2.57 & 4.86 \\ 
308746785 & 19.2092 &  &  &  & 0.064 & 2.95 & 5.28 \\ 
308852608 & 19.2092 & 0.0858 & 222.8 & 0.6 & 0.057 & 2.11 & 4.87 \\ 
308852618 & 19.2092 &  &  &  & 0.064 & 2.58 & 4.95 \\ 
308851582 & 19.2092 & 1.2385 & 206.2 & 5.1 & 0.064 & 2.84 & 4.97 \\ 
287428184 & 19.2091 &  &  &  & 0.064 & 2.76 & 5.26 \\ 
308991822 & 19.2092 &  &  &  & 0.064 & 2.73 & 4.97 \\ 
309146836 & 19.2092 & 4.0865 & 69.9 & 2.2 & 0.063 & 2.67 & 6.02 \\ 
307490686 & 19.2092 &  &  &  & 0.064 & 2.67 & 5.27 \\ 
287773752 & 19.2091 &  &  &  & 0.064 & 2.53 & 5.28 

\end{longtable}

\begin{longrotatetable}
\textbf{Notes for Table \ref{tab:gaiabinaries}. -- }Column (1) is the TOI number and Column (2) is the TIC number. Columns (3) and (4) is the TIC number for the primary and secondary stars. Columns (5) and (6) gives the distance from the star from \citet{bailerjones18}. Columns (7) to (8) give the proper motion in Gaia DR2 for the primary and secondary stars. Column (9) gives the on-sky separation based on the Gaia DR2 coordinates, and Column (10) gives the projected physical separation using the average of the distances to the two stars. Column (11) gives the Gaia $G$-band contrast of the stellar pair.

\footnotesize
\tabcolsep=0.12cm

\begin{longtable*}{ccccccccccc}
\caption{Gaia DR2 binaries to TESS targets not detected by SOAR \label{tab:gaiabinaries}}\\
\hline
\hline
\noalign{\vskip 3pt}  
\text{TOI} & \text{TIC} & \multicolumn{2}{c}{Gaia DR2 ID}  & \multicolumn{2}{c}{Distance} & \multicolumn{2}{c}{Proper motion} & \text{Sep.} & \text{Projected sep.} & \text{Gaia contrast} \\ [0.1ex]
 &  & Primary & Secondary & Primary & Secondary & Primary & Secondary &  &  &  \\ [0.1ex]
 \hline
  &  &  &  & (pc) & (pc)  & (mas/yr)  & (mas/yr)  & (\arcsec) & (AU) & (mag) \\ [0.1ex]
\hline
\noalign{\vskip 3pt}  
\endfirsthead
\multicolumn{11}{c}
{\tablename\ \thetable\ -- \textit{Continued}} \\
\hline \hline
\noalign{\vskip 3pt} 
\text{TOI} & \text{TIC} & \multicolumn{2}{c}{Gaia DR2 ID}  & \multicolumn{2}{c}{Distance} & \multicolumn{2}{c}{Proper motion} & \text{Sep.} & \text{Projected sep.} & \text{Gaia contrast} \\ [0.1ex]
 &  & Primary & Secondary & Primary & Secondary & Primary & Secondary & (\arcsec) & (AU) & (mag) \\ [0.1ex]
\hline
\noalign{\vskip 3pt}  
\endhead
\endfoot
\hline
\endlastfoot

106 & 38846515 & 4675352109658261376 & 4675352109658261120 & 364.4 & 363.4 & 10.9 & 9.9 & 3.96 & 1443.0 & 6.88 \\ 
110 & 281459670 & 4906145613282734208 & 4906145608987769216 & 342.1 & 349.5 & 22.8 & 23.2 & 7.41 & 2535.0 & 6.99 \\ 
221 & 316937670 & 6495299760663617664 & 6495299764959269632 & 50.2 & 50.3 & 306.2 & 303.7 & 8.3 & 416.7 & 0.25 \\ 
241 & 77031414 & 2323985539482908416 & 2323985535188372480 & 233.0 & 233.6 & 88.4 & 87.9 & 6.12 & 1426.0 & 0.59 \\ 
404 & 166833457 & 4859136199796131200 & 4859136195500112256 & 281.7 & 296.3 & 35.8 & 35.7 & 12.23 & 3445.2 & 7.29 \\ 
418 & 178284730 & 5094154336330994688 & 5094154336332482176 & 117.6 & 116.8 & 27.6 & 27.1 & 7.24 & 851.4 & 2.37 \\ 
488 & 452866790 & 3094290054327367168 & 3094290019967631360 & 27.4 & 27.5 & 554.7 & 551.8 & 49.26 & 1349.7 & 3.48 \\ 
489 & 455096220 & 3094733088793018240 & 3094733088793018496 & 507.0 & 517.7 & 12.1 & 11.9 & 9.02 & 4573.1 & 5.29 \\ 
490 & 455135327 & 3096441729861716224 & 3096441729861715968 & 214.0 & 211.8 & 29.5 & 30.4 & 3.83 & 819.6 & 4.91 \\ 
566 & 1133072 & 5645867968120705664 & 5645866490651959040 & 83.6 & 83.0 & 66.8 & 65.9 & 10.65 & 890.3 & 4.37 \\ 
567 & 13349647 & 5750936092375254016 & 5750936092376995840 & 386.3 & 352.3 & 9.6 & 9.3 & 4.87 & 1881.3 & 5.11 \\ 
578 & 423275733 & 5674618444832114304 & 5674618444832114560 & 730.9 & 731.3 & 21.9 & 21.9 & 5.11 & 3734.9 & 1.95 \\ 
732 & 36724087 & 3767281845873242112 & 3767281639714811648 & 22.0 & 22.1 & 421.9 & 422.2 & 15.79 & 347.4 & 2.64 \\ 
747 & 15445551 & 6077185317188247936 & 6077185317188247552 & 298.4 & 310.9 & 4.1 & 4.5 & 8.09 & 2414.1 & 2.05 \\ 
774 & 294301883 & 3603529272750802560 & 3603529277045332608 & 297.5 & 285.6 & 12.7 & 12.7 & 4.36 & 1297.1 & 5.85 \\ 
841 & 238932509 & 5500698107867793536 & 5500698103572577920 & 306.1 & 295.7 & 21.4 & 20.6 & 12.46 & 3814.0 & 6.98 \\ 
874 & 232025086 & 5554044522261661440 & 5554044724125040000 & 130.5 & 130.1 & 13.2 & 13.7 & 6.64 & 866.5 & 2.95 \\ 
896 & 102283403 & 3341977210515243648 & 3341977210515241856 & 156.1 & 158.3 & 10.8 & 11.4 & 22.42 & 3499.8 & 0.45 \\ 
951 & 449050247 & 3285409673726608768 & 3285409669430237824 & 204.3 & 206.3 & 43.6 & 43.3 & 4.39 & 896.9 & 0.23 \\ 
967 & 445586472 & 5295349601116616064 & 5295349601118468864 & 522.1 & 515.6 & 19.0 & 18.6 & 5.25 & 2741.0 & 5.74 \\ 
1017 & 182943944 & 5540514065315160576 & 5540514069621433728 & 407.0 & 394.2 & 14.8 & 14.4 & 8.51 & 3463.6 & 6.07 \\ 
1037 & 363260203 & 5249932525297323008 & 5249932525282268928 & 279.7 & 269.4 & 22.5 & 21.3 & 11.55 & 3230.5 & 8.73 \\ 
1047 & 370745311 & 5247292116826741504 & 5247291910673100416 & 134.9 & 135.5 & 42.8 & 42.5 & 14.38 & 1939.9 & 3.77 \\ 
1047 & 370745311 & 5247292116826741504 & 5247291945028054528 & 134.9 & 134.7 & 42.8 & 43.4 & 13.95 & 1881.9 & 6.03 \\ 
1052 & 317060587 & 6357524189130820992 & 6357524189130821376 & 129.8 & 127.9 & 63.4 & 63.9 & 11.51 & 1494.0 & 5.4 \\ 
1061 & 253990973 & 6715848255460271232 & 6715848255463929088 & 133.2 & 142.0 & 76.0 & 77.2 & 15.12 & 2014.0 & 10.01 \\ 
1108 & 295599256 & 6435813230961681152 & 6435813196604325120 & 417.1 & 389.6 & 15.2 & 15.7 & 7.72 & 3220.0 & 2.68 \\ 
1201 & 29960110 & 5157183324996790272 & 5157183324996789760 & 37.9 & 37.9 & 170.6 & 180.2 & 8.35 & 316.5 & 0.29 \\ 
1208 & 273985865 & 4618022851832408832 & 4618022851832408704 & 134.8 & 133.9 & 41.4 & 39.6 & 5.56 & 749.5 & 1.94 \\ 
1209 & 30037565 & 4654747845890497920 & 4654747880241586432 & 173.8 & 184.4 & 64.0 & 64.7 & 22.86 & 3973.1 & 10.71 \\ 
1220 & 319259194 & 4624368305235123328 & 4624368300939410432 & 273.7 & 281.0 & 16.9 & 17.0 & 7.23 & 1978.9 & 6.24 \\ 
1704 & 95129101 & 878303842617591296 & 878303838322225280 & 343.5 & 344.0 & 5.9 & 6.0 & 8.98 & 3084.6 & 3.27 \\ 
2009 & 243187830 & 2791782794564103808 & 2791782794564103680 & 20.5 & 20.5 & 501.0 & 504.7 & 9.41 & 192.9 & 4.6 \\ 
2215 & 425561347 & 6756796546685748864 & 6756796512326012160 & 70.9 & 70.6 & 10.2 & 10.4 & 62.61 & 4439.0 & 0.25 \\ 
2231 & 100504381 & 6476174516109022464 & 6476174477453442944 & 174.3 & 169.3 & 19.0 & 18.4 & 20.33 & 3543.5 & 8.37 \\ 
2233 & 421455387 & 6747949326363848192 & 6747949322068159360 & 193.0 & 193.2 & 24.9 & 25.5 & 4.47 & 862.7 & 3.92 \\ 
2246 & 349786918 & 5287259433217021568 & 5287260189131352960 & 226.1 & 224.6 & 53.3 & 53.5 & 5.85 & 1322.7 & 0.49 \\ 
2303 & 197959526 & 6500411184062140928 & 6500411257078392320 & 644.0 & 659.2 & 18.7 & 18.6 & 3.15 & 2028.6 & 5.02 \\ 
2325 & 152800085 & 6548157167260713344 & 6548157167261649792 & 223.9 & 218.1 & 19.5 & 19.6 & 4.52 & 1012.0 & 4.86 \\ 
2327 & 177308364 & 5262401743054129280 & 5262398788116631296 & 100.4 & 100.8 & 47.1 & 46.8 & 38.8 & 3895.5 & 4.04 \\ 
2340 & 29959761 & 5156648751892607488 & 5156648687468486784 & 179.0 & 190.7 & 103.9 & 103.0 & 9.98 & 1786.4 & 2.43 \\ 
2359 & 436102447 & 3333179085971026304 & 3333179090266494848 & 594.3 & 555.7 & 16.6 & 17.1 & 6.43 & 3821.3 & 2.54 \\ 
2374 & 439366538 & 6828814283414902912 & 6828814283414902784 & 134.7 & 135.2 & 33.4 & 34.0 & 22.35 & 3010.5 & 2.61 \\ 
2383 & 20897611 & 2365517907594715520 & 2365517907595582464 & 174.6 & 171.8 & 50.0 & 50.2 & 24.09 & 4206.1 & 0.56 \\ 
2409 & 321068176 & 4948043530355426560 & 4948043603371169152 & 187.6 & 189.3 & 27.0 & 26.9 & 23.84 & 4472.4 & 1.44 \\

\end{longtable*}
\end{longrotatetable}
\clearpage

\begin{longrotatetable}
\textbf{Notes for Table \ref{tab:whitelist}. -- }Columns (1) and (2) give the TOI and TIC numbers, respectively. Column (3) designates the components of the resolved binaries according to the WDS style (mostly 'AB'). This matters for resolved triple systems, indicating their hierarchy. The equatorial coordinates for J2000, in degrees, are given in Columns (4) and (5). Column (6) give the filter (mostly $I$, with a few targets observed also in $V$), Column (7) the date of the observation (in Julian years). For resolved binaries, Columns (8) and (10) give the position angle $\theta$ and the separation $\rho$, while Columns (9) and (11) contain estimates of the measurement errors in tangential ($\rho \sigma_\theta$) and radial ($\sigma_\rho$) directions, in mas. The measured magnitude difference $\Delta m$ is given on Column (12). Some targets have multiple measurements. For unresolved sources (single stars), the Columns (8) to (12) are empty. Flags for the photometry are provided in Column (13). These flags are ':' for a companion with a low signal-to-noise ratio, 'q' for an identified quadrant from the shift-and-add images, '*' if the photometry is corrected for anisoplanatism using the average image. The estimated resolution limit is listed in Column (14) for all stars; Columns (15) and (16) give the estimated maximum detectable $\Delta m$ at separations of 0\farcs15 and 1\arcsec.

\tiny
\tabcolsep=0.2cm
\begin{longtable}{cccccccccccccccc}
\caption{Full SOAR Speckle Observation List and Binary Properties \label{tab:whitelist}}\\
\hline
\hline
\noalign{\vskip 3pt}  
\text{TOI} & TIC & Comp. & \multicolumn{2}{c}{R.A., Dec. (J2000)} & Filter & \text{Obs.} & $\theta$ & $\rho*$$\sigma_{\theta}$ & $\rho$ & $\sigma_{\rho}$ & $\Delta m$ & Flag & Min. $\rho$ & \multicolumn{2}{c}{Limiting $\Delta m$} \\ [0.1ex]
 &  & & (deg) & (deg) &  & Year & ($^{\circ}$) & (mas) & (\arcsec) & (mas) & (mag) & & (\arcsec) & 0\farcs15 & 1\arcsec \\ [0.1ex]
\hline
\noalign{\vskip 3pt}  
\endfirsthead
\multicolumn{16}{c}
{\tablename\ \thetable\ -- \textit{Continued}} \\
\hline \hline
\noalign{\vskip 3pt} 
\text{TOI} & TIC & Comp. & \multicolumn{2}{c}{R.A., Dec. (J2000)} & Filter & \text{Obs.} & $\theta$ & $\rho*$$\sigma_{\theta}$ & $\rho$ & $\sigma_{\rho}$ & $\Delta m$ & Flag & Min. $\rho$ & \multicolumn{2}{c}{Limiting $\Delta m$} \\ [0.1ex]
 &  & & (deg) & (deg) &  & Year & ($^{\circ}$) & (mas) & (\arcsec) & (mas) & (mag) & & (\arcsec) & 0\farcs15 & 1\arcsec \\ [0.1ex]
\hline
\hline
\noalign{\vskip 3pt}  
\endhead
\endfoot
\hline
\endlastfoot

 0 & 1433 & BC & 36.86796 & -27.63592 & I & 2020.927 & 233.0 & 4.1 & 1.463 & 4.1 & 5.6 & * & 0.041 & 3.11 & 5.61 \\ 
 0 & 154671430 & --- & 122.31090 & -29.55236 & I & 2019.950 &  &  &  &  &  & &  0.041 & 1.89 & 3.95 \\ 
 0 & 1596 & AB & 153.81100 & -58.76625 & I & 2020.996 & 341.0 & 0.2 & 1.425 & 0.2 & 1.4 & & 0.041 & 2.53 & 3.25 \\ 
 0 & 1596 & BC & 153.81100 & -58.76625 & I & 2020.996 & 195.7 & 0.3 & 0.230 & 0.8 & 2.1 & & 0.041 & 2.53 & 3.25 \\ 
 0 & 220568520 & --- & 46.28538 & -62.85667 & I & 2019.857 &  &  &  &  &  & &  0.064 & 2.56 & 3.82 \\ 
 0 & 278956474 & --- & 101.42192 & -57.13811 & I & 2020.018 &  &  &  &  &  & : & 0.074 & 0.88 & 1.79 \\ 
 0 & 308538095 & --- & 120.85142 & -60.66550 & I & 2019.857 &  &  &  &  &  & &  0.072 & 2.45 & 3.81 \\ 
 0 & 4892 & AB & 128.45477 & -49.94047 & I & 2020.018 & 100.4 & 0.4 & 0.427 & 0.3 & 1.4 & q & 0.041 & 3.37 & 4.12 \\
\end{longtable}

\end{longrotatetable}

\end{document}